%% file: main.tex
\documentclass{lmcs}
\pdfoutput=1

\usepackage{silence}
\usepackage{lastpage}
\lmcsdoi{16}{1}{15}
\lmcsheading{}{\pageref{LastPage}}{}{}%
{Sep.~07,~2018}{Feb.~22,~2022}{}

\pdfoutput=1
\newif\ifonline\onlinetrue%
\onlinefalse%
\newif\ifLONG\LONGtrue%
\newif\iflmcs\lmcstrue%
\newif\ifANONYMOUS\ANONYMOUSfalse%

\input{macros}
\ifonline%
\renewcommand{\comment}[1]{}
\fi

\usepackage[overlay,absolute,
]{textpos}
\usepackage{url}
\usepackage{xcolor}
\xdefinecolor{bisque}{rgb}{1,.894,.77}
\xdefinecolor{darkblue}{rgb}{.275,.51,.705}
\xdefinecolor{darkred}{rgb}{.648,.165,.165}
\usepackage{alltt}
\usepackage{amssymb}
\usepackage{stmaryrd}
\usepackage{bbold}
\usepackage{xspace}
\usepackage{tikz}

\pdfoutput=1

\begin{document}
\DeclareFontShape{OT1}{cmtt}{bx}{n}{
  <5><6><7><8><9><10><10.95><12><14.4><17.28><20.74><24.88>cmttb10}{}

\interfootnotelinepenalty=10000

\title[Covariance and Contravariance: a fresh look at an old issue]{Covariance and Contravariance:\texorpdfstring{\\}{} a fresh look at an old issue\texorpdfstring{\\[0.5em]}{}
{\normalfont{(a primer in advanced type systems \texorpdfstring{\\}{}for learning functional programmers)}}}

\author[G. Castagna]{Giuseppe Castagna}
\address{CNRS, Institut de Recherche en Informatique Fondamentale, Universit\'e de Paris, France}

\ifonline
\iflmcs%
\begin{textblock}{0.9}[0.5,0.5](.500,0){\color{gray}
\begin{center}
\Large\sf\noindent UNFINISHED DRAFT:\@ this text is made available only for teaching purposes.\\[-0.7mm] Please do not redistribute and contact the author if you want to reuse it.
\end{center}
}\end{textblock}
\else
\begin{textblock}{0.7}[0.5,0.5](.613,.12){\color{gray}
\begin{center}
\Large\sf\noindent UNFINISHED DRAFT:\@ this text is made available only for teaching purposes.\\[-2mm] Please do not redistribute and contact the author if you want to reuse it.
\end{center}
}\end{textblock}
\fi
\fi
\begin{abstract}
\input{abstract}
\end{abstract}


\maketitle

\section{Introduction}\label{intro}
\input{intro}

\input{plan}

\section{Types primer for the learning programmer}\label{primer}
\input{primer}

\section{The covariance and contravariance (false) problem}\label{covcon}
\input{covcon}

%

\section{Type algorithms for the language designer \break(i.e., the electrical blueprints)}\label{electric}

\input{electric}

\section{A roadmap to the theory that makes all this stuff work}\label{theory}
\input{theory}

\section{Philosophy and Lessons}\label{conclusion}
\input{conclusion}

\subsubsection*{Acknowledgements}
\input{acks}

\bibliographystyle{alpha}
\bibliography{main}
\appendix

\section{Exercise solutions}\label{exercises}
\input{exercises}

\end{document}

%% file: macros.tex
\newenvironment{enum}{\begin{enumerate}\vspace{-1pt}\topsep0pt\parskip0pt\partopsep0pt\itemsep1pt}{\end{enumerate}\vspace{-1pt}}
\def\{{\char123\relax}
\def\}{\char125\relax}
\newcommand{\perl}{}
\ifANONYMOUS%
\newcommand{\IC}{Castagna\xspace}
\newcommand{\Ih}{he\xspace}
\newcommand{\myt}{that\xspace}
\newcommand{\myh}{his\xspace}
\newcommand{\myC}{Castagna's\xspace}
\else
\newcommand{\IC}{I\xspace}
\newcommand{\Ih}{I\xspace}
\newcommand{\myt}{my\xspace}
\newcommand{\myh}{my\xspace}
\newcommand{\myC}{my\xspace}
\fi

\newcommand{\col}{\texttt{\,:\,}}
\newcommand{\pator}[2]{#1 \Or #2} 
\newcommand{\patand}[2]{#1 \And #2} 
\newcommand{\patrec}[1]{\texttt{\{}#1\texttt{\}}}
\newcommand{\patorec}[1]{\patrec{#1 \,\texttt{,}\,\textbf{..}}}
\newcommand{\toprec}{\patrec{\textbf{..}}}
\newcommand{\dom}[1]{\textsf{dom}(#1)}
\newcommand{\DNF}{\textsf{dnf}}
\newcommand{\rb}{\ensuremath{\rbrace}}
\newcommand{\lb}{\ensuremath{\lbrace}}
\newcommand{\p}[1]{\texttt{#1}}
\renewcommand{\k}[1]{{\tt\textbf{\color{darkblue}#1}}}
\newcommand{\s}[1]{{\tt{\color{darkred}#1}}}
\newcommand{\zero}{\ensuremath{\mathbb{0}}}
\newcommand{\one}{\ensuremath{\mathbb{1}}}
\newcommand{\bdd}[3]{\ensuremath{#1\texttt{?}#2\texttt{:}#3}}
\newcommand{\tdd}[4]{\ensuremath{#1\texttt{?}#2\texttt{:}#3\texttt{:}#4}}
\newcommand{\sem}[1]{\llbracket#1\rrbracket}
\newcommand{\C}{\texttt{C}}
\newcommand{\D}{\texttt{D}}
\renewcommand{\S}{\texttt{S}}
\newcommand{\Sone}{\ensuremath{\texttt{S}_1}}
\newcommand{\Stwo}{\ensuremath{\texttt{S}_2}}
\newcommand{\stag}{_{\texttt{\tiny tags}}}
\newcommand{\sint}{_{\texttt{\tiny ints}}}
\newcommand{\sprod}{_{\texttt{\tiny prod}}}
\newcommand{\sarrw}{_{\texttt{\tiny arrw}}}
\newcommand{\T}{\texttt{T}}
\newcommand{\R}{{\texttt{R}}}
\newcommand{\Tone}{\ensuremath{\texttt{T}_1}}
\newcommand{\Ttwo}{\ensuremath{\texttt{T}_2}}
\newcommand{\U}{\texttt{U}}
\newcommand{\Sub}{\texttt{<:}}
\newcommand{\To}{\texttt{-{}->}}
\newcommand{\Any}{\texttt{Any}}
\newcommand{\Empty}{\texttt{Empty}}
\newcommand{\Bool}{\texttt{Bool}}
\newcommand{\Nat}{\texttt{Nat}}
\newcommand{\Even}{\texttt{Even}}
\newcommand{\Odd}{\texttt{Odd}}
\newcommand{\Int}{\texttt{Int}}
\newcommand{\Point}{\texttt{Point}}
\newcommand{\ColPoint}{\texttt{ColPoint}}
\newcommand{\Or}{\ensuremath{\texttt{|}}}
\newcommand{\IV}[2]{\texttt{[#1..#2]}}
\newcommand{\pair}[2]{\texttt{(}#1\,\texttt{,}\,#2\texttt{)}} 
\newcommand{\spair}[1]{\texttt{\scriptsize(}\mathtt{#1}_1\texttt{\scriptsize,}\,\mathtt{#1}_2\texttt{\scriptsize)}} 
\newcommand{\sarrow}[1]{\mathtt{#1}_1\,\texttt{\scriptsize-{}->}\,\mathtt{#1}_2} 
\newcommand{\Not}[1]{\texttt{not(}#1\texttt{)}} 
\newcommand{\true}{\texttt{true}}
\newcommand{\false}{\texttt{false}}
\newcommand{\cduce}{$\mathbb{C}$Duce\xspace}
\newcommand{\comment}[1]{[{\textcolor{gray}{\emph{#1}}}]}
\newcommand{\minus}{\texttt{\symbol{92}}}
\newcommand{\eqdef}{\stackrel{\textrm{\tiny def}}{=}}
\iflmcs%
\renewcommand{\tag}[1]{\texttt{`#1}}
\renewcommand{\And}{\ensuremath{\texttt{\&}}}
\else
\newcommand{\eqref}[1]{(\ref{#1})\xspace}
\newcommand{\tag}[1]{\texttt{`#1}}
\newcommand{\And}{\ensuremath{\texttt{\&}}}
\fi


%% file: abstract.tex
Twenty years ago, in an article titled ``Covariance and
contravariance: conflict without a cause'', \IC argued that covariant
and contravariant specialization of method parameters in
object-oriented programming had different purposes and deduced that,
not only they could, but actually they should both coexist in the
same language.

In this work I reexamine the result of that article in the light of
recent advances in (sub-)typing theory and programming languages,
taking a fresh look at this old issue.

Actually, the revamping of this problem is just an excuse for writing
an essay that aims at explaining sophisticated type-theoretic concepts,
in simple terms and by examples, to undergraduate computer science
students and/or willing functional programmers.

Finally, I took advantage of this opportunity to describe some
undocumented advanced techniques of type-systems implementation
that are known only to few insiders that dug in the code of some compilers: therefore,
even expert language designers and implementers may find this work worth of
reading.

This is a corrected version of the paper \href{https://arxiv.org/abs/1809.01427v7}{\tt arXiv:1809.01427v7} published originally on Feb.\ 13, 2020.


%% file: intro.tex
Twenty years ago \IC wrote an article titled ``Covariance and contravariance: conflict without a cause''~\cite{Cas94} where \Ih argued that the heated debate that at the time opposed the faction of covariant overriding of methods in object-oriented languages against the congregation of the contravariant specialization had no real ground, since the two policies had different orthogonal purposes that not only could but actually should coexist in the same language. The article was, I would dare to say,
fairly successful even outside the restricted research circles. For instance, for many years at the entry ``\texttt{Contra-/Co-\,variance}'' of FAQ of the \texttt{comp.object} Usenet newsgroup (question 71) the answer was just a pointer to \myt article (actually, to  the tech-rep that preceded the publication). In spite of that, I think that the message of the article did not (or, at least, could not) reach practitioners and thus influence actual programming. Probably all that may have got to (some) programmers was that there was some theoretical paper that explained what each of covariance and contravariance was good for (with the associated reaction: ``\ldots so what?'').
One  reason for that, I think, is that at the time both type theory and programming languages were not developed enough to well explain the issue. I will not explain here again the whole article but the point is that in order to expose \myh argumentation \IC had to invent some constructions that did not look close to anything present in  programming languages at the time: in particular \Ih had to use weird ``overloaded types'' (these were sets of function types with two eerie formation conditions), write functions by using a strange ``\&''-infix notation, and even in their most simple form  functions had to distinguish two kinds of parameters by separating them by a vertical bar. I am sure that alone any of these oddity was enough to put off any serious programmer.

Twenty years have passed, both programming languages and, even more, type theory have much evolved to a point that I think it is now possible to explain covariance and contravariance to practitioners%
\ifLONG
, a task the first half of this article is devoted to.

\fi
To do that I will use the type theory of \emph{semantic subtyping}~\cite{FCB08}, while to illustrate all the examples I chose to  use the programming language \emph{Perl\,6}~\cite{perl6}, but an important aspect of this paper is that you can read it without any preliminary knowledge of them.

A reader aware of the theory of semantic subtyping may be astonished that I use such a theory to target practitioners. As a matter of fact, semantic subtyping is a sophisticated theory that relies on complex set-theoretic properties that, for sure, do not belong  to the education and practice background of most practitioners. The point is that while the underlying theory of types is probably out of reach of an average programmer, its types are very easy to use and understand for this programmer since, as I show in this paper, they can be explained in terms of very simple notions such as sets of values and set containment.
\ifLONG
It is like cars. Forty years ago most cars had such a simple conception that nearly everybody with some experience and few common tools could open the trunk and fix them.\footnote{~Though at that time I did not succeed to convince my parents to let me repair our family car (well, I was twelve).} Nowadays cars are so full of electronics that for many of them you must go to authorized dealers to have it repaired since this is out of reach for generic repairers. However, all this complexity is hidden to the end-user, and cars today are much simpler to drive than they were forty years ago. So it is for type systems, whose definitions are getting more and more involved but (in several cases) they are getting simpler and simpler for the programmer to use.
\fi

For what concerns the Perl\,6 language, I am not a great user or
supporter of it (the examples I give here count among the most
complicated programs I wrote in it). Although it probably does not
have the most elegant (and surely not the most streamlined) syntax I
ever saw in a programming language, I chose it because it has the
double advantage of having enough syntax to explain the
covariance/contravariance problem and of having a
fuzzy-yet-to-be-fixed type system. So while all the expressions I will
write can be run on any of the several Perl\,6 implementations
currently being developed, I will keep of Perl types just their syntax
(and with several liberties), and give to these types my very personal interpretation. Although Perl\,6 is not the
best candidate to present this paper (the perfect candidate would be
the programming language \cduce~\cite{BCF03,cduce} whose type system is
here borrowed and grafted on Perl) one of the challenges of this paper
was to use a mainstream programming language that was not designed with types in
mind---far from that---, whence the choice of Perl\,6. I am aware that
this choice will make some colleague researchers sneer
and some practitioners groan%
: please give me the benefit
of the doubt%
\ifLONG\ till Section~\ref{electric}.
\fi


%% file: plan.tex
\paragraph{Plan of the article.}
I organize the rest of the paper as if it were the documentation bundled with a TV set. When you buy a TV you want to read the instructions on how to tune channels and connect it to your Wii{\scriptsize\texttrademark}: these are in the user manual. You usually skip the electrical schemes that come with it, unless you want to repair it, or to build your own TV from them. Besides, if you are curious to know why these electrical schemes show you nightly news instead of exploding and killing everybody in the room, then you probably are a researcher and you need to read few articles and books on electronics and electromagnetism for which the bundled documentation is useless.

Section~\ref{primer} is my ``user manual'': it is a primer on types
for a Perl programmer who never used types (seriously) before.  There
I use (a personalized version of) Perl\,6 types to explain what types
are and how they are related in the ``semantic subtyping''
framework. Although most of the notions will be known to most of the
readers, the purpose of the presentation is to demonstrate that with
few easy-to-grasp key concepts, it is possible to bring programmers to
sophisticated reasoning about programs and types. Section~\ref{covcon}
applies the notions introduced in the primer to the covariance vs.\
contravariance issue. In that section I will present object-oriented
programming in Perl\,6, show why one can consider objects and classes
as syntactic sugar to define some ``multi subroutines'', explain the
issue of covariance and contravariance and their use in terms of these
multi subroutines.  Section~\ref{electric} is my ``electrical
blueprint''. It aims at language designers and implementers and
explains how  semantic subtyping can be efficiently implemented, by
describing the algorithms and data structures to be used to check type
inclusion and perform type assignment. Section~\ref{theory} is the one
that you never find in a TV documentation and it gives a roadmap to
the references that allow a researcher to understand the principles
underlying semantic subtyping and why the algorithms of the preceding
sections work: if you do not believe me and you need pointers to the
formal proofs, then this is the section you need to look at. In
Section~\ref{conclusion} I try to summarize the lessons that can be
drawn from this work and, in particular, what I learnt from writing
it. The primer section contains several exercises whose solutions are
given in appendix. All the article long I use (admittedly, too many) footnotes to complete
the main text with practical observations, missing definitions, and
technical considerations for the advanced reader; the last kind of
footnotes are marked by an asterisk and can be safely skipped (actually, on second thought, I suggest skipping all footnotes unless you are lost).

This article has several possible keys of reading; an obvious one is
that this paper can be taken as a proposal for a statically safe
type-system for the functional core of Perl 6 and a specification of
the algorithms to implement it.  However, the actual motivation to
write this article came from two distinct sources. The first
motivation is that semantic subtyping explains the
covariance/contravariance issue of~\cite{Cas94} in much a cleaner way and without
resorting to a somehow ``ad hoc'' formalism (as the
$\lambda\&$-calculus of~\cite{Cas94} was). So for a long time I have
been wanting to reframe my old work in terms of semantic
subtyping. But the motivation that spurred me to start writing it, is
that few years ago I started to teach an undergraduate course on
advanced programming at \'Ecole Normale Sup\'erieure de
Cachan. Formerly I had been teaching issues about covariance and
contravariance only at master level, and I then realized that while
the problem is still relevant, the original explanation was too
high-level for undergraduate students (i.e., it still contained too many
$\lambda$'s). The real target reader of this work is, thus, the
undergraduate student of an advanced programming course, and this
explains why all examples of this paper are written in a popular
language and it does not contain a single ``$\lambda$'', theorem, or
inference rule (just two formal definitions). So the not so hidden challenge tackled by this work is
to explain sophisticated type theoretic concepts to a willing
functional programmer. These two motivations also explain why I
consider this work as both a \emph{theoretical} and
an \emph{educational} pearl, in the sense of the ICFP and POPL conferences call
for papers.


%% file: primer.tex
\subsection{What is a type?}\label{type}

If you know what a value is (i.e., any result that can be returned by an expression), then you know what a type is: it is a set of values (though not all sets of values are types\footnote{* Formally, this can be seen by considering cardinalities: types and values are inductively defined and, therefore, they form recursively enumerable sets, while the powerset of values is not.
  More prosaically, the set of all functions that when applied to, say, an integer formed by an odd number of digits return a palindrome integer, is a set of values that hardly seems to conform to the idea of a type.}) Then you also know what \emph{subtyping} is since it coincides with set containment: a type is a subtype of a second type if all values of/in the former type are values of/in the latter. What a type system does is to define a \emph{typing relation} that is a relation between expressions and types. This relation must have the property that an expression is given (i.e., it is related to) a type only if whenever it returns a value, then the value is of/in that type.

Let me give some more details. I will consider a very restricted syntax for Perl types (and actually some of these types were just proposed but never implemented) as described by the following grammar:%
\ifLONG
\footnote{~For the reader not aware of the notation that follows (called EBNF grammar), this is a standard way to define the syntax of some language (in our case, the language of types). It can be understood by reading the symbol ``::='' as ``is either'' and the symbol ``\texttt{|}'' as ``or''. This yields the following reading. ``A type \T\ \emph{is either} \Bool{} \emph{or} \Int{} \emph{or} \Any{} \emph{or} \pair{\T$_1$}{\T$_2$} (where   {\T$_1$} and {\T$_2$} are types) \emph{or} {\T$_1$}\Or{\T$_2$} (where   {\T$_1$} and {\T$_2$} are types) \emph{or} \ldots''. Notice that to define, say,  \pair{\T$_1$}{\T$_2$} we assumed to know what the types {\T$_1$} and {\T$_2$} that compose it are: this roughly corresponds to defining types \emph{by induction}.}
\fi
\[\T::=  \Bool\,\mid\,\Int\,\mid\,\Any\,\mid\,\pair\T\T \,\mid\,\T\Or\T\,\mid\,\T\And\T \,\mid\,\Not\T\,\mid\,\T\To\T\]
What does each type mean? To define the meaning of a type, we define the set of values it denotes. Since the types above are defined inductively we can define their precise meaning by induction, namely:
\begin{description}
  \item[\underline\Bool] denotes the set that contains just two values \lb\true\,,\,\false\rb%
  \item[\underline\Int]  denotes the set that contains all the numeric constants: \lb\texttt{0, -1, 1, -2, 2, -3,\ldots{}}\rb.
  \item[\underline\Any] denotes the set that contains all the values of the language.
  \item[\underline{$\pair{\T_1}{\T_2}$}] denotes the set that contains all the possible pairs $(v_1,v_2)$ where $v_1$ is a value in $\T_1$ and $v_2$ a value in $\T_2$, that is $\lb (v_1,v_2)\mid v_1\in\T_1\,,\,v_2\in\T_2\rb $.
  \item[\underline{$\T_1\Or\,\T_2$}] denotes the \emph{union} of the sets denoted by $\T_1$ and $\T_2$, that is the set $\lb v\mid v\in\T_1\textrm{ or } v\in\T_2\rb $
  \item[\underline{$\T_1\And\,\T_2$}] denotes the \emph{intersection} of the sets denoted by $\T_1$ and $\T_2$, that is the set $\lb v\mid v\in\T_1\textrm{ and } v\in\T_2\rb $.\medskip
  \item[\underline{\Not\T}] denotes the set of all the values that are not in the set denoted by \T, that is $\lb v\mid v\not\in\T\rb $. So in particular \Not\Any denotes the empty set.
  \item[\underline{$\T_1\To\T_2$}] is the set of all function values that when applied to a value in $\T_1$, if they return a value (i.e., if the application does not loop), then this value is in $\T_2$.
\end{description}
Of course, the last case is the most delicate one and deserves more explanation. First of all we must define what a ``functional value'' is. A functional value is a (closed%
\ifLONG
\footnote{~A function is closed if all the variables that appear in its body are either the parameters of the function or they have been defined inside the body}%
\fi
) expression that defines a function. In $\lambda$-calculus we would say it is a lambda-abstraction.\footnote{~Lambda-abstractions have been recently integrated in languages such as \texttt{C\#} and Java, where they are called ``lambda-expressions''.} In Perl\,6 it is any expression of the form ``\texttt{\tt\textbf{sub}} \verb+(+\emph{parameters}\verb+)+ \verb+{+ \emph{body}\verb+}+ ''. So for instance \texttt{\tt\textbf{sub}\,}\verb|(Int $x){ return $x + 1 }| is a Perl\,6 value that denotes the successor function.\footnote{~In Perl variables are prefixed by the dollar sign, and \texttt{sub} is the apocope for \texttt{sub}routine.} Functions can be named (which will turn out to be quite handy in what follows) such as for \texttt{\tt\textbf{sub}} \verb|succ(Int $x){ return $x + 1 }| which gives the name \p{succ} to the successor function. It is easy to see that \texttt{succ} is a value in $\Int\To\Int$: it suffices to apply the definition I gave for arrow types, that is, check that when \texttt{succ} is applied to a value in \Int, it returns a value in \Int{} (\texttt{succ} is total).\footnote{* Notice that the definition of the set of values denoted by an arrow type does not depend on the inputs on which a function diverges. As a consequence, the function that diverges on all inputs belongs to every arrow type. This is the reason why the set denoted by an arrow type (or by any intersection of arrow types) is never empty: it always contains the function that diverges on all arguments.}

\subsection{What is a subtype?}\label{subtype}
Perl\,6 provides the constructor \p{subset} to define types such as \Nat, \Even, and \Odd, respectively denoting the set of natural, even, and odd numbers. For instance, the last two types can be defined as follows (where \texttt{\%} is in Perl\,6 the modulo operator):
\begin{alltt}
   \k{subset} Even \k{of} Int \k{where} \{ $_ % 2 == 0 \}
   \k{subset} Odd  \k{of} Int \k{where} \{ $_ % 2 == 1 \}
\end{alltt}
(e.g., the first is Perl syntax for $\textit{Even}=\lbrace x\in\Int\mid x \;\mathbf{ mod }\; 2 = 0\rbrace$).
Both these types are subsets of \Int.\footnote{~More precisely, I should have said ``they denote subsets'', but from now on I identify types with the sets of values they denote.} Let us use these types to show that a same value may belong to different types. For instance, it is easy to see that \texttt{succ} is also a value in \Even\To\Int: if we apply \texttt{succ} to a even number, it returns a value in \Int. Since \texttt{succ} is a value both in  \Even\To\Int{} and in  \Int\To\Int, then, by definition of intersection type, it is a value in  (\Even\To\Int)\,\And\,(\Int\To\Int).  This is true not only for \texttt{succ} but also for all values in \Int\To\Int: whenever we apply a function in \Int\To\Int{} to an even number, if it returns a value, then this value is in \Int. We say that \Int\To\Int{} is a \emph{subtype} of \Even\To\Int{} and write  it as \Int\To\Int\,\Sub\,\Even\To\Int. So, as I informally said at the beginning of the section, subtyping is just set containment on values.
\begin{defi}[\textbf{Subtype}]\label{subtyping}
A type $\T_1$ is a \emph{subtype} of a type $\T_2$, written $\T_1\Sub\T_2$, if all values in $\T_1$ are values in $\T_2$.
\end{defi}
\noindent
While the subtyping relation \Int\To\Int\Sub\Even\To\Int\ holds, the converse (i.e., \Even\To\Int\break\,\Sub\Int\To\Int) does not. For instance, the division-by-two function
\k{sub} \verb|(Int $x){ $x / 2 }|
\noindent is a value in \Even\To\Int{} but not in \Int\To\Int{} since when applied to an integer does not always return an integer (for
Perl's purists: notice that I omitted the \texttt{\textbf{return}}
keyword: since in Perl\,6 it is optional, in this section I will
systematically omit it). Actually, since we have that
\Int\To\Int\Sub\Even\To\Int, then stating that \texttt{succ} has type
(\Even\To\Int)\,\And\,(\Int\To\Int) does not bring any further
information: we are intersecting a set with a superset of it,
therefore this intersection type is \emph{equivalent to} (i.e., it denotes the
same set of values as) \Int\To\Int. To see an example of a meaningful intersection type, the reader can easily
check that the \texttt{succ} function is in
(\Even\To\Odd)\,\And\,(\Odd\To\Even): \texttt{succ} applied to
an even number returns an odd number (so it is in \Even\To\Odd) and
applied to an odd number it returns an even one (so it is also in
\Odd\To\Even). And this brings useful information since
(\Even\To\Odd)\,\And\,(\Odd\To\Even)\Sub\Int\To\Int{} is a
\emph{strict} containment, that is, there are values of the type on the right hand-side of ``\Sub'' that do not have the type on the left hand-side of the relation (e.g., the identity function is in
\Int\To\Int, but it has neither type \Even\To\Odd, nor type
\Odd\To\Even).\footnote{\label{forreviewer2}%
***$^{(!)}$ My fellow type theorists---but probably just them---may have noticed that I was a tad sloppy here. The theory I use in this paper gives expressions an \emph{intrinsic semantics} (in the sense of Reynolds~\cite{Rey03}) since the semantics of functions depends on the explicit type annotations of their parameters. So, strictly speaking, it is not the \p{succ} function I defined in the previous page that has type (\Even\To\Odd)\,\And\,(\Odd\To\Even) (since, its parameter is annotated by \Int, then the function has type \Int\To\Int\ldots and every supertype of it, but not the above intersection type) but, rather, a version of it explicitly annotated with the intersection type at issue. Likewise, the definition of the division-by-two function above is not well-typed and should rather be written as \k{sub} \p{(Even \$x)\{ \$x / 2 \}}. In particular, the intrinsic semantics explains why later in Section~\ref{2.3} the second version of \p{sumC} has type~\eqref{subex4} but not~\eqref{tre}.
}
The subtyping relation above holds not only for \Even{} and \Odd, but it can be generalized to any pair of arbitrary types. Since \Int{} is equivalent to \Odd\Or\Even{} (i.e., the two types denote the same set of values), then the last subtyping relation can be rewritten as
(\Even\To\Odd)\,\And\,(\Odd\To\Even) \Sub (\Odd\Or\Even{})\To(\Odd\Or\Even{}). This is just an instance of the following relation
\begin{equation}\label{sei}
(\S_1\To\T_1)\,\And\,(\S_2\To\T_2)~~\Sub~~(\S_1\Or\S_2)\To(\T_1\Or\T_2)
\end{equation}
that holds for all types, $\S_1$, $\S_2$, $\T_1$, and $\T_2$, whatever these types are. A value in the type on the left hand-side of~(\ref{sei}) is a function that if applied to an $\S_1$ value it returns (if any) a value in $\T_1$, and if it is applied to  a value in $\S_2$ it returns (if any) a value in $\T_2$. Clearly, this is also a function that when it is applied to an argument that is either in $\S_1$ or in $\S_2$, it will return a value either in  $\T_1$ or in $\T_2$ (from now on I will omit the ``if any'' parentheses). I leave as an exercise to the reader [EX1] the task to show that the inclusion is strict, that is, that there exists a value in the type on the right-hand side of~\eqref{sei} that is not in the type on left-hand side (see exercise solutions in the appendix).

This shows how some simple reasoning on values allows us to deduce sophisticated subtyping relations. By a similar reasoning we can introduce two key notions for this paper, those of covariance and of contravariance. These two notions come out as soon as one tries to deduce when a type  $\S_1\To\T_1$ is included in another function type $\S_2\To\T_2$. Let us first try to deduce it on a specific example: consider the function \p{\textbf{sub}}\verb| double(Int $x){ x+x }|. This function is of type $\Int\To\Even$. It is easy to see that \p{double} (but also any other function in $\Int\To\Even$)  is also a function in \Int\To\Int, since whenever it returns an \Even, it therefore returns an \Int. If \p{double} returns an \Int\ when applied to an \Int, then in particular \p{double} returns an \Int\ when it is  applied to, say, an \Odd{} number: that is, \p{double} is a function in $\Odd\To\Int$. Since this reasoning holds true for every function in \Int\To\Even, then we can deduce that  \Int\To\Even\Sub\Odd\To\Int. This relation holds not only for \Int, \Even, and \Odd, but also for every pair of types whose sub-components are in a similar relation. Every function value in  $\S_1\To\T_1$ transforms values in $\S_1$ into values in $\T_1$; therefore, it is also a function that transforms values in any type $\S_2$ smaller than $\S_1$ into values that are contained in $\T_1$ and, thus, in any type $\T_2$ greater than $\T_1$. Since these two conditions are also necessary for containment, then we have following rule\footnote{* For the advanced reader, in the presence of an empty type, the two conditions on the right-hand side of~\eqref{arrows} are sufficient but not necessary. If we use \Empty{} to denote \Not\Any---i.e., the empty type---, then the correct rule is: $(\S_1\To\T_1)\,\Sub\,(\S_2\To\T_2)$ if and only if $(\S_2\Sub\S_1\mbox{ and }\T_1\Sub\T_2)$ \emph{or} $(\S_2\Sub\Empty)$. Likewise in the presence of an empty type the right-hand side condition of~\eqref{prodsub} later on are not necessary: $\pair{\S_1}{\T_1}\,\Sub\,\pair{\S_2}{\T_2}$ if and only if $(\S_1\Sub\Empty)\mbox{ \emph{or} }(\T_1\Sub\Empty)\mbox{ \emph{or} }(\S_1\Sub\S_2 ~\mbox{ \emph{and} }~\T_1\Sub\T_2)$.}:
\begin{equation}\label{arrows}
(\S_1\To\T_1)\,\Sub\,(\S_2\To\T_2)~~~\Longleftrightarrow~~~ \S_2\Sub\S_1 ~\mbox{ and }~\T_1\Sub\T_2
\end{equation}
We notice that while the ``\T'' types have the same positions with respect to \Sub\ in both sides of equation~\eqref{arrows}, the ``\S'' types have a different position on the right and left sides of~\eqref{arrows}.
Borrowing the terminology from category theory, it is customary to say that the function type constructor ``\To'' is \emph{covariant} on the codomain type---since it preserves the direction of the \Sub{} relation---, and is \emph{contravariant} on the domain type---since it inverts the direction of the \Sub{} relation.

As a final remark, note that while in general it is easy to check whether a value is in some type (just apply the rules of Section~\ref{type}), it is usually quite harder to decide whether a type is contained into another. However, while a programmer should clearly master the former problem, she should be just mildly worried by the latter. Of course, she must be able to use in her programming practice some generic subtyping rules such as (\ref{arrows}) and at least the following (pretty straightforward) ones:
\begin{eqnarray}
  \pair{\S_1}{\T_1}\,\Sub\,\pair{\S_2}{\T_2}&\Longleftrightarrow&\S_1\Sub\S_2 ~\mbox{ and }~\T_1\Sub\T_2\label{prodsub}\\
\S_1\Sub\S_2 ~\mbox{ and }~\T_1\Sub\T_2 &\Longrightarrow& \S_1\And\T_1\,\Sub\,\S_2\And\T_2\\
\S_1\Sub\S_2 ~\mbox{ and }~\T_1\Sub\T_2 &\Longrightarrow& \S_1\Or\T_1\,\Sub\,\S_2\Or\T_2\\
\S\Sub\T &\Longleftrightarrow&\Not\T\Sub\Not\S\label{ruleneg}
\end{eqnarray}
However she is not required to be able to decide subtyping for every possible pair of types, since this requires a knowledge deeper than the few above rules (as an aside, notice that the rules~\eqref{arrows}--\eqref{ruleneg} are not enough to deduce the subtyping relation in~(\ref{sei}):
technically one says that the above set of rules provides a sound but
incomplete axiomatization of the subtyping relation\footnote{\label{extrarules}* To deduce such a kind of relations one should also use axioms such as the following ones (excerpted from~\cite{BDL95})
\begin{eqnarray}
(\S\To\T)\And(\S\To\U) &\Sub&\S\To\T\And\U\label{sette}\\[-1mm]
(\S\To\U)\And(\T\To\U) &\Sub&\S\Or\T\To\U\label{otto}
\end{eqnarray}
The reader may try as an exercise to prove that both the relations in~\eqref{sette}~\eqref{otto} above and their converse hold [EX2].}). The problem of
deciding subtyping between two generic types is a task for the
language designer and implementer. This is the one who must implement
the algorithms that not only check the containment of two generic
types, but also generate informative error messages that explain the programmer the reasons why some
containment she used does not hold.
\ifLONG
In Section~\ref{electric} I will
explain for the language designer how to do it.
\fi

\subsection{Intersections, overloading, and dispatching}\label{2.3}
The cognoscente reader will have recognized in the intersection of arrow types such as $(\S_1\To\T_1)\,\And\,(\S_2\To\T_2)$ the type typical of overloaded functions: a function of this type returns results of different types according to the type of its argument.  Strictly speaking, an overloaded function is a function that executes different code for arguments of different types: the semantics of the function is ``overloaded'' since formed of different semantics each corresponding to a different piece of code. The only function we have seen so far with an intersection type---i.e., \texttt{succ}---is not really ``overloaded'' since it always executes the same code both for arguments in \Even{} and for arguments in \Odd. In this case it would be more correct to speak of \emph{behavioral type refinement} since intersections of arrows do not correspond to different pieces of code (each implementing different functions all denoted by the same operator) but, instead, they provide a more precise description of the function behavior.\footnote{* Such a usage of intersection types is known---in my opinion, somehow misleadingly---as  \emph{coherent overloading}: see Section~1.1.4 in~\cite{PierceThesis}.} However Perl\,6 allows the programmer to define ``real'' overloaded functions by giving multiple definitions of the same function prefixed by the \p{multi} modifier:
\begin{alltt}
 \k{multi} \k{sub} sum(Int $x, Int $y) \{ $x + $y \}
 \k{multi} \k{sub} sum(Bool $x, Bool $y) \{ $x && $y \}
\end{alltt}
Here the \p{sum} function has two different definitions: one for a pair of integer parameters (in which case it returns their sum) and one for a pair of Boolean parameters (in which case it returns their logical ``and'' denoted in Perl by \verb|&&|). For instance, \texttt{sum(37,5)} returns \p{42} and \p{sum(True,False)} returns \p{False}. Clearly, the function \p{sum} above has the following type
\begin{equation}\label{uno}
(\pair{\Int}{\Int}\To\Int)\ \And\ (\pair\Bool\Bool\To\Bool),
\end{equation}
which states that if the function is applied to a pair of integers, then it returns integer results, and that if it is applied to a pair of Boolean arguments it returns Boolean results.
The actual code to be executed is chosen at run-time according to the type of the actual parameters.

Although in the case above both parameters are taken into account to select the code, it is clear that checking the type of just one parameter---e.g., the first one---would have sufficed. So I could have equivalently written the sum function in a \emph{curried}%
\footnote{~In mathematics and computer science, \emph{currying} is the technique of transforming a function that takes multiple arguments (or a tuple of arguments) in such a way that it can be called as a chain of functions, each with a single argument.} version I call \p{sumC}:
\begin{alltt}
 \k{multi} \k{sub} sumC(Int  $x)\{ \k{sub} (Int  $y)\{$x + $y \} \}
 \k{multi} \k{sub} sumC(Bool $x)\{ \k{sub} (Bool $y)\{$x && $y\} \}
\end{alltt}
The syntax above makes it clear that only the type of the parameter \p{\$x} is used to select the code, while the second parameter is just an argument to be passed to the function returned by the selection.
Perl\,6 provides a very handy double-semicolon syntax ``\texttt{\k{;;}}'' to distinguish parameters that are used for the selection of a multi-function (those on the left of ``\texttt{\k{;;}}'') from those that are just passed to the selected code (those on the right of ``\texttt{\k{;;}}''). So by using the syntactic sugar ``\texttt{\k{;;}}'', the code above can be ``equivalently'' rewritten as:\footnote{* Strictly speaking this and the previous definition of \p{sumC} are not equivalent since the first application is curried and therefore it must be fed by one argument at a time (e.g., \p{sumC(3)(39)}) while in the latter it is not, so it receives all the arguments at once (e.g., \p{sumC(3,39)}): the second notation forbids partial application. Here I just focus on the types and not on the particular syntax, so I will sweep such differences under the carpet.}

\begin{alltt}
 \k{multi} \k{sub} sumC(Int $x \texttt{\k{;;}} Int $y) \{ $x + $y \}
 \k{multi} \k{sub} sumC(Bool $x \texttt{\k{;;}} Bool $y) \{ $x && $y \}
\end{alltt}

The type of the \p{sum} function has changed since in both \p{multi} definitions (either curried or with``\texttt{\k{;;}}'')  \p{sumC} has type:
\begin{equation}\label{due}
({\Int}\To({\Int}\To\Int))\ \And\ (\Bool\To(\Bool\To\Bool)).
\end{equation}
The functions of this type are different from those of the type in~\eqref{uno}:
the type in (\ref{uno}) is the set of all functions that when applied to a pair of integers return an integer and when applied to a pair of Booleans they return a Boolean. The functions in the type of (\ref{due}) are functions that when applied to an integer return a function from integers to integers, and when applied to a Boolean they return a function from Booleans to Booleans.

Currying requires some care when all arguments are necessary to select the code to execute. Imagine that I wanted to add the code to handle the cases when the two arguments are of different types and that, therefore, I added two extra \p{multi} definitions for \p{sum}, which is then defined in the following (weird) recursive way:
\begin{alltt}
 \k{multi} \k{sub} sum(Int $x, Int $y) \{ $x + $y \}
 \k{multi} \k{sub} sum(Bool $x, Bool $y) \{ $x && $y \}
 \k{multi} \k{sub} sum(Bool $x, Int $y) \{ sum($x , $y>0) \}
 \k{multi} \k{sub} sum(Int $x, Bool $y) \{ sum($y , $x) \}
\end{alltt}
As a consequence \p{sum} has now the following type (the reader is invited to check the typing as an exercise [EX3])
\begin{equation}\label{quattro}
\begin{array}{rcl}
&&(\pair{\Int}{\Int}\To\Int)\\[-.4mm]
  &\And& (\pair{\Bool}{\Bool}\To\Bool)
  \\[-.4mm]
&\And& (\pair{\Bool}{\Int}\To\Bool)\\[-.4mm]
&\And& (\pair{\Int}{\Bool}\To\Bool).
\end{array}
\end{equation}
If, as we did with the previous version, we want define a variant of this function that \emph{dispatches} (i.e., performs the selection) only on the first argument, then every selection for the first argument must return a nested multi function that will dispatch on the second argument. That is, we are looking for a function \p{sumC} of the following type:
\begin{equation}\label{tre}
\begin{array}{rcl}
&&(\Int\;\To\;((\Int\To\Int)\ \And\ (\Bool\To\Bool)))\\[-.3mm]
&\And\!\!\!\!& (\Bool\To((\Bool\To\Bool)\ \And\ (\Int\To\Bool)))
\end{array}
\end{equation}
The type above is the type of functions that when they are applied to an
integer return another function of type
\texttt{(\Int\To\Int)\ \And\ (\Bool\To\Bool)} and that when they are
applied to a Boolean return another function of type
\p{(\Bool\To\Bool)\ \And\ (\Int\To\Bool)}. In other terms they are
functions that can be applied to a sequence of two arguments of type \Int\ or \Bool:
if the first argument is an integer, then the result of the double
application will be of the same type as the type of the second
argument; if the first argument is a Boolean, then also the result will
be a Boolean.\footnote{\label{ex4}~By equation~\eqref{otto} and the exercise in Footnote~\ref{extrarules}, the type in~\eqref{tre} is equivalent to:%
\[(\Int\;\To\;((\Int\To\Int)\ \And\ (\Bool\To\Bool)))\ \And\ (\Bool\To((\Bool\Or\Int)\To\Bool));\]
prove it as an exercise [EX4].
}

How is a function of the type in~\eqref{tre} defined? In particular how can we define \p{sumC}---the curried version of the second definition of \p{sum}---which should have this type? A first attempt consists in defining two auxiliary multi functions \p{sumI} and \p{sumB} that perform the dispatch on the second argument of \p{sumC} and that are called according to whether the first argument of \p{sumC} is an integer  (\p{sumI}) or a Boolean (\p{sumB}), respectively:
\renewcommand{\sp}{\hspace{1mm}}
\newcommand{\negvspace}{\vspace{-.8mm}}
\begin{alltt}
 \k{multi}\sp\k{sub}{\sp}sumI(Int{\sp}$y{\sp}\k{;;}{\sp}Int{\sp}$x)\{{\sp}$x{\sp}+{\sp}$y{\sp}\}
 \k{multi}{\sp}\k{sub}{\sp}sumI(Bool{\sp}$y{\sp}\k{;;}{\sp}Int{\sp}$x){\sp}\{{\sp}sumC($y,$x)\}\medskip
 \k{multi}{\sp}\k{sub}{\sp}sumB(Bool{\sp}$y{\sp}\k{;;}{\sp}Bool{\sp}$x){\sp}\{{\sp}{\sp}$x{\sp}&&{\sp}$y{\sp}\}
 \k{multi}{\sp}\k{sub}{\sp}sumB(Int{\sp}$y{\sp}\k{;;}{\sp}Bool{\sp}$x){\sp}\{{\sp}sumC($x,$y>0){\sp}\}\medskip
 \k{multi}{\sp}\k{sub}{\sp}sumC(Int{\sp}$x{\sp}\k{;;}{\sp}Int\Or{}Bool{\sp}$y){\sp}\{{\sp}sumI($y,$x)\,\}
 \k{multi}{\sp}\k{sub}{\sp}sumC(Bool{\sp}$x{\sp}\k{;;}{\sp}Int\Or{}Bool{\sp}$y){\sp}\{{\sp}sumB($y,$x)\,\}
\end{alltt}
In a nutshell, \texttt{sumC} calls \texttt{sum\emph{X}(\$y,\$x)} where \texttt{\emph{X}} is either \texttt{I} or \texttt{B} according to whether the first argument of \texttt{sumC} is of type \Int\ or \Bool. When called, each \texttt{sum\emph{X}} subroutine dispatches to the appropriate code according to the type of its first argument, that is, the second argument of the calling \texttt{sumC}.  This technique gives us a definition of \p{sumC} with the expected behavior but not the expected type. The first \p{multi} definition of \p{sumC} has type $\Int\;\To\;(\Int\Or\Bool)\To(\Int\Or\Bool)$ (notice in particular that the second argument has a union type), while the second \p{multi} definition of \p{sumC}  has type $\Bool\To(\Int\Or\Bool)\To\Bool$. Thus the whole \p{sumC} function has type (see also Footnote~\ref{forreviewer2}):
\begin{equation}\label{subex4}
\begin{array}{rcl}
&&(\Int\;\To\;(\Int\Or\Bool)\To(\Int\Or\Bool))\\[-.3mm]
&\And\!\!\!\!& (\Bool\To(\Int\Or\Bool)\To\Bool)\vspace{-.5mm}
\end{array}
\end{equation}
and this type is less precise than (i.e., it is a supertype of) the
type in~\eqref{tre} (see also the type in
Footnote~\ref{ex4} that is equivalent to the~\eqref{tre} type). For instance, the type in~\eqref{tre} states that
when a function of this type is applied to two arguments of type
\Int, then the result will be of type \Int, while the type in~\eqref{subex4} states, less precisely, that the result will be of type
\Int\Or\Bool.

The solution to define a version of \p{sumC} that has the type in~\eqref{tre} is not so difficult: instead of defining separate auxiliary functions (to which the second argument of \p{sumC} must be explicitly passed as an argument) I will use  nested \p{multi} subroutines. Unfortunately Perl\,6 does not allow the programmer to write anonymous \p{multi} functions. But if we suppose that anonymous \p{multi} functions were allowed (e.g., by considering anonymous \p{multi} definitions in the same block to define the same anonymous function---a suggestion to the designers of Perl), then the correspondence between the following definition of \p{sumC} and the type in~(\ref{tre}) should be, I think, pretty clear:
\iflmcs
\begin{alltt}
 \k{multi} \k{sub} sumC(Int $x)\{
         \k{multi} \k{sub} (Int $y) \{ $x + $y \}
         \k{multi} \k{sub} (Bool $y) \{ sumC($y)($x) \}
       \}\negvspace
                                                                       \refstepcounter{equation}{\rm(\theequation)}\label{mycode}\negvspace
 \k{multi} \k{sub} sumC(Bool $x)\{
         \k{multi} \k{sub} (Bool $y) \{ $x && $y \}
         \k{multi} \k{sub} (Int $y) \{sumC($x)($y>0)\}
       \}
\end{alltt}
\else
\begin{alltt}
 \k{multi} \k{sub} sumC(Int $x)\{
         \k{multi} \k{sub} (Int $y) \{ $x + $y \}
         \k{multi} \k{sub} (Bool $y) \{ sumC($y)($x) \}
       \}
                                               \refstepcounter{equation}{\rm(\theequation)}\label{mycode}
 \k{multi} \k{sub} sumC(Bool $x)\{
         \k{multi} \k{sub} (Bool $y) \{ $x && $y \}
         \k{multi} \k{sub} (Int $y) \{sumC($x)($y>0)\}
       \}
\end{alltt}
\fi
This code states that \texttt{sumC} is a multi-subroutine formed by two definitions each of which returns as result another (anonymous) multi-subroutine.%
\footnote{~\label{encoding}\perl Although the code~\eqref{mycode} cannot be executed in Perl\,6 because of the absence of anonymous multi-subroutines, it can be faked by non-anonymous multi-subroutines as follows:\\[1mm]
\begin{minipage}{\textwidth}{\tt
\hspace*{3mm}\k{multi} \k{sub} sumC(Int \$x)\{ \\[-.3mm]
\hspace*{20mm}             \k{multi} \k{sub} foo(Int \$y) \{ \$x + \$y \} \\[-.3mm]
\hspace*{20mm}             \k{multi} \k{sub} foo(Bool \$y) \{ sumC(\$y)(\$x) \}\\[-.3mm]
\hspace*{20mm}           \&foo\\[-.3mm]
\hspace*{16.5mm}           \}\\[1mm]
\hspace*{3mm}\k{multi} \k{sub} sumC(Bool \$x)\{\\[-.3mm]
\hspace*{20mm}             \k{multi} \k{sub} foo(Bool \$y) \{ \$x \&\& \$y \}\\[-.3mm]
\hspace*{20mm}             \k{multi} \k{sub} foo(Int \$y) \{sumC(\$x)(\$y>0)\}\\[-.3mm]
\hspace*{20mm}           \&foo\\[-.3mm]
\hspace*{16.5mm}           \}}
\end{minipage}
\\[1mm]
where \p{\&foo} is Perl's notation to return the subroutine named \p{foo}.
}

I have not done anything new here. As it will be clear later on, to define the code in~\eqref{mycode} I just applied the ``double dispatching'' technique proposed by Ingalls in 1986 at the first OOPSLA conference~\cite{Ingalls86}, though I doubt that anybody at the time saw it as a special case of currying in the presence of intersection types: as a matter of fact, it is not so
straightforward to see that the type in~\eqref{tre} is the curried version
of the type in~\eqref{quattro} (the code in~\eqref{mycode} is actually a ``proof''---in the sense of~\cite{typeiso91}---that currying is the relation between the type in (\ref{tre}) and the one in (\ref{quattro})).

\subsection{Formation rules for multi-subroutines}\label{formation}
In order to conclude the discussion about multi-subroutines I have to give some more details about how their ``dynamic dispatch'' is performed. When I first introduced multi-subroutines I said that the code to execute was chosen at run-time according to the type of the actual parameter: this is called \emph{dynamic dispatch}, since the argument of the function is \emph{dispatched} to the appropriate code \emph{dynamically}, that is, at run-time. However, in all the examples I showed so far delaying the choice of the code until run-time does not seem a smart choice since in all of them it is possible to choose the code at compile time according to whether the arguments are Booleans or Integers. In other terms, the examples I gave are not conceptually different from operator overloading as found in Algol 68 where it is resolved at static time (i.e., during the compilation). Of course, making the dispatch at run-time is computationally much more expensive than making it at compile time. Therefore, dynamic dispatch is useful and justified only if the selection of the code may give at run time a result different from the one that would be obtained at compile time, that is to say, if the type of function arguments may evolve during the computation. In statically-typed languages\footnote{~In a typed programming language, types can be checked either at compile-time or a run-time. Languages that check types at compile time are called statically-typed languages, while those that check types at run-time are said to be dynamically typed. A combination of static and dynamic type-checking is also possible. In this work I focus on how to define static type-checking for Perl, which currently is dynamically typed.} this happens in the presence of a subtyping relation: without  subtyping an expression in a statically-typed language must have the same type the whole computation long; with a subtyping relation, instead, the type of an expression may change during the computation, notably decrease:
\begin{quote}
\emph{
 In a statically-typed language with subtyping, the type of the
 expression occurring at a given point of a program may decrease during the computation. }
\end{quote}
This is
the reason why in the presence of subtyping it is sensible to distinguish
between the \emph{static} and \emph{dynamic} type of an expression,
the former being the type of the expression at compile time, the
latter the type that this expression has after being completely
evaluated.

As a simple example take the successor function I defined at the beginning of this presentation and apply it to, say, \texttt{(3+2)}:
\begin{alltt}
  (\k{sub}(Int $x)\{$x+1\})(3+2)
\end{alltt}
At compile time the function in the expression above has type \Int\To\Int\ (i.e., the usual type of the successor function) while the argument has
type \Int\ (i.e., the type of the application of the plus operator to two integers); therefore the whole expression has (static) type \Int. The first step of execution computes the argument
expression to \texttt{5}; the type of the argument has thus decreased
from \Int\ to \Odd, while the whole application has still type \Int{}
(the application of a function of type \Int\To\Int\ to an argument of
type \Odd{} has type \Int). The second step of execution computes the
application and returns \texttt{6}: the type of the whole expression
has just decreased from \Int\ to \Even.

Now that we have seen that in Perl\,6 types may dynamically evolve, let us see an example of how to use dynamic dispatch.
For instance, we can define the sum of two integers modulo 2 as follows.
\begin{eqnarray}
 \texttt{ \k{multi} \k{sub} mod2sum(Even \$x , Odd \$y) \{ 1 \}}\nonumber\\[-1.1mm]
 \texttt{ \k{multi} \k{sub} mod2sum(Odd \$x , Even \$y) \{ 1 \}}\label{mod2sum}\\[-1.1mm]
 \texttt{ \k{multi} \k{sub} mod2sum(Int \$x , Int \$y) \{ 0 \} } \nonumber
\end{eqnarray}
When \p{mod2sum} is applied to a pair of integers, then the most specific definition that matches the (dynamic) types of the arguments is selected. So if the argument is a pair composed by an even and an odd number then, according to their order, either the first or the second branch is selected. If the integers are instead both even or both odd, then  only the last definition of \p{mod2sum} matches the type of the argument and thus is selected. Since in general it is not possible to determine at compile time whether an integer expression will evaluate to an even or an odd number, then a selection delayed until runtime is the only sensible choice. Let us study the type of \p{mod2sum}. Obviously,  the function has type \pair\Int\Int\To\Int. However, it is easy to deduce  for this function a type much more precise. First, notice that the \p{subset} construction allows the programmer to define types that contain just one value. For instance the singleton type \p{Two} that contains only the value ``\p{2}'' is defined as:
\begin{alltt}
   \k{subset} Two \k{of} Int \k{where} \{ $_ = 2 \}.
\end{alltt}
Let me adopt the convention to use the same syntax to denote both
a value and the singleton type containing it. So I will use \p 2 rather than \p{Two} to denote the above singleton type. More generally, for every value \texttt{\emph{v}} of type \T{} in the language I assume the following definition as given
\begin{alltt}
   \k{subset} \emph{v} \k{of} \T \k{where} \{ $_ = \emph{v} \}.
\end{alltt}
With the above conventions, it is easy to deduce for  \p{mod2sum} the type  \texttt{\pair\Int\Int\To0{\Or}1}, that is, the type of a function that when applied to a pair of integers returns either \p{0} or \p{1}. But a slightly smarter type system could deduce a much more informative type:
\ifLONG
\begin{equation}\label{ty0}\tt
\hspace*{-3mm}\begin{array}{cl}
&(\pair{\Even}{\Odd}\To\p 1)\\[-.4mm]
\And\!\!\!\!\!&(\pair{\Odd}{\Even}\To\p 1)\\[-.4mm]
\And\!\!\!\!\!&(\pair{\Int\!}{\!\Int}\To\p 0\Or\p 1)\hspace{-3mm}
\end{array}
\end{equation}
\fi
First of all, notice that if in the last arrow of the intersection above we had specified just \p 0 as return type, then this would have been an error. Indeed, \p{mod2sum} does \emph{not} have \pair{\Int}{\Int}\To\p 0 type: it is not true that when \p{mod2sum} is applied to a pair of integers it always returns \p 0. But the type checker can be even smarter and precisely take into account the code selection policy. The code selection policy states that the code that is selected is the most specific one that is  compatible with the type of the argument(s).
Since the first two \p{multi} definitions  of  \p{mod2sum} in~\eqref{mod2sum} are more specific (i.e., they have smaller input types) than the third and last \p{multi} definition, then the code in this last \p{multi} definition will be selected only for arguments whose types are not compatible with the first two \p{multi} definitions. That is, this last definition will be selected only for pairs of integers that are neither in \pair{\Odd}{\Even}{} nor in \pair{\Even}{\Odd}. So a smarter type checker will deduce for \p{mod2sum} the following type:\footnote{*** Yet another caveat for my fellow type theorists. The sloppiness hinted at by Footnote~\ref{forreviewer2} does not apply here: even though the third \k{multi} definition of \p{mod2sum} is declared for \pair\Int\Int{} arguments, we can consider all the \k{multi} expressions of  \p{mod2sum} as forming a single definition that contains all the information needed to deduce the type~\eqref{ty1}. This is similar to what happens when typing pattern-matching expressions, where the types of the capture variables of a pattern are computed by taking into account the patterns that are matched before it.}
\begin{equation}\label{ty1}\tt
\hspace*{-3mm}\begin{array}{cl}
&(\pair{\Even}{\Odd}\To\p 1)\\[-.4mm]
\And\!\!\!\!\!&(\pair{\Odd}{\Even}\To\p 1)\\[-.4mm]
\And\!\!\!\!\!&((\pair{\Int\!}{\!\Int}\And\texttt{not}\pair{\Even\!}{\!\Odd}\And\texttt{not}\pair{\Odd\!}{\!\Even})\To\p 0)\hspace{-3mm}
\end{array}
\end{equation}
This type is strictly more precise than (i.e., it is a strict subtype of) the type in~\eqref{ty0} (as an exercise [EX5], we invite the reader to find a function value that is in the type in~\eqref{ty0} and not in type in~\eqref{ty1}). Actually the type is so precise that it completely defines the function it types. Indeed it is easy by some elementary set-theoretic reasoning to see that  \pair{\Int}{\Int}\And\texttt{not}\pair{\Even}{\Odd}\And\texttt{not}\pair{\Odd}{\Even}{} is equivalent to (i.e., it denotes the same set of values as) the type \pair{\Even}{\Even}\Or\pair\Odd\Odd. Equally easy is to see that for every type $\T_1$, $\T_2$, $\S$, the type  $\T_1\Or\T_2\To\S$ is equivalent to   $(\T_1\To\S)\And(\T_2\To\S)$, since every function in the former type has both the arrow types in the intersection, and vice versa (\emph{cf.} the exercise of Footnote~\ref{extrarules}). By applying these two equivalences to~(\ref{ty1}) (in particular to the last factor of the top-level intersection) we obtain that~\eqref{ty1} is equivalent to
\begin{equation}\label{ty2}\tt
\begin{array}{rcl}
&&(\pair{\Even}{\Odd}\To\p 1)\\[-.4mm]
&\And\!\!\!&(\pair{\Odd}{\Even}\To\p 1)\\[-.4mm]
&\And\!\!\!&(\pair{\Odd}{\Odd}\To\p 0)\\[-.4mm]
&\And\!\!\!&(\pair{\Even}{\Even}\To\p 0)
\end{array}
\end{equation}
from which it is easy to see that this last type is the most precise
type we can deduce for \p{mod2sum}. Before proceeding, let me stop an
instant to answer a possible doubt of the reader. The reader may wonder
whether it is permitted to apply a function of type~\eqref{ty2}
(equivalently, of type~\eqref{ty1}) to an argument of type
\pair{\Int}{\Int}: as a matter of fact, no arrow in~\eqref{ty2} has
\pair{\Int}{\Int} as domain. Of course it is permitted: the \emph{domain}
of a function typed by an intersection of arrows is the \emph{union}
of the domains of all the arrows. In the specific case the union of
the domains of the four arrows composing~\eqref{ty2} is $\pair{\Even\!}{\!\Odd}\Or\pair{\Odd\!}{\!\Even}\Or\pair{\Odd\!}{\!\Odd}\Or\pair{\Even\!}{\!\Even}$, that is,
\pair{\Int}{\Int}. So a function with type~\eqref{ty2} can be applied
to any argument whose type is (a subtype of) \pair{\Int}{\Int}. The
precise rule to type such an application is explained later on.

\subsubsection{Ambiguous selection}

Dynamic dispatch does not come for free. Besides the cost of
performing code selection at run-time, dynamic dispatch requires a lot
of care at static time as well, which is the whole point here. In
particular, dynamic dispatch introduces two new problems, those of
\emph{ambiguous selection} and of \emph{covariant specialization}. In
order to illustrate them I must better describe how dynamic selection
is performed at runtime. I said that always the most specific code is
selected. This is the code of the \texttt{multi} expression whose
parameter types best approximate the types of the arguments. In
practice consider a function defined by $n$ \texttt{multi}
expressions---call them \emph{branches}---, the $i$-th branch being
defined for inputs of type $\T_i$. Apply this function to an argument
that evaluates to a value of type $\T$. The possible candidates for
the selection are all the branches whose input type $\T_i$ is a
supertype of $\T$. Among them we choose the one defined for the least
$\T_i$, that is, the most specific branch.

Of course, the system must verify---possibly at compile time---that
whenever the function is applied to an argument in the function domain, then a most specific branch will always exist. Otherwise we bump into an
ambiguous selection problem. For instance, imagine that in order to
spare a definition I decided to write the function \p{mod2sum} in the
following (silly) way:
\begin{equation}\label{ambigous}
\hspace*{-4mm}\begin{array}{l}
\texttt{\k{multi sub }} \texttt{mod2sum(Even \$x , Int \$y)\{ \$y \% 2 \}}\\
\texttt{\k{multi sub }} \texttt{mod2sum(Int \$x , Odd \$y)\{\;\,(\$x+1)\;\%\;2\;\,\}}\hspace*{-2mm}
\end{array}
\end{equation}
If the first argument is even, then we return the second argument
modulo 2; if the second argument is odd, then we return the successor
of the first argument modulo 2. While from a mathematical viewpoint
this definition is correct, computationally this definition is problematic since it
may lead an ambiguous dynamic selection. Let \texttt{\emph{e}} be an
expression of type \Int, then \texttt{mod2sum(\emph{e},1)} is
well-typed (a static selection would choose the code of the second
branch). But if \texttt{\emph{e}} reduces to an even number, then both
codes can be applied and there is not a branch more specific than the
other: according to the current Perl semantics the execution is
stuck.%
\ifLONG
\footnote{* Two solutions alternative to being stuck are to choose one of the two
  branches either at random or according to a predetermined
  priority. Both solutions are compatible with the typing discipline
  we are describing in this work (and with the silly definition~\eqref{ambigous}) but they are seldom used: the use of
  some priority corresponds to using \emph{class precedence lists} in
  object-oriented languages with multiple inheritance (e.g., this is what is used in Common Lisp Object System), while I am not
  aware of any language that uses random selection.}
\fi
Clearly such
a problem can and must be statically detected (though, current implementations
of Perl\,6 detect it only at run-time). This can be done quite easily
as explained hereafter. Consider again our silly definition~\eqref{ambigous} of
\p{mod2sum} which is composed by two branches (i.e., pieces of code) one
for inputs of type \pair\Even\Int, the other for inputs of type
\pair\Int\Odd. Consider the intersection of these two types,
\pair\Even\Int\,\And\,\pair\Int\Odd, which is equal to
\pair\Even\Odd. The function \p{mod2sum} can be applied to values in
this intersection, but there does not exist a most specific branch to
handle them. In order to avoid selection ambiguity, we have to perform
systematically such a check on the intersections of the input types, as stated by the
following definition:
\begin{defi}[\textbf{Ambiguity}]\label{ambiguity}
Let \emph{$\T_h$} denote the input type of the $h$-th branch of a multi-subroutine composed of $n$ branches. The multi-subroutine is \emph{free from ambiguity} if and only if for all $i$, $j\in[1..n]$ and $i\not=j$, either  \emph{$\T_i\And\T_j$} is empty (i.e., \emph{$\T_i\And\T_j\Sub\Not\Any$}), or the
set \emph{$\lb\T_h\mid\T_i\And\T_j\Sub\T_h, h\in[1..n]\rb$} has a unique least element.\footnote{~The condition of uniqueness ensures that all the $\T_h$ are (semantically) distinct.}
\end{defi}
In words, although we must test that for all possible types of the arguments, there always exists a most specific branch, in practice it suffices to test such existence for the types that are a non-empty intersection of the input types of two distinct branches. A compiler must reject any multi-subroutine that is not free of ambiguity.

Ambiguity freedom is a formation condition for the definition of multi-subroutines that constrains the input types of the various branches and ensures that the computation will never be stuck on a selection. It is important to understand that \emph{this is not a problem related to the type system}, but just a problem about giving the semantics of multi-subroutine definitions. The type system has no problem of having functions of type
\begin{equation}\label{amb2}
  \texttt{(\,\,(Even,Int)\,\To\,0\Or 1\,\,) \& (\,\,(Int,Odd)\,\To\,0\Or 1\,\,)}
\end{equation}
which, intuitively, is the type that corresponds to the multi
subroutine in~\eqref{ambigous}: a function of this type can be applied
to arguments of type \pair{\Int}{\Int} and the result will be given
the type \verb:0|1:; the definition of \texttt{mod2sum} given at the
beginning of Section~\ref{formation} has indeed this type. The problem
only occurs when there are two multi \emph{definitions} with
parameters \pair{\Even}{\Int} and \pair{\Int}{\Odd} since they make
the run-time selection (thus, the semantics) of the multi-subroutine
undefined. In~\cite{Cas94} \IC explained how the problem of ambiguity
corresponds in object-oriented languages to the problem of method
selection with multiple inheritance: I invite the interested reader to
refer to that work for more information.

\subsubsection{Overriding}

In order to ensure the soundness of the type system (i.e., that all
type errors will be statically detected) we must impose a
second condition on the definition of multi-subroutines, which
constrains the return types of the branches. We have already seen
that since types evolve along the computation, so does the selection
of the branches. For instance, take again the original definition of
\p{mod2sum} as I defined it in~\eqref{mod2sum}. If we apply the
function to a pair of expressions of type \Int---e.g., \texttt{mod2sum(3+3,3+2)}---, then the compiler
statically deduces that the third branch will be selected. But if the
pair of expressions reduces to a pair composed of, say, an even and an
odd number---e.g., \texttt{mod2sum(6,5)}--, then at run-time the code of the first branch will be
executed. Now, different branches may have different return
types. This is clearly shown by the functions \p{sum} and \p{sumC}
which in the first version I gave for them are formed of two branches
returning different types (\Int{} and \Bool{} for the former, and
\Int\To\Int{} and \Bool\To\Bool{} for the latter, as  respectively stated by the types in~\eqref{uno} and~\eqref{due}). Since the selected
branch may change along the computation, so can the corresponding
return type, that is, the type of the application. Type soundness can
be statically ensured only if the type of every expression decreases with the
computation. Why is it so? Imagine that a function \p{f} with input
type \Int{} is applied to an expression of type \Int. Before
performing the application we evaluate the argument. Therefore
the value returned at runtime by this expression must be of type
\Int{} or of a smaller type, say \Even, but not of a distinct type
such as \texttt{Bool}, since the function \p f expects arguments of
type \Int, not of type \Bool. This must hold also when the argument of \texttt{f} is the result of the application of a multi-subroutine, which implies that if the branch selected for the multi-subroutine
changes along the computation, then the newly selected branch must
have a return type smaller than or equal to the return type of the
previously selected branch. Consider now a multi-function with $n$
branches (as in Definition~\ref{ambiguity}) whose $i$-th branch has
input type $\T_i$ (for $1\leq i\leq n$), apply it to an argument, and
statically determine that the selected branch is the $h$-th one. What
are the branches that can potentially be selected at run-time? Since
the type of the argument cannot but decrease along the computation,
then the candidates for selection are all the branches whose input
type is a subtype of $\T_h$. These are the branches that
\emph{specialize} (I will also say \emph{override} or \emph{refine})
the branch $\T_h$ and, as can be evinced by the discussion above,
soundness is ensured only if the return types of these branches are smaller than or
equal to the return type of the $h$-th branch. This yields the
following definition.
\begin{defi}[\textbf{Specialization}]\label{specialization}
Let \emph{$\T_h$} and \emph{$\S_h$} respectively denote the input type and the return type of the $h$-th branch of a multi-subroutine composed of $n$ branches. The multi-subroutine is \emph{specialization sound} if and only if for all  $i$, $j\in[1..n]$,  \emph{$\T_i\Sub\T_j$} implies  \emph{$\S_i\Sub\S_j$}.
\end{defi}
In words, the return type of each branch must be smaller than or equal to the return type of all the branches it specializes.

Definition~\ref{specialization} for specialization-soundness is, as
Definition~\ref{ambiguity} for ambiguity-freedom, a formation rule for
programs in Perl 6. These rules concern more the definition of the
language, than its type theory. As the type system has no problem
to have a type of the form as in~\eqref{amb2} which intersects two arrows whose domains have a common lower bound, so the type system has no problem in considering functions of type:
\begin{equation}\label{amb4}
(\Int\To\Odd)\;\And\;(\Even\To\Int)
\end{equation}
even if the return type of the arrow with the smaller domain is not
smaller than the other return type. We will return on this type later
on. For the time being, notice that while ambiguity-freedom is needed
to ensure unambiguous semantics of the programs,
specialization-soundness is necessary to ensure the soundness of the
system. Thus even if both are formation rules for Perl~6 programs, specialization-soundness can be explained and justified purely in terms of the type
system. In order to do that we need to better explain how to deduce
the type of the application of a function whose type is an
intersection of arrow types to some argument in its domain. To that end let us consider again the
type in~\eqref{amb2}. This type is not very interesting since both
arrows that compose it return the same type: now we know (see
Footnote~\ref{extrarules}) that $(\S_1\To \T)\And(\S_2\To\T)$ is
equivalent to $\S_1\Or\S_2\To \T$ and therefore~\eqref{amb2} is
equivalent to a single arrow without intersections (i.e., \texttt{((\Even,\Int)|(\Int,\Odd))\To (0|1)}). In order to make the type 
more interesting let us modify it by using two distinct output types
whose intersection and union are non trivial. For instance, let us
consider the following type:
\begin{equation}\label{amb3}
  \texttt{((Even,Int)\To(0,Int))\,\&\,((Int,Odd)\To(Int,1))}
\end{equation}
A function of this type accepts any pair of integers that is not in \texttt{(Odd,Even)}: the domain of
the function is the union of the domains of the arrows, that is
\texttt{(Even,Int)|(Int,Odd)}; if we use ``\minus'' to denote set-theoretic difference  (i.e., $\S{\minus}\T$ is syntactic sugar for $\S\And\Not\T$) then this type is equivalent to
\texttt{(Int,Int)\minus(Odd,Even)}. The type in~\eqref{amb3} specifies that if the first projection of the argument of a function of this type is even, then the first projection of the result will be \p 0, and if the
second projection of the argument is odd, then the second projection
of the result will be \p 1. A simple example of a function with this
type is the function that maps $(x,y)$ to $(x\,\mathbf{ mod
}\,2,y\,\mathbf{ mod }\,2)$.

Let us now examine all the possible cases for an application of a function of the type in~\eqref{amb3}. According to the type of the argument of the function, four different cases are possible:
\begin{enumerate}

\item The argument is of type \p{(Even,Odd)}: both arrows \emph{must} apply; the first arrows tell us that the first projection of the result will be \p 0, while the second arrow tell us that the second projection will be \p 1; so we deduce the result type \p{(0,1)}, that is the intersection of the result types;
\item The argument is of type \p{(Even,Even)}: then only the first arrow applies and we deduce the type \p{(0,Int)};
\item The argument is of type \p{(Odd,Odd)}: then only the second arrow applies and we deduce the type \p{(Int,1)};
\item The argument is of type \p{(Int,Int)\minus(Odd,Even)} (i.e., we just know that it is in the domain of the function): both arrows \emph{may} apply; since we do not know which one will be used then we take the union of the result types, that is, \p{(0,Int)|(Int,1)}.
\end{enumerate}
We can generalize this typing to the application of functions of a generic type
\begin{equation}\label{twoarrows}\tt
(\S_1\To\T_1)\,\And\,(\S_2\To\T_2)
\end{equation}
to an argument of type $\T$. Only four cases are possible:\footnote{~There is a mildly interesting fifth case when the type \T\ of the argument is empty, that is when the argument is statically known to diverge.}
\begin{enumerate}
\item If the argument is in $\S_1\And\S_2$ (i.e., if $\T\Sub\S_1\And\S_2$), then the application has type $\T_1\And\T_2$.
\item If the argument is in $\S_1$ and case 1 does not apply (i.e., $\T\Sub\S_1$ and $\T\minus\S_2$ is not empty), then the application has type $\T_1$.
\item If the argument is in $\S_2$ and case 1 does not apply (i.e., $\T\Sub\S_2$ and $\T\minus\S_1$ is not empty), then the application has type $\T_2$.
\item If the argument is in $\S_1\Or\S_2$ and no previous case applies, then the application has type $\T_1\Or\T_2$.
\end{enumerate}

\noindent
Of course things become more complicated with functions typed by an
intersection of three or more arrows: the complete formalization of
the typing of their applications is outside the scope of this primer
and it
\ifLONG
will be given in Section~\ref{electric}.
\fi
But an intersection of
two arrows is all we need to explain the specialization soundness
rule. So let us consider the type in~\eqref{twoarrows} and suppose
that, say, $\S_2\Sub\S_1$. What happens when we apply a function
of this type to an argument of type $\S_2$? Since $\S_2\Sub\S_1$ then
$\S_2\,\Sub\,\S_1\And\S_2$, therefore it is the first of the four possible cases that
applies: the argument is in $\S_1\And\S_2$ and thus the result will be
of type $\T_1\And\T_2$ (\ldots which is smaller than $\T_2$).

Let us rephrase what we just discovered: if we have a function of type
\[(\S_1\To\T_1)\,\And\,(\S_2\To\T_2)\] with $\S_2\Sub\S_1$, and we
apply it to an argument of the smaller type, that is $\S_2$, then the
result will be in $\T_1\And\T_2$. The fact that $\T_2$ is not smaller than
$\T_1$ does not matter because the typing rules tell us that the
application of this function to an argument of type $\S_2$ will return
a result in a type smaller than $\T_1$.

Let us consider a concrete example
for $\S_i$, $\T_i$ by going back to the type in~\eqref{amb4}. A
function is of type $(\Int\To\Odd)\;\And\;(\Even\To\Int)$  \emph{only if} for arguments of type \Even{} it
returns results in $\Odd\And\Int$, that is in $\Odd$: although the arrow
with domain \Even{} does not specify a result type smaller than the
one for the arrow of domain \Int{}, the typing rules ensure that all the results will be included in the latter type. In the previous sentence I italicized ``only if'' to stress its importance, since this is a point that may lead to confusion. For instance, the function \p{foo} defined as:
\begin{alltt}
   \k{multi} \k{sub} foo(Int $x) \{ 3 \}
   \k{multi} \k{sub} foo(Even $x)\{ $x % 3 \}
\end{alltt}
does \emph{not} have type $(\Int\To\Odd)\;\And\;(\Even\To\Int)$,
despite the fact that the first branch returns an
odd number  when applied to an integer and that the second branch
returns an integer  when applied to an even number (recall that ``\,\texttt{\%}\,'' is the modulo operator, so the result of applying
``\,\texttt{\% 3}'' to an \Even{} number can be either odd or even). The
function \p{foo} does not have this intersection type because when it is
applied to an argument of (static) type \Int, it may return a result
that is \Even{} rather than \Odd. For an example of that consider
\p{foo(foo(8))}: if \p{foo} were of type
$(\Int\To\Odd)\;\And\;(\Even\To\Int)$, then the type deduced for
\p{foo(foo(8))} would be \Odd{} (\p{8} is of type \Even, so \p{foo(8)}
is of type \Int{} and, thus, \p{foo(foo(8))} is of type \Odd), but the
application would return \p{2} (both the inner and the outer application of \p{foo} return
\p{2}). The definition above of \p{foo} has (just) type $\Int\to\Int$; for \p{foo} to have
type $(\Int\To\Odd)\;\And\;(\Even\To\Int)$, the second branch of \p{foo} should
always return a result in \Odd.

In the view of what I just wrote, I can next show that Definition~\ref{specialization} is, as for the condition of ambiguity, just a formation rule and \emph{not a problem related to the type system}. It is a design choice that ensures that the type of a well-typed multi definition is what a programmer expects it to be. Consider a definition of the form
\begin{alltt}
   \k{multi} \k{sub} bar(\Sone $x) \k{returns} \Tone \{ ... \}
   \k{multi} \k{sub} bar(\Stwo $x) \k{returns} \Ttwo \{ ... \}
\end{alltt}
In Perl~6 the \k{returns} keyword declares the result type of a subroutine. In the example above, it declares that the first definition has type $\S_1\To \T_1$ and the second
one type $\S_2\To \T_2$. If each definition is well typed, then the programmer expects the function
\p{bar} defined the multi definitions above to have type \p{($\S_1\To
  \T_1$)\And($\S_2\To \T_2$)}. Suppose that $\S_2\Sub\S_1$. In this case the 
typing rules state that a function has type \p{($\S_1\To
  \T_1$)\And($\S_2\To \T_2$)} only if whenever it is applied to an 
argument of type $\S_2$ it returns a result in $\T_1\And\T_2$. By
the semantics of multi definitions, if \p{bar} is applied to an
argument of type $\S_2$, then it executes the second definition of
\p{bar} and, therefore, it returns a result in $\T_2$. Putting the two observations together we conclude that \p{bar}
has type \p{($\S_1\To \T_1$)\And($\S_2\To \T_2$)} if and only if
$\T_2\,\Sub\,\T_1\And\T_2$ and---by a simple set-theoretic reasoning---if and only if $\T_2\Sub\T_1$: exactly the condition enforced by
Definition~\ref{specialization}.

To say it otherwise: the type system does not have any problem in
managing a type of the form \p{($\S_1\To \T_1$)\And($\S_2\To \T_2$)} even if
\Sone\Sub\Stwo\ and \Tone\ and \Ttwo\ are not related. However, such a
type might confuse the programmer since a function of this type
applied to an argument of type $\Sone$ returns a result in
\Tone\And\Ttwo. In order to avoid this confusion the language designer
can (without loss of generality) force the programmer to specify that
the return type for a \Sone\ input is (some subtype of)
\Tone\And\Ttwo. This can be obtained by accepting only specialization
sound definitions and greatly simplifies the presentation of the type
discipline of the language.
Let me show this again on the function \p{foo} I defined earlier in this section. As said, the type deduced for its current defintion is \Int\To\Int. A definition with explicit return
types as
\begin{alltt}
   \k{multi} \k{sub} foo(Int $x) \k{returns} Odd \{ 3 \}
   \k{multi} \k{sub} foo(Even $x)\k{returns} Int \{ $x % 3 \}
\end{alltt}
would be rejected for the reason that it is not specialization sound since the second branch must return subtype
of \Odd, and \Int{} is not one: this justification is simpler
than explaining that the function has not type
$(\Int\To\Odd)\;\And\;(\Even\To\Int)$ for the reason that the second branch does not
always return an odd result. If I modify the second branch so that it always returns odd numbers,
but I do not change the type annotations:
\begin{alltt}
   \k{multi} \k{sub} foo(Int $x) \k{returns} Odd \{ 3 \}
   \k{multi} \k{sub} foo(Even $x)\k{returns} Int \{ ($x+1) % 4 \}
\end{alltt}
then the specialization sound policy would reject also this
definition, for the same reason as the previous one, even though now
\p{foo} is of type $(\Int\To\Odd)\;\And\;(\Even\To\Int)$. But, once
more, this is probably easier for the programmer to understand than
the fact that for every application in which the second branch is
selected the deduced type would be \Odd{} rather than \Int.

This concludes the presentation of the primer on type theory. In this
section I have introduced substantial type-theoretic concepts and
explained them in terms of the set-theoretic interpretation of
types. We have seen how the set-theoretic interpretation of types
allows us to deduce non-trivial subtyping relations on complex
types. I have also tried to distinguish concepts that concern the
semantics of the language, such as the two conditions on the
definition of multi-subroutines, from the type-theoretic concepts that
explain and justify them. It is now time to put all we have learnt so
far in practice and apply it to understand and solve a controversy that has been  the core of a heated debate in the object-oriented languages community for several years in the late eighties, early nineties: the so-called ``covariance vs.\ contravariance problem''.

\subsection{Lessons to retain}\label{lesson1}
For the busy programmer who did not have time to read this long section, all I explained here can be summarized by the following six simple rules:
\begin{enumerate}
  \item Types are sets of values. In particular:
  \begin{enum}
   \item the type $\S\To\T$ is the set of all functions that when
     applied to values in \S{} return only results in \T;\@
   \item union, intersection, and negation types are the sets of values obtained
     by applying the corresponding set operations.
  \end{enum}
  \item \S{} is a subtype of \T{} if and only if all values in \S{}
    belong also to~\T.
  \item Two types are equal if they are the same set of values.

  \item A multi-subroutine is typed by the intersection of the types
    of each multi definition; but, for that to hold, these definitions must
    satisfy the property of covariant specialization
    (Definition~\ref{specialization})
  \item When a multi-subroutine is applied to an argument, the most precise multi
    definition for that argument is used; but, for that to happen, the definitions that compose the multi
    subroutine must be free from ambiguity
    (Definition~\ref{ambiguity}).
  \item The definition of a subroutine in which only some arguments are used
    for code selection (the ``\k{;;}'' notation) corresponds to defining a function
    whose parameters are those preceding the ``\k{;;}'' and that returns another
    function  whose parameters are those following the ``\k{;;}''.

\end{enumerate}



%% file: covcon.tex
If I have not disgusted my average programmer reader, yet, she should
have acquired enough acquaintance with types and dynamic dispatching
to be ready to tackle what in the study of object-oriented languages
is called \emph{the covariance vs.\ contravariance problem}.
Since this controversy took place in the object-oriented community, I therefore
start explaining it by using Perl 6 objects and then reframe it in
the context introduced in the previous section. I assume the reader to
be familiar with the basic concepts of object-oriented programming.

\subsection{Objects in Perl 6}\label{objects}
 The classic example
used in the late eighties to explain the problem at issue was that of
programming a window system in which the basic objects were
pixels, represented by the class \p{Point} written in Perl\,6 as:
\begin{alltt}

 \k{class} Point \{
     \k{has} $.x \k{is} rw;
     \k{has} $.y \k{is} rw;

     \k{method} origin() \{
         ($.x==0)&&($.y==0)
     \}

     \k{method} move(Int $dx, Int $dy) \{
         $.x += $dx;
         $.y += $dy;
         \k{return} \k{self};
     \}
 \};

\end{alltt}
Objects (or \emph{instances}) of the class \p{Point} have two \emph{instance variables}
(\emph{fields}, in Java parlance) \p{x} and \p{y}, and two methods
associated to the messages \p{origin} and \p{move}, respectively. The
former takes no argument and returns whether the receiver point has
null coordinates or not, the latter takes two integers, modifies
the instance variables of the receiver of the message \p{move} (i.e.,
the \emph{invocant} of the method \p{move}, in Perl parlance), and returns
as result the receiver itself, which in Perl\,6 is
denoted by the keyword \p{self}. New instances are created by the
class method \p{new}.  Methods are invoked by sending the corresponding message to objects,  by dot
notation. For example,
\begin{alltt}
  \k{my} Point $a = Point.new(x => 23, y => 42);
  $a.move(19,-19);
\end{alltt}
creates an instance of the class \p{Point} whose instance variables
have values \p{23} and \p{42} and then moves it by inverting the values of the
instance variables. Notice that in the definition of \p{\$a} (first line of our code snippet: in Perl, \k{my} is the keyword to define a variable in the current block) the name \p{Point} has a double usage: while the second occurrence of \p{Point} denotes a \emph{class} (to create a new instance) the first occurrence is a \emph{type} (to declare the type of \p{\$a}). In our set-theoretic interpretation the \emph{type} \Point\
denotes the set of all the objects that are instances of the
\emph{class} \p{Point}.

If later we want to extend our window system with colors (in late
eighties black and white screens were the norm), then we define \emph{by inheritance} a
subclass \p{ColPoint} that adds a \p{c} field of type string (in
Perl, \p{Str}) for the color and a method \p{iswhite}, it \emph{inherits} the fields \p x and \p y and the method \p{origin} from \p{Point}, and, finally, it \emph{specializes} (or \emph{overrides}) the
method \p{move} of \p{Point} so that white colored points are not modified.
\begin{alltt}
 \k{class} ColPoint \k{is} Point \{
     \k{has} Str $.c \k{is} rw;

     \k{method} iswhite() \{
         \k{return} ($.c=="white");
     \}

     \k{method} move(Int $dx, Int $dy) \{
         \k{if} not(\k{self}.iswhite()) \{
            $.x += $dx;
            $.y += $dy;
         \}
         \k{return} \k{self};
     \}
 \};
\end{alltt}
In many object-oriented languages, Perl 6 included, inheritance is
associated with subtyping: declaring a subclass by the keyword \p{is},
as in the example above, has the double effect of making the new class definition
inherit the code of the super-class definition \emph{and} of declaring the
type of the objects of the subclass to be a subtype of the type of the
objects of the super-class. In our example it declares
\p{ColPoint} to be a subtype of \p{Point}. Therefore
every object of type/class \p{ColPoint} is also of type \p{Point} and, as such, it
can be used wherever a \p{Point} object is expected. For
instance, the following code
\begin{alltt}
 \k{my} Point $a = ColPoint.new(x=>2,y=>21,c=>"white");
 $a.move(3,3);
\end{alltt}
is legal and shows that it is legitimate to assign a \p{ColPoint} object to a
variable \p{\$a} declared of type \p{Point}. Notice that, even
though \p{\$a} has static type \p{Point}, it is dynamically bound to
an object of type/class \p{ColPoint} and, therefore, the code executed
for \verb|$a.move| will be the one defined in the
class \p{ColPoint}. In particular, during the execution of this method,
the message \p{iswhite} will be dynamically sent to (the object
denoted by) \verb|$a| even though \verb|$a| has static type \p{Point},
for which no \p{iswhite} method is defined (the  wordings ``\emph{dynamic dispatch}'' and ``\emph{dynamic binding}'' come from there).  On the contrary, an
invocation such as \verb|$a.iswhite()| will (or, rather, should, since
current implementations of Perl\,6 do not check this point) be statically
rejected since \verb|$a|, being of type \p{Point}, cannot in general
answer to the message \p{iswhite}, and it is in general out of reach of a type
system to determine at static time whether \p{\$a} will be bound to
a \p{ColPoint} or not (although in this particular case it is pretty obvious).

\subsection{Inheritance and subtyping}\label{inheritance}

Since the objects of a subclass can be used where objects of the
super-class are expected, then  definitions by
inheritance must obey some formation rules to ensure type soundness. This is sensible for overridden
methods: while it is harmless to add new instance variables or methods
in a subclass, specialization of methods requires the use of a subtype. The
reasons for this is that, as an object of a subclass can be used where
an object of the super-class is expected, so the overriding method of
the subclass can be used where the overridden method of the super-class is expected. This is clearly shown by the code
snippet \p{\$a.move(3,3)} I wrote a few lines above: since \p{\$a} is of
(static) type \p{Point}, then the type system assumes that the
message \p{move} will execute the method defined in the
class \p{Point}, which expects a pair of integers and returns
a \p{Point} (i.e., it is a function of type \pair\Int\Int\To\Point),
even though in reality it is the method in the definition of \ColPoint\ that is used
instead. Using the latter definition where the former is expected is
type safe since the type of the method in \ColPoint\ is a
subtype of the type of the method in \Point:  rule~\eqref{arrows}
states that \pair\Int\Int\To\ColPoint\,\,\Sub\,\pair\Int\Int\To\Point, and
indeed \p{\$a.move(3,3)} will return a color point, that is, a value of
a subtype of the expected \Point\ type.

\paragraph{Contravariant overriding:} In general, to ensure type safety, when in a definition by inheritance we specialize (i.e., override)
a method the new method must have a subtype of the type of the
original method. By the co-/contra-variant rule in~\eqref{arrows} this means that the return
type of the new method must be a subtype of the return type of the old
method (covariant specialization) while its arguments must have a
supertype of the type of the arguments of the old method
(contravariant specialization). Since the latter is taken as characteristics of
this kind of specialization, the whole rule has taken the name
of \emph{contravariant overriding}.

\paragraph{Covariant overriding:} So far so good. Troubles start
when one considers \emph{binary methods}, that is methods
with arguments of the same type as the type of the receiver~\cite{BruEtAl96}. The paradigmatic example is the \p{equal} method which, for the class \p{Point}, can be defined as follows.
\begin{alltt}
 \k{class} Point \{
     \k{has} $.x \k{is} rw;
     \k{has} $.y \k{is} rw;

     \k{method} equal(Point $p) \{
         ($.x==$p.x)&&($.y==$p.y)
     \}
 \};
\end{alltt}
When later we introduce the \ColPoint\ class it is natural to want to redefine the method \p{equal} so as $(i)$ it takes arguments of type \ColPoint\ and $(ii)$ it compares also the colors, that is:
\begin{alltt}
 \k{class} ColPoint \k{is} Point \{
     \k{has} Str $.c \k{is} rw;

     \k{method} equal(ColPoint $p) \{
         ($.x==$p.x)&&($.y==$p.y)&&($.c==$p.c)
     \}
 \};
\end{alltt}
This is called \emph{covariant} specialization, since in the subclass a method overrides the previous definition of the method by using parameters whose types are subtypes of the types of the corresponding parameters in the overridden method.
Unfortunately, the definition above (and covariant specialization in general) is unsound as shown by the following snippet which is statically well typed but yields a type error at run-time:
\begin{alltt}
 \k{my} Point $a = ColPoint.new(x=>2,y=>21,c=>"white");
 \k{my} Point $b = Point.new(x=>2,y=>21);
 $a.equal($b);
\end{alltt}
The code above passes static type-checking: in the
first line a \ColPoint\ object is used where a \Point\ instance is
expected---which is legitimate---; in the second line we simply create a new object; while in the last line we send the message
\p{equal} to the object \p{\$a} with argument \p{\$b}:
since \p{\$a} is (statically) a \Point\ object, then it is
authorized to send to it the message \p{equal} with a \Point\
argument. However, at run-time \p{\$a} is bound to a color point and
by, dynamic binding, the method in the definition of \ColPoint\ is used. This tries to access the \p c instance variable of the
argument \p{\$b} thus yielding a run-time error: the static type system is
unsound.\footnote{%
\ifLONG
* \perl For Perl 6 purists. 
\fi
Since Perl 6 performs dynamic type checking, then the run-time error actually happens at the call of the \p{equal} method rather than at its execution.
\ifLONG
Here I described the behaviour shown when types are checked only at compile time, since a sound static type checking makes dynamic type checking useless. Also notice that, for the sake of simplicity, I used ``\p{==}'' (rather than ``\p{eq}'' as required in Perl) for the string equality operator.%
\fi
}

To have soundness and use color points where points are expected the type of the parameter of the \p{equal} method in \ColPoint\ must be either \Point\ or a supertype of it.

Despite this problem, covariant overriding had (has?) its strenuous defenders
who advocated that they wanted both to use color points
where points were expected and to define an \p{equal} method that
compared color points with other color points and not just with
points. The various solutions proposed (perform a dynamic check for
method arguments---as in O$_2$---, adopt covariant specialization
despite its unsoundness---as in Eiffel---or simply do not care about
parameter types in overriding methods---as in Perl 6) were all
incompatible with static soundness of the type system (or, as for
LOOM~\cite{LOOM}, they gave up subtyping relation between \Point\
and \ColPoint). Thus contravariant overriding and covariant
specialization looked as  mutually exclusive typing
disciplines for inheritance, and the dispute about the
merits and demerits of each of them drifted towards a religious
war. In what follows I will use the type theory we learned in the primer
contained in Section~\ref{primer} to argue that covariance and contravariance
are not opposing views, but distinct concepts that can coexist in
the same object-oriented language. But first I have to transpose the
whole discussion in the setting of Section~\ref{primer} where objects
were absent. This is quite easy since if we abstract from some details
(e.g., encapsulation, implementation, access), then the object-oriented  part of Perl\,6 is all
syntactic sugar.

\subsection{It is all syntactic sugar}\label{synsugar}

Consider a class definition in Perl\,6. It is composed of two parts: the
part that describes the class's objects (i.e., the fields that compose
them) and the part that describes the operations on the class objects
(i.e., the methods). Methods are nothing but functions associated to a
name (the message) and with an implicit parameter denoted
by \p{self}. Thus a method such as the one defined for \p{move} in
class \Point\ is a function with 3 parameters, one of type \Point\ (denoted by \texttt{self}) and
two of type \Int. As a matter of fact, if we write:
\begin{alltt}
  \k{my} Point $a = Point.new(x => 23, y => 42);
  $a.move();
\end{alltt}
Perl\,6 complains by
\[
\texttt{Not enough positional parameters passed; got 1 but expected 3},
\]
since in the second line of the above code it detected the point argument \p{\$a} of \p{move},
but the two integer arguments are missing. Now, if we consider a single
message, such as \p{move}, then it is associated to different
function definitions (the methods in \Point\ and \ColPoint\ for \p{move}) and
the actual code to execute when a message such as \p{move} is sent is chosen according
to the type of the receiver of the message, that is, according to the
type of the methods' hidden argument. In Section~\ref{primer} we
already saw that functions with multiple definitions are
called \p{multi} subroutines, and that the arguments used to choose the code are those listed before the ``\k{;;}'' (if it occurs) in the
parameter list.

If we remove method definitions from classes and replace them by multis
where the receiver parameter has become an explicit parameter, then
we cannot observe any difference from an operational point of view. In
other terms we can rewrite the first two class definitions of
Section~\ref{objects} as follows and obtain something that is observationally equivalent
\begin{alltt}
  \k{class} Point \{
     \k{has} $.x \k{is} rw;
     \k{has} $.y \k{is} rw;
  \};

  \k{multi sub} origin(Point \s{$self}) \{
     (\s{$self}.x==0)&&(\s{\$self}.y==0)
  \};

  \k{multi sub} move(Point \s{$self} \k{;;} Int $dx, Int $dy) \{
     $self.x += $dx;
     $self.y += $dy;
     \k{return} \s{$self};
  \};

  \k{class} ColPoint \k{is} Point \{
     \k{has} Str $.c \k{is} rw;
  \};

  \k{multi sub} iswhite(ColPoint \s{$self}) \{
     \k{return} (\s{$self}.c=="white");
  \};

  \k{multi sub} move(ColPoint\,\s{$self}\,\k{;;} Int\,$dx, Int\,$dy)\,\{
     \k{if} not(iswhite(\s{$self})) \{
        $self.x += $dx;
        $self.y += $dy;
     \}
     \k{return} \s{$self};
  \};
\end{alltt}
Notice that class definitions now contain only instance variable
declarations. In practice, classes have become record types
(i.e., respectively \emph{hashes} and \emph{structures} in Perl and C) whose
subtyping relation is explicitly defined.\footnote{* Technically, in this case one speaks
of \emph{name} subtyping, insofar as we \emph{declared} \ColPoint\ to
be a subtype of \Point.} Method definitions are external to class
definitions and have become multi subroutines enriched with an extra parameter \s{\$self} whose type is used to select the
code to execute and, as such, it is separated from the other parameters by ``\k{;;}''. Finally, method invocation has become function application. This is shown by the body of the method \p{move} for \ColPoint\ where the method invocation \p{\k{self}.iswhite()} has been transformed into \p{iswhite(\s{\$self})}.

The transformation we just defined tells us that from a point of view
of types the two class definitions at the beginning of
Section~\ref{objects} are equivalent to defining two (record)
types \Point\ and \ColPoint\ (the latter subtype of the former) and
three multi-subroutines respectively of type:
\iflmcs%
\begin{alltt}
origin:  Point\To\Bool
iswhite: \ColPoint\To\Bool
move:    (\Point\To(\pair{\Int\!}{\!\Int}\To\Point))\,\And\,(\ColPoint\To(\pair{\Int\!}{\!\Int}\To\ColPoint))
\end{alltt}
\else
\begin{alltt}
origin:  Point\To\Bool
iswhite: \ColPoint\To\Bool
move:    (\Point\To(\pair\Int\Int\To\Point))
       \And (\ColPoint\To(\pair\Int\Int\To\ColPoint))
\end{alltt}
\fi
The formation rules of multi-subroutines are satisfied, specifically for \p{move} where
\[\ColPoint\Sub\Point\] requires by
Definition~\ref{specialization} that
(\pair\Int\Int\To\ColPoint)\Sub(\pair\Int\Int\To\Point). Since the
latter holds, then the above definitions---thus, the class definitions
at the beginning of Section~\ref{objects}---are type sound.

If we apply the same transformation to the \p{equal} method we obtain
\begin{alltt}
  \k{multi sub} equal(Point \s{$self} \k{;;} Point $p) \{
    (\s{$self}.x==$p.x)&&(\s{$self}.y==$p.y)  \};

  \k{multi sub} equal(ColPoint \s{$self} \k{;;} ColPoint $p) \{
    (\s{$self}.x==$p.x)&&(\s{$self}.y==$p.y)&&(\s{$self}.c==$p.c)\};
\end{alltt}
The definitions above define a function \p{equal} that \emph{should} have the following type:
\begin{equation}\label{equaltype}
\begin{array}{rcl}
\p{equal:} \!\!\!\!&&  (\Point\To(\Point\To\Bool))\\
       &{\And}\!\!\!\!\!\!& (\ColPoint\To(\ColPoint\To\Bool))\hspace*{.5cm}
\end{array}
\end{equation}
However, the two multi definitions yield a function that does not have this type inasmuch as they do not comply with
the specialization formation rule: since \ColPoint\Sub\Point, then by
Definition~\ref{specialization} we need
(\ColPoint\To\Bool)\Sub(\Point\To\Bool) which \emph{by contravariance}
does not hold. Thus the definition is unsound. An equivalent way
to see this problem is that a function \p{equal} is of the type
in~\eqref{equaltype} only if when applied to a \ColPoint\ object it
returns a value in the following intersection type:
\begin{equation}\label{encapsulatedtype}
(\Point\To\Bool)\And(\ColPoint\To\Bool)
\end{equation}
(\emph{cf.} the first of the cases listed right after
equation~\eqref{twoarrows} and the discussion thereby); but the multi
definitions above---in particular the second multi definition of \texttt{equal}---do not (the second definition does not have
type \Point\To\Bool\ and, \emph{a fortiori}, it does not have the intersection above).

So far we did not learn anything new since we already knew that the definition above was not sound. If however in the definition of the multi subroutine \p{equal} we separate the two parameters by a ``\p{,}'' rather than a ``\k{;;}'', that is
\begin{alltt}
  \k{multi sub} equal(Point {$self} \s{,} Point $p) \{
    ({$self}.x==$p.x)&&({$self}.y==$p.y)  \};

  \k{multi sub} equal(ColPoint {$self} \s{,} ColPoint $p) \{
    ({$self}.x==$p.x)&&({$self}.y==$p.y)&&({$self}.c==$p.c)\,\};
\end{alltt}
then the definition becomes well typed, with type
\begin{equation}\label{currytype}
\begin{array}{rcl}
\p{equal:} \!\!\!\!&&  (\pair\Point\Point\To\Bool)\\
       &{\And}\!\!\!\!\!\!& (\pair\ColPoint\ColPoint\To\Bool)\hspace*{.5cm}
\end{array}
\end{equation}
In particular, the specialization formation rule of
Definition~\ref{specialization} is
now satisfied: since we have  \pair\ColPoint\ColPoint\Sub\pair\Point\Point, then this
requires \Bool\Sub\Bool, which clearly holds. So the definition
of \p{equal} is now sound: where is the trick?  The
consequence of replacing ``\p{,}'' for ``\k{;;}'' is that the new
version of \p{equal} uses the dynamic types of \emph{both} arguments to choose
the code to execute. So the second definition of \p{equal} is
selected only if both arguments are (dynamically) instances of \ColPoint. If any
argument of \p{equal} is of type \Point, then the first definition is
executed.  For instance, the method invocation \p{\$a.equal(\$b)} I
gave in Section~\ref{inheritance} to show the unsoundness of covariant
specialization now becomes \p{equal(\$a,\$b)}; and since the dynamic type
of \p{\$b} is also used to select the code to execute, then when this type is
\Point\ the first
definition of the multi subroutine \p{equal} is selected.

With hindsight the solution is rather obvious: this
transformation tells us that it is not possible to choose the code for
binary methods, such as \p{equal}, by using the type of just one
parameter (i.e., the type of the receiver); the only sound solution is
to choose the code to execute based on the types of both arguments.

Finally, observe that we are not obliged to check the type of both
arguments at the same time: we can first check one argument and then,
if necessary (e.g., for \p{equal} if the first argument is
a \ColPoint), check the other. This tells us how to solve the problem in the original setting, that is where methods are defined in classes: it suffices to use multi subroutines also for methods. In Perl\,6 this is obtained by adding the modifier \k{multi} in front of a method definition. This yields a solution in which the definition of the class \Point\ does not change:
\begin{alltt}
 \k{class} Point \{
     \k{has} $.x \k{is} rw;
     \k{has} $.y \k{is} rw;

     \k{method} equal(Point $p) \{
         ($.x==$p.x)&&($.y==$p.y)
     \}
 \};
\end{alltt}
while the class \ColPoint\ now defines a ``multi method'' for \p{equal}:
\begin{alltt}
 \k{class} ColPoint  \k{is} Point \{
     \k{has} Str $.c \k{is} rw;

     \k{multi method} equal(Point $p) \{
         ($.x==$p.x)&&($.y==$p.y)
     \}

     \k{multi method} equal(ColPoint $p) \{
         ($.x==$p.x)&&($.y==$p.y)&&($.c==$p.c)
     \}
 \};
\end{alltt}
The body of the \p{equal} method with a \ColPoint\ parameter is as
before. We just added an extra method in the class \ColPoint\ to handle
the case in which the argument of \p{equal} is a \Point,
that is, it is an argument that was statically supposed to be compared
with another \Point: in this case we do not check the \p{c} instance
variable. In words, if \p{equal} is sent to a \Point, then the method
defined in \Point\ is executed; if \p{equal} is sent to a \ColPoint,
then the type of the argument is used to select the appropriate multi
method.

While the type of \p{equal} defined in \Point\
is \Point\To\Bool{} (as it was before), the type of \p{equal}
in \ColPoint\ is now the intersection type
(\Point\To\Bool)\And(\ColPoint\To\Bool). The latter is a subtype of
the former (obviously, since for all types \S, \T, we have \S\And\T\Sub\S).
So at the end we did not discover anything really new, since it all sums up to using the old classic rule for subclassing:
\begin{quote}\em
\textbf{Overriding rule:} the type of an overriding method must be a subtype of the type of the
method it overrides
\end{quote}
The only novelty is that now intersection types give us the possibility to
have some form of covariant specialization: in a class \p{D} subclass of
a class \p{C} we can safely override a method of type \p{C}\To\S\ by a
new method of type \texttt{(\p{C}\To\S)\And(\p D\To\T)} where the code
associated to \p D\To\T\ represents the covariant specialization of
the old method.

Two concluding remarks on the \Point-\ColPoint\ example. First, notice
that the type of \p{equal} in \ColPoint\ is exactly the type we found
in~\eqref{encapsulatedtype}, that is, the type suggested by the
intersection type theory for the overriding method. Hence, if
we abstract from the ``syntactic sugar'' of the objects and we
consider methods as multi-subroutines with implicit parameters, then
the last two classes define a
multi-subroutine \p{equal} that has  type~\eqref{equaltype}, the type we
were looking for. Second, it is worth noticing that the solution based
on multi methods is modular. The addition of the class \ColPoint\ does
not require any modification of the class \Point, and a class such
as \Point\ can be defined independently of whether it will be later
subclassed with a covariant method specialization or not.%
\ifLONG
\footnote{\label{interesting}* Interestingly, the same behaviour as the solution above can be obtained simply by adding the modifier \p{multi} to the method \p{equal} in \p{Point} to the definitions given in Section~\ref{inheritance}, since in that case the (multi) method of \p{Point} is not overridden in \p{ColPoint}:
\\[1mm]\parbox[b]{\textwidth}{\tt\smallskip
\hspace*{2mm}\k{class} Point \{ \\[-.3mm]
\hspace*{7mm}     \k{has} \$.x \k{is} rw; \\[-.3mm]
\hspace*{7mm}     \k{has} \$.y \k{is} rw; \\[1mm]
\hspace*{7mm}     \s{multi} \k{method} equal(Point \$p) \{  \\[-.3mm]
\hspace*{10mm}         (\$.x==\$p.x)\&\&(\$.y==\$p.y)  \\[-.3mm]
\hspace*{7mm}     \} \\[-.3mm]
\hspace*{2mm} \}; \\[1mm]
\hspace*{2mm}\k{class} ColPoint  \k{is} Point \{ \\[-.3mm]
\hspace*{7mm}     \k{has} Str \$.c \k{is} rw; \\[1mm]
\hspace*{7mm}     \k{multi method} equal(ColPoint \$p) \{  \\[-.3mm]
\hspace*{10mm}         (\$.x==\$p.x)\&\&(\$.y==\$p.y)\&\&(\$.c==\$p.c) \\[-.3mm]
\hspace*{7mm}     \} \\[-.3mm]
\hspace*{2mm} \};}
however this solution is less modular than the one above since it may require to modify the definition of \p{Point} when the class \p{ColPoint} is added at a later time.}
\fi
\smallskip
\begin{center}
\iflmcs%
\def\boxlength{13.39cm}
\else
\def\boxlength{7.59cm}
\fi
\parbox{\boxlength}{{\sc Excursus.}
In the preceding example  the method that must be added to handle covariant specialization is explicitly defined by the programmer. However, it is possible to imagine a different solution in which this ``missing'' method is instead added by the compiler. All the method added by the compiler has to do is to call the overridden method (i.e., to dispatch the message to ``super''). The choice of whether the method added to handle covariant specialization is to be written by the programmer or inserted by the compiler, is a choice that concerns language design. The reader can refer to~\cite{BC97} and~\cite{Cas95} to see how the second choice can be implemented in Java and O$_2$, respectively.}
\end{center}\smallskip
To summarize, we have just seen how the type theory we studied in
Section~\ref{primer} allows us to propose a solution to the covariant
specialization of methods. Although the rule for subclassing is still
the usual one---you can override methods only by methods of smaller
type---, the presence of intersection types tells us that it is fine to
covariantly specialize a method as long as we do it by a multi-method
that dynamically handles the case in which the argument has a
supertype of the expected type (i.e., an argument that was intended
for the overridden method). Although the detailed explanation of how
we arrived to this solution is to some degree convoluted, the solution
for the final programmer amounts to retaining the \emph{Overriding
Rule} I stated above and to be able to apply it when complex types are
involved, in particular when functions are typed by intersection
types. This is the reason why it is important for a programmer to
grasp the basic intuition of the subtyping relation and, in that
respect, the set-theoretic interpretation of types as sets of values
should turn out to be, I believe, of great help.

\subsection{Lessons to retain}\label{lesson2}

As for the previous section, I summarize  for the busy programmer the content of this section in two rules:
\begin{enumerate}
  \item The type of an overriding method must be a subtype of the type
        of the method it overrides, whether these types are arrows or
        intersections of arrows.

  \item As a consequence of the previous point, it is sound to
        covariantly specialize a method as long as we do it by a
        multi-method in which at least one definition can handle
        arguments intended for the overridden method.
\end{enumerate}


%% file: electric.tex
In this section I am going to describe the algorithms and
data-structures needed to implement the type system of the
previous section. I will give a bare description of the algorithms and data-structures and justify them just informally. Detailed justifications and proof of
formal properties such as correctness can be found in the references commented in
Section~\ref{theory}.

I will proceed first by defining the algorithm that decides whether
two types are in subtyping relation as defined in
Definition~\ref{subtyping}. Next I will describe the data structures
used to efficiently implement types and their operations. Then, I
will describe the typing of expressions, focusing on those that
are more difficult to type, namely, subroutines, projections,
applications, and classes. I conclude the section with the algorithms for records whose presentation is quite technical and can be skipped at first reading.

\subsection{Type syntax}\label{typesyntax}
But first I must define the types I will be working with.
These are
the types defined in Section~\ref{type} with two slight
improvements: more precise base types and recursive types.

More precisely I replace \Bool\ and \Int\ respectively
by ``tags'' (ranged over by $t$) and ``intervals'', two base types that cover a wide range of possible implementations. This
yields the following grammar
\[\T::= \textit{t}\,\mid\,\IV{$i$}{$j$}\,\mid\,\Any\,\mid\,\pair\T\T \,\mid\,\T\To\T\,\mid\,\T\Or\T\,\mid\,\T\And\T \,\mid\,\Not\T\]
An interval is denoted by \IV{$i$}{$j$} where $i$ and $j$ are numeric
constants or ``\texttt{*}'' (which denotes infinite). For
instance, \IV{2}{4} is the type that denotes the
set \lb\texttt{2}, \texttt{3}, \texttt{4}\rb, \IV{1}{*} is the type of
positive integers, while \Int\ now becomes a shorthand
for \IV{*}{*}. A tag is a sequence of letters starting by a
back-quote, such as \tag{li} or \tag{title}, that denotes a
user-defined value and are akin to Perl's ``barewords'' (actually, barewords are
deprecated in Perl). When used as a type, a tag denotes the singleton
containing that tag. In particular, I can encode \Bool\ as the union
of two tag types, \tag{true}\Or\tag{false}. The remaining types have
the same interpretation as before. Tags, intervals, products and
arrows are called \emph{type constructors}, while unions,
intersections, and negations are called \emph{type connectives}. I
will use \Empty\ to denote the type \Not\Any, and $S\setminus T$ to
denote the set-theoretic difference of two sets $S$ and $T$ (i.e.,
$S\cap\neg T$).

To cope with recursive types, I will consider the set of possibly infinite
syntax trees generated by the grammar above (in technical jargon, it is the
language ``coinductively'' produced by the grammar) that satisfy the following
two conditions: $(a)$ they are \emph{regular} trees (i.e., each tree has
finitely many distinct subtrees) and $(b)$ on every infinite branch of
a tree there are infinitely many occurrences of product or arrow type
constructors (these two conditions are called ``contractivity conditions''). The first restriction ensures that recursive types have
a finite representation (e.g., finite sets of equations or recursive
definitions). The second restriction ensures that recursion
always traverses at least one type \emph{constructor}, thus barring out
ill-formed types such as $\T = \T \Or \T$ (which does not carry any
information about the set of values it denotes) or $\T = \Not\T$ (which cannot
represent any set). A consequence of the second restriction is also
that we can express only finite unions and intersections.

The reader disconcerted by infinite trees can
consider, instead, the following inductive definition of types, that
contains an explicit term for recursive definitions and that is equivalent to the
previous definition stated in terms of infinite trees:
\begin{eqnarray}\label{concretetypes}
\!\textbf{Atoms}\ \  a &\!\!\!\!\!\!\!::=\!\!\!\!\!\!\!& \textit{t}\,\mid\,\IV{$i$}{$j$}\,\mid\,\pair\T\T \,\mid\,\T\To\T \label{atoms}\\[.3mm]
\!\textbf{Types }\ \ \T &\!\!\!\!\!\!\!::=\!\!\!\!\!\!\!& a\,\mid\, X\,\mid\, \texttt{rec\,}X\texttt{\,=\,}a \,\mid\,\T\Or\T\,\mid\,\T\And\T \,\mid\,\Not\T\,\mid\,\Any\qquad
\end{eqnarray}
Although this last definition is probably easier to understand, I will
mainly work with the one with infinite trees (since it
yields simpler formulations of the algorithms and is closer to actual implementation) and will reserve the
recursive types notation for the examples. Nevertheless, the last
definition is important because it clearly separates type constructors (i.e., the meta-operators that construct atoms%
\footnote{
We call them atoms in analogy to propositional logic, since they are
the building blocks on which the logical connectives of union,
intersection, and negation are applied (note that the our
arrows \emph{do not} denote logical implication).})
from type connectives and shows that there are four different kinds of
constructors: tags, intervals, products, and arrows. An important
property that I will use in what follows is that for each of these
four kinds it is possible to
define a top type that contains all and only the types
of the same kind, namely.
\begin{enumerate}
\item $\pair{\Any}{\Any}$ is the greatest product type. It contains all
the pairs of values. I use $\Any\sprod$ to denote it.
\item $\Empty\To\Any$ is the greatest function
type. It contains all the functions. I use $\Any\sarrw$ to
denote it.
\item $\IV{*}{*}$ (i.e., \Int) is the largest
interval. It contains all the integers. I use $\Any\sint$ to
denote it.
\item
$\Not{\Any\sprod\Or\Any\sarrw\Or\Any\sint}$ contains all the tag values. I use $\Any\stag$ to
denote it.
\end{enumerate}
A final consideration before describing the subtyping algorithm. Types
are possibly infinite trees that denote possibly infinite sets of
values, but the theory presented here accounts only for \emph{finite}
values: there is not such a value as, say, an infinite list. So while the
type \texttt{rec$\,X\,$=\,}\tag{nil}\Or\pair{\Int}{$X$} is the type of the
finite lists of integers, the type  \texttt{rec$\,X\,$=\,}\pair{\Int}{$X$} is the
empty type.

\subsection{Subtyping algorithm}\label{subalgo}
The key property the subtyping algorithm is based on, is that types are
sets. Since subtyping is set-containment, then the algorithm uses
classic set-theore\-tic transformations to simplify the problem. For
instance, to prove that
$\pair{\T_1\!}{\T_2}\And\pair{\S_1\!}{\S_2}$ is empty (i.e.,
$\pair{\T_1}{\T_2}\And\pair{\S_1}{\S_2}\Sub\Empty$), the
algorithm uses set-theoretic equivalences and decomposes the
problem into simpler subproblems, namely:
$\pair{\T_1\!\!}{\!\T_2}\And\pair{\S_1\!\!}{\!\S_2}$ is empty if and only if
$\pair{\T_1\And\S_1\!}{\T_2\And\S_2}$ is empty, if and only if either
$\T_1\And\S_1$ is empty or ${\T_2\And\S_2}$ is empty. With that in
mind the subtyping algorithm can be summarized in 4 simple steps.

\begin{description}
\item[\emph{Step 1}] {\it transform the subtyping problem into an
emptiness decision problem.} Deciding whether $\S\Sub\T$ holds is equivalent
to deciding whether the difference of the two types is empty. So the
first step of the algorithm is to transform the problem $\S\Sub\T$
into $\S\And\Not\T\Sub\Empty$.

\item[\emph{Step 2}] {\it put the type whose emptiness is to be decided in a
\iflmcs%
disjunctive
\else
disjunc\-tive
\fi
normal form.} Our types are a propositional logic
whose atoms are defined by grammar~\eqref{atoms}. A \emph{literal}
(ranged over by $\ell$) is either an atom or the negation of an atom: $\ell::=
a\,\mid\,\Not{a}$. Every type is equivalent to a type in \emph{disjunctive normal form}, that is, a
union of intersections of literals:
\[
\bigvee_{i\in I}\bigwedge_{j\in J}\ell_{ij}
\]
with the convention that \Any{} and \Empty{} are, respectively,  the empty
intersection and the empty union. Therefore, the second step of the
algorithm consists in transforming the type $\S\And\Not\T$  whose
emptiness is to be checked, into a disjunctive normal form.

\item[\emph{Step 3}] {\it simplify mixed intersections.} The algorithm
has to decide whether a disjunctive normal form, that is, a union of
intersections, is empty. A union is empty if and only if every member
of the union is empty. Therefore the problem reduces to deciding
emptiness of an intersection of literals:
$\bigwedge_{i\in I} \ell_i$. Notice that there are four kinds of atoms and thus
of literals, one for each type constructor. Intersections that contain literals of different kinds
can be straightforwardly simplified: if in the intersection there are
two atoms of different constructors, say, $\pair{\T_1}{\T_2}$ and
$\S_1\To\S_2$, then their intersection is empty and so is the whole
intersection; if one of the two atoms is negated and the other is not,
say,  $\pair{\T_1}{\T_2}$ and $\Not{\S_1\To\S_2}$, then the negated
atom is useless and can
be removed since it contains the one that is not negated; if both
atoms are negated, then the intersection can be split in two
intersections of atoms of the same kind by intersecting the atoms
with the respective top types (e.g.,
$\Not{\pair{\T_1}{\T_2}}\And\Not{\S_1\To\S_2}$ can be split into
the union of $\Not{\pair{\T_1\!}{\T_2}}\And\pair{\Any\!}{\!\Any}$ and $\Not{\S_1\To\S_2}\And(\Empty\To\Any)$).

Therefore, the third step of the algorithm performs these simplifications so that the problem is reduced to
deciding emptiness of intersections that are formed by literals of the
same kind, that
is one of the following cases (where $P$ stands for ``positives'' and
$N$ for ``negatives''):
\begin{eqnarray}
& \displaystyle\bigwedge_{p\in P} t_p~\And~\bigwedge_{n\in N} \Not{t_n} \label{atomtag}\\
& \displaystyle\bigwedge_{p\in P} \IV{$i_p$}{$j_p$}~\And~\bigwedge_{n\in N} \Not{\IV{$i_n$}{$j_n$}} \label{atomint}\\
& \displaystyle\bigwedge_{\spair{S}\in P} \!\!\!\!\pair{\S_1}{\S_2}\quad~\And~\bigwedge_{\spair{T}\in N} \!\!\!\!\Not{\pair{\T_1}{\T_2}}\label{atomprod}\\
& \displaystyle\bigwedge_{\sarrow{S}\in P} \!\!\!\!({\S_1}\To{\S_2})\quad~\And~\bigwedge_{\sarrow{T}\in N} \!\!\!\!\!\Not{{\T_1}\To{\T_2}}\label{atomfun}
\end{eqnarray}

\item[\emph{Step 4}] {\it solve uniform intersections and recurse.} At this point we
have to decide whether every intersection generated at the previous step is
empty. When the intersection is formed by atoms of base type---i.e., cases~\eqref{atomtag} and~\eqref{atomint}---, then emptiness can be immediately decided: an intersection as in
equation~\eqref{atomtag} is empty if and only if the same tag appears
both in a positive and a negative position (i.e., there exists $p\in P$
and $n\in N$ such that $t_p=t_n$) or two distinct tags appear in positive position (i.e., there exists $p_1,p_2 \in P$ such that $t_{p_1}\not= t_{p_2}$), while whether  an intersection as in
equation~\eqref{atomint} is empty can be decided by simple
computations on the interval bounds.

If instead the intersection is composed of products or arrows, then first we
check whether we already proved that the intersection type at issue is empty (we have
recursive types so we memoize intermediate results). If we did not,
then we memoize the type and decompose it using its set-theoretic
interpretation. Precisely, to decide emptiness
of the type in~\eqref{atomprod}, we decompose the problem into
checking that for all $N'{\subseteq} N$
\begin{equation}\label{prodsubtyping}
\left(\bigwedge_{\spair{S}\in P}\!\!\!\!\!\S_1\;\;\Sub\!\!\!\!\!\bigvee_{\spair{T}\in N'}\!\!\!\!\!\T_1\right)\textsf{ or }  \left(\bigwedge_{\spair{S}\in P}\!\!\!\!\!\S_2\;\;\Sub\!\!\!\!\!\bigvee_{\spair{T}\in N\setminus N'}\!\!\!\!\!\T_2\right)
\end{equation}
holds, which is done by recursively proceeding to Step 1.

To decide emptiness of the type in~\eqref{atomfun},
we decompose the problem into checking whether there exists
$\T_1\To\T_2\in N$ such that $\T_1\leq\bigvee_{\sarrow{S}\in
P}\S_1$ and for all $P'\subsetneq P$  (notice that containment is strict)
\begin{equation}\label{arrowsubtyping}
\left(\T_1\quad\ \Sub\!\!\!\!\!\bigvee_{\sarrow{S}\in
P'}\!\!\!\!\!\S_1\right)\textsf{ or }  \left(\bigwedge_{\sarrow{S}\in
P\setminus P'}\!\!\!\!\!\S_2\;\;\Sub\quad \T_2\right)
\end{equation}
holds, which is checked by recursively proceeding to Step 1.\\
\textbf{[\emph{End of the Algorithm}]}
\end{description}

\medskip\noindent
In order to complete the presentation of the algorithm above, let me
show how to define the recursive functions that compute the Boolean
function defined in~\eqref{arrowsubtyping}
in Step 4. I focus on the algorithm for~\eqref{arrowsubtyping} since
it is the most difficult one and leave the case for products~\eqref{prodsubtyping} as an
exercise. Given an atom $\T_1\To\T_2\in N$ I first need to check whether
$\T_1\leq\bigvee_{\sarrow{S}\in P}\S_1$ and, if so, then compute a
function, say, $\Phi$ that given $\T_1$, $\T_2$ and $P$ checks whether
for every strict subset $P'$ of $P$, the formula
in~\eqref{arrowsubtyping} holds. The definition of $\Phi$ is not
straightforward, therefore I will introduce it gradually.

Let me first slightly simplify the problem by relaxing the
strictness of the containment: I will define a function $\Phi$ that
checks~\eqref{arrowsubtyping} for every (possibly non-strict) subset $P'$ of $P$, that is, for $P$ too. Indeed, notice that the requirement of strictness in $P'\subsetneq P$
is just an optimization to avoid a useless check: we already know
that for $P'=P$ the formula~\eqref{arrowsubtyping} is true (because
$\T_1\leq\bigvee_{\sarrow{S}\in P}\S_1$ holds). Therefore strictness is not needed for the correctness the algorithm. I leave the implementation of the strictness optimization for $\Phi$ as an exercise.

The basic idea to
program the function $\Phi$ is the following: pick an element $e$ in
$P$ (the choice of $e$ can be arbitrary) and let $Q$ be $P\setminus\lb
e\rb$; then $\Phi$ does two things: $(i)$ it recursively solves the problem for all
subsets of $Q$ (i.e., the subsets of $P$ that do not contain $e$) and
$(ii)$ it recursively solves the problem for all subsets of
$Q$ to which $e$ is added (i.e., all the subsets of $P$ that
contain $e$). To avoid repeating calculations in different recursive calls,
$\Phi$ keeps track of three sets of elements of $P$: the set of elements that are still to be selected to form some subset $P'$ of $P$ (denoted by $P^\circ$); the set of elements that were already selected to be in $P'$  (denoted by $P^+$); and
the set of elements that were already selected not to be in $P'$ (denoted by $P^-$). A generic call for $\Phi$ will then be
of the form $\Phi(\T_1\;,\;\T_2\;,\;P^\circ\;,\;P^+\;,\;P^-)$ such
that $P=P^\circ\cup P^+\cup P^-$; such a call will pick an element
$\S_1^\circ\To\S_2^\circ$ in $P^\circ$ and $(i)$ recursively solve the problem for all subsets of
$Q = P^\circ\setminus \lb\S_1^\circ\To\S_2^\circ\rb$ by adding the element
to $P^-$ and thus recording that the element must not be considered,
that is
$\Phi(\T_1\;,\;\T_2\;,\;Q\;,\;P^+\;,\;P^-\cup \lb\S_1^\circ\To\S_2^\circ\rb)$
and $(ii)$ recursively solve the problem for all subsets of
$Q$ to which $\S_1^\circ\To\S_2$ is added, which is done by adding the element to
$P^+$, that is
$\Phi(\T_1\;,\;\T_2\;,\;Q\;,\;P^+\cup \lb\S_1^\circ\To\S_2^\circ\rb\;,\;P^-)$. In
other terms, the recursive calls of $\Phi(\T_1\;,\;\T_2\;,\;P^\circ\;,\;P^+\;,\;P^-)$ will
solve the problem for all subsets of $P^+\cup P^\circ$ that contain (at least) $P^+$, that is, these recursive calls will check equation~\eqref{arrowsubtyping} for all subsets $P'$ of $P$ such that $P^+\subseteq
P' \subseteq P^+\cup P^\circ$. When $P^\circ$ is empty, and there no longer is an element to pick to make the recursive calls, then $P^+$ is the $P'$ at issue and $P^-$ is $P\setminus P'$.  In that case
formula~\eqref{arrowsubtyping} tells us that we have to check that
$\T_1$ is smaller than the union of the domains of all the arrows in
$P^+$ and that $\T_2$ is larger than the intersection of the codomains
of the arrows in $P^-$. This leads to the last observation we need
before giving the actual definition of $\Phi$: we do not need to
keep the whole $P^+$ and $P^-$ sets; we just need the union of the
domains for $P^+$ and the intersection of the codomains for
$P^-$. This yields the following recursive definition of the subtyping
relation for the arrow case $\bigwedge_{\sarrow{S}\in P}
({\S_1}\To{\S_2})\;\Sub\;{{\T_1}\To{\T_2}}$:
\begin{equation}\label{bof}
(\T_1\;\Sub\bigvee_{\sarrow{S}\in
P}\S_1)\;\textsf{ and }\;\Phi(\T_1,\T_2,P,\Empty,\Any)
\end{equation}
where
\[
\begin{array}{l}
\Phi(\T_1\;,\;\T_2\;,\;\varnothing\;,\;\D\;,\;\C) = (\T_1\Sub \D)\textsf{ or } (\C\Sub \T_2)\\[2mm]
\Phi(\T_1\;,\;\T_2\;,\;P\cup\lb\S_1^\circ\To\S_2^\circ\rb\;,\;\D\;,\;\C) = \\
\qquad\Phi(\T_1\;,\;\T_2\;,\;P\;,\;\D\;,\;\C\And\S_2^\circ)\textsf{ and }
\Phi(\T_1\;,\;\T_2\;,\;P\;,\;\D\Or\S_1^\circ\;,\;\C)
\end{array}
\]
The definition $\Phi$ above is not very interesting: all it does is
just to explore the whole space of the subsets of $P$. But it becomes
interesting when we try  to get rid of the two extra parameters
$\C$ and $\D$. Notice that $\C$ and $\D$ are two accumulators
respectively for  the intersection of the
\underline{$\C$}odomains and the union of the \underline{$\D$}omains of the arrows singled out in the previous recursive calls.
As a matter of fact, we do not need these two extra parameters since
we can store them in the first two parameters, by using suitable
set-theoretic operations.  If we define the function $\Phi'$ as
follows:
\[
\begin{array}{l}
\Phi'(\Empty\;,\;\T_2\;,\;\varnothing) =  \texttt{true}\\[2mm]
\Phi'(\T_1\;,\;\Empty\;,\;\varnothing) =  \texttt{true}\\[2mm]
\Phi'(\T_1\;,\;\T_2\;,\;\varnothing) =  \texttt{false}\hspace{33mm}\textrm{otherwise}\\[2mm]
\Phi'(\T_1\;,\;\T_2\;,\;P\cup\lb\S_1^\circ\To\S_2^\circ\rb) = 
\Phi'(\T_1\;,\;\T_2\And\S_2^\circ\;,\;P)\textsf{ and }
\Phi'(\T_1\setminus\S_1^\circ\;,\;\T_2\;,\;P)
\end{array}
\]
then it is not difficult to see that for all $\T_1$, $\T_2$, $P$, \C, and $\D$,
\[\Phi(\T_1\;,\;\T_2\;,\;P\;,\;\D\;,\;\C) = \Phi'(\T_1\setminus\D\;,\;\C\setminus\T_2,P)\]
and that, therefore, we can use
\begin{equation}\label{implem}
(\T_1\;\Sub\bigvee_{\sarrow{S}\in
P}\S_1)\;\textsf{ and }\;\Phi'(\T_1,\Not{\T_2},P)
\end{equation}
instead of~\eqref{bof}.

I invite the reader to verify that both $\Phi$ and $\Phi'$ compute the
Boolean function described in~\eqref{arrowsubtyping} and leave the definition of a similar function for~\eqref{prodsubtyping} as exercise [EX6]. In
particular, I suggest trying to use the properties
$\bigwedge_{\spair{S}\in P}\pair{\S_1}{\S_2}
= \pair{\bigwedge_{\spair{S}\in P}\S_1}{\bigwedge_{\spair{S}\in
P}\S_2}$ and $\pair{\S_1}{\S_2}\setminus\pair{\T_1}{\T_2}
= \pair{\S_1\setminus\T_1}{\S_2}\Or\pair{\S_1}{\S_2\setminus\T_2}$ to
define an algorithm potentially more efficient.
I also leave as an exercise [EX7] the modification of $\Phi$ and $\Phi'$ to implement the optimization corresponding to the strict containment  $P'\subsetneq P$ when checking~\eqref{arrowsubtyping}.

$\Phi'(\T_1\;,\;\T_2\;,\;P)$ is pretty efficient when the subtyping
relation holds, since there is no choice: all strict subsets of $P$ must be
tested. However, it does not perform any test before emptying the parameter $P$ by a sequence of repeated recursive calls. As a possible optimization, it may be interesting to perform some checks earlier, when $P$ is yet not empty, thus possibly
avoiding some recursive calls when the subtyping relation does not hold.
A way to do so for a generic call $\Phi(\T_1\;,\;\T_2\;,\;P^\circ\;,\;P^+\;,\;P^-)$ is first to check formula~\eqref{arrowsubtyping} for $P'=P^\circ\cup P^+$ and, only  if it succeeds, then perform the two recursive calls. This corresponds to modifying the definition
of  $\Phi$ as follows:
\[
\begin{array}{l}
\Phi(\T_1\;,\;\T_2\;,\;\varnothing\;,\;\D\;,\;\C) = (\T_1\Sub \D)\textsf{ or } (\C\Sub \T_2)\\[2mm]
\Phi(\T_1\;,\;\T_2\;,\;P\cup\lb\S_1^\circ\To\S_2^\circ\rb\;,\;\D\;,\;\C) = \\
\qquad((\T_1\Sub \S_1^\circ\Or\D)\textsf{ or } (\bigwedge_{\sarrow{S}\in
P}\!\S_2\And\C\;\Sub\;\T_2))\makebox[0cm]{\hspace*{18mm}(*)}\\
\qquad\textsf{and }\Phi(\T_1\;,\;\T_2\;,\;P\;,\;\D\;,\;\C\And\S_2^\circ)\\
\qquad\textsf{and }\Phi(\T_1\;,\;\T_2\;,\;P\;,\;\D\Or\S_1^\circ\;,\;\C)
\end{array}
\]
which yields the following modification for $\Phi'$:
\[
\begin{array}{l}
\Phi'(\Empty\;,\;\T_2\;,\;\varnothing) =  \texttt{true}\\[2mm]
\Phi'(\T_1\;,\;\Empty\;,\;\varnothing) =  \texttt{true}\\[2mm]
\Phi'(\T_1\;,\;\T_2\;,\;\varnothing) =  \texttt{false}\hspace{13mm}\textrm{otherwise}\\[2mm]
\Phi'(\T_1\;,\;\T_2\;,\;P\cup\lb\S_1^\circ\To\S_2^\circ\rb) = \\
\qquad((\T_1\Sub \S_1^\circ)\textsf{ or } (\bigwedge_{\sarrow{S}\in
P}\!\S_2\Sub\;\Not{\T_2}))\makebox[0cm]{\hspace*{18mm}(*)}\\
\qquad\textsf{and }\Phi'(\T_1\;,\;\T_2\And\S_2^\circ\;,\;P)\\
\qquad\textsf{and }\Phi'(\T_1\setminus\S_1^\circ\;,\;\T_2\;,\;P)
\end{array}
\]
The extra checks performed in the $(*)$-marked lines are interesting in the case they fail, since this allows
us to stop the computation of $\Phi$ and $\Phi'$ even when the
parameter $P$ is not empty, thus potentially saving several recursive
calls. The price to pay is that when the subtyping relation holds, the functions
will perform $n$ useless checks (since they will be done twice) where $n$ is the cardinality of
$P$. But in the case of $\Phi'$ these useless checks have a very low
cost, since they amount to checking whether $\T_1$ of $\T_2$ are empty
types. As we show in the next section, checking the emptiness of a
type just amounts at checking the content of four fields in a record;
this must be contrasted with the fact that every recursive call
requires to perform either an intersection or a difference of types,
which are much costlier operations.

\subsection{Data structures and their
operations}\label{datastructures}

The subtyping algorithm (as well as the typing algorithm  I describe
further on) works with types in disjunctive normal form, which are
transformed by applying unions, intersections, and differences. Since I want to avoid to
normalize types at each step of the algorithm, then the algorithm will work with types
stored in disjunctive normal form. Thus I need to find an efficient
representation for disjunctive normal forms and define the operations
of union, intersection, and difference so that they preserve this
representation. Recall that a type in disjunctive normal form can be
represented as:
\begin{equation}\label{dnf}
\bigvee_{i\in I}(\bigwedge_{p\in P_i}a_{p}~\And\bigwedge_{n\in N_i}\Not{a_{n}})
\end{equation}
where $a_i$'s are atoms. A naive representation of~\eqref{dnf} (such as lists of pairs of lists) does not fit our needs since it is not
compact and makes it difficult to efficiently implement set-theoretic operations.

\begin{figure}[t]
\small\center%
\begin{tikzpicture}[level distance=7mm]
  \tikzstyle{level 1}=[sibling distance=22mm]
  \tikzstyle{level 2}=[sibling distance=12mm]
  \tikzstyle{level 3}=[sibling distance=8mm]\node[minimum size=5mm,circle,draw](z){$a_1$}
   child{
     node[minimum size=5mm,circle,draw]{$a_2$}
       child{
          node[minimum size=5mm,circle,draw]{\one}
       }
       child{
          node[minimum size=5mm,circle,draw]{$a_3$}
          child{
             node[minimum size=5mm,circle,draw]{\one}
          }
          child{
             node[minimum size=5mm,circle,draw]{\zero}
          }
       }
   }
   child{
     node[minimum size=5mm,circle,draw]{$a_3$}
       child[sibling distance=8mm]{
             node[minimum size=5mm,circle,draw]{\zero}
       }
       child[sibling distance=8mm]{
             node[minimum size=5mm,circle,draw]{\one}
       }
   };
\end{tikzpicture}
\caption{BDD for  $(a_1\wedge a_2)\vee(a_1\wedge\neg{a_2}\wedge a_3)\vee(\neg{a_1}\wedge\neg{a_3})$\label{fig-bdd}}
\end{figure}
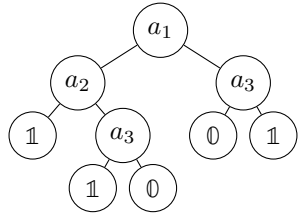

\paragraph{Binary decision diagrams.}
A classic
technique to store compactly Boolean functions is to use Binary
Decision Diagrams (BDD). In the context I am studying a BDD is a labeled binary tree whose
nodes are labeled by atoms and whose leaves are labeled by \zero\ or \one.
In a BDD every path starting from the root and terminating by \one\ represents the
intersection of the atoms that label the nodes of the path, each atom occurring negated or
not according to whether the path went through the left or right child of the atom's
node. For instance, the BDD representing the disjunctive normal form $(a_1\And a_2)\Or(a_1\And\Not{a_2}\And a_3)\Or(\Not{a_1}\And\Not{a_3})$ is given in Figure~\ref{fig-bdd} (the first and third summands of the union correspond respectively to the leftmost and rightmost paths of the BDD).
Formally, BDD are defined by the following grammar:
\[B \quad::=\quad \zero\;\mid\;\one\;\mid\; \bdd{a}{B}{B}\]
and have the following interpretation:
\[
\begin{array}{rcl}
\sem{\zero} & = & \Empty\\
\sem{\one}  & = & \Any\\
\sem{\bdd{a}{B_1}{B_2}} & = & (a\And \sem{B_1})\Or(\Not a\And\sem{B_2})
\end{array}
\]
The interpretation above maps a BDD into the disjunctive normal form it represents (of
course, after having simplified the intersections with \Any\
and \Empty), that is, into the union of the intersections that correspond
to paths ending by a $\one$. To ensure that the atoms occurring on a
path are distinct, we define a total order on the atoms and impose
that on every path the order of the labels strictly increases. It is
possible to implement all set-theoretic operations directly on BDD\@.
Let $B$, $B_1$, and $B_2$ denote generic BDDs, $B_1= \bdd{a_1}{C_1}{D_1}$ and  $B_2= \bdd{a_2}{C_2}{D_2}$. Unions, intersections, and differences of BDDs are defined as follows:
\[\begin{array}{lll}
\one\vee B = B\vee\one = \one\hspace*{2cm}&\one\wedge B = B\wedge\one = B \\
\zero\vee B = B \vee\zero = B &\zero\wedge B = B \wedge\zero = \zero \\
B\setminus\one=\zero\setminus B =\zero &  B\setminus\zero = B\\
\end{array}
\]
\[
\begin{array}{rcl}
\one\setminus\bdd{a}{B_1}{B_2} & = & \bdd{a}{\one\setminus B_1}{\one\setminus B_2}
\\ \\
B_1\vee B_2  &= &
\left\lb
  \begin{array}{ll}
    \bdd{a_1}{C_1\vee C_2}{D_1\vee D_2} & \textrm{for }a_1=a_2\\
    \bdd{a_1}{C_1\vee B_2}{D_1\vee B_2} & \textrm{for }a_1<a_2\\
    \bdd{a_2}{B_1\vee C_2}{B_1\vee D_2} & \textrm{for }a_1>a_2\\
  \end{array}
\right.
\\ \\
B_1\wedge B_2  &= & 
\left\lb%
  \begin{array}{ll}
    \bdd{a_1}{C_1\wedge C_2}{D_1\wedge D_2} & \textrm{for }a_1=a_2\\
    \bdd{a_1}{C_1\wedge B_2}{D_1\wedge B_2} & \textrm{for }a_1<a_2\\
    \bdd{a_2}{B_1\wedge C_2}{B_1\wedge D_2} & \textrm{for }a_1>a_2\\
  \end{array}
\right.
\\ \\
B_1\setminus B_2  &= & 
\left\lb%
  \begin{array}{ll}
    \bdd{a_1}{C_1\setminus C_2}{D_1\setminus D_2} & \textrm{for }a_1=a_2\\
    \bdd{a_1}{C_1\setminus B_2}{D_1\setminus B_2} & \textrm{for }a_1<a_2\\
    \bdd{a_2}{B_1\setminus C_2}{B_1\setminus D_2} & \textrm{for }a_1>a_2\\
  \end{array}
\right.\end{array}
\]
Notice that $\one\setminus B$ computes the negation of $B$ and is obtained by exchanging the $\zero$ leaves of $B$ into $\one$ and vice versa.

After having performed any of these operations, we can simplify a BDD
by replacing every subtree of the form $\bdd{a}{B}{B}$ by just $B$.

\paragraph{BDD with lazy unions.}
A well-known problem of BDDs is that by repeatedly applying unions we
can have an exponential blow-up of their size. To obviate this problem
the \cduce{} compiler uses a lazy implementation for unions. These are
evaluated just when they are needed, that is, when computing differences and
intersections. To obtain it we represent BDDs as ternary trees, of the
form \tdd{a}{B_1}{B_0}{B_2}, where the middle child represents a lazy
union:
\[
\begin{array}{rcl}
\sem{\tdd{a}{B_1}{B_0}{B_2}} & = & (a\And \sem{B_1})\Or\sem{B_0}\Or(\Not a\And\sem{B_2})
\end{array}
\]
Let $B_1= \tdd{a_1}{C_1}{U_1}{D_1}$ and  $B_2= \tdd{a_2}{C_2}{U_2}{D_2}$. When two atoms are merged, unions are lazily recorded in the middle child:
\[
\begin{array}{rcl}
B_1\vee B_2  &= &
\left\lb
  \begin{array}{ll}
    \tdd{a_1}{C_1\vee C_2}{U_1\vee U_2}{D_1\vee D_2} & \textrm{for }a_1=a_2\\
    \tdd{a_1}{C_1}{U_1\vee B_2}{D_1} & \textrm{for }a_1<a_2\\
    \tdd{a_2}{C_2}{B_1\vee U_2}{D_2} & \textrm{for }a_1>a_2\\
  \end{array}
\right.
\end{array}
\]
The intersection $B_1\wedge B_2$ materializes the lazy unions when the top-level atoms are the same and trees are merged, and is defined as:
\[
\left\lb
  \begin{array}{ll}
    \tdd{a_1}{(C_1{\vee} U_1)\wedge (C_2{\vee} U_2)}{\zero}{(D_1{\vee} U_1)\wedge (D_2{\vee} U_2)} & \textrm{for }a_1=a_2\\
    \tdd{a_1}{C_1\wedge B_2}{U_1\wedge B_2}{D_1\wedge B_2} & \textrm{for }a_1<a_2\\
    \tdd{a_2}{B_1\wedge C_2}{B_1\wedge U_2}{B_1\wedge D_2} & \textrm{for }a_1>a_2\\
  \end{array}
\right.
\]
while the difference $B_1\setminus B_2$ materializes unions also when top-level atoms are distinct, and is defined as:
\[
\left\lb
  \begin{array}{ll}
    \tdd{a_1}{(C_1{\vee}U_1)\setminus(C_2{\vee}U_2)}{\zero}{(D_1{\vee}U_1)\setminus(D_2{\vee}U_2)} & \textrm{for }a_1=a_2\\
    \tdd{a_1}{(C_1{\vee}U_1)\setminus B_2}{\zero}{(D_1{\vee}U_1)\setminus B_2} & \textrm{for }a_1<a_2\\
    \tdd{a_2}{B_1\setminus (C_2{\vee}U_2)}\zero{B_1\setminus (D_2{\vee}U_2)} & \textrm{for }a_1>a_2\\
  \end{array}
\right.
\]
After having performed these operations it is possible to perform two simplifications, namely, replace $(\tdd{a}{B_1}{\one}{B_2})$ by $\one$, and replace $(\tdd{a}{B}{C}{B})$ by $B\vee C$.

\paragraph{Disjoint atoms.} The representation above does not exploit
the property used by \emph{Step 3} of the subtyping algorithm, namely,
that it is possible to consider disjoint normal forms in which the
intersections do not mix atoms of different kinds.  This means that
the union given in~\eqref{dnf} can be seen as the union of four
different unions, one for each kind of atom:
\begin{eqnarray}
\bigvee_{i\in I\stag}(\bigwedge_{p\in P_i}t_{p}~\And\bigwedge_{n\in
N_i}\Not{t_{n}})&\vee\label{dnftag}\\
\bigvee_{i\in I\sint}(\bigwedge_{p\in P_i}\IV{$h_{p}$}{$k_p$}~\And\bigwedge_{n\in
N_i}\Not{\IV{$h_{n}$}{$k_n$}})&\vee\label{dnfint}\\
\bigvee_{i\in I\sprod}(\bigwedge_{p\in P_i}\pair{\S_p}{T_{p}}~\And\bigwedge_{n\in
N_i}\Not{\pair{\S_n}{\T_{n}}})&\vee\label{dnfprod}\\
\bigvee_{i\in I\sarrw}(\bigwedge_{p\in P_i}\S_p\To\T_{p}~\And\bigwedge_{n\in
N_i}\Not{\S_n\To\T_{n}})\label{dnfarrow}
\end{eqnarray}
Instead of representing a disjunctive normal form by a unique BDD---which  mixes atoms of different kinds---, it is more
compact to store it in four distinct unions.  A type will be represented by a record with a field for each distinct kind of atom, each field containing a disjunctive normal of form of the corresponding atom, namely:
\begin{alltt}
   \{ tags: \ensuremath{\DNF\stag} ;           {\color{gray}// \textrm{\emph{stores a union as in}~\eqref{dnftag}}}
     ints: \ensuremath{\DNF\sint} ;           {\color{gray}// \textrm{\emph{stores a union as in}~\eqref{dnfint}}}
     prod: \ensuremath{\DNF\sprod} ;           {\color{gray}// \textrm{\emph{stores a union as in}~\eqref{dnfprod}}}
     arrw: \ensuremath{\DNF\sarrw}             {\color{gray}// \textrm{\emph{stores a union as in}~\eqref{dnfarrow}}}
   \}
\end{alltt}
 Let \texttt{Kinds} denote the set {\lb\texttt{tags}, \texttt{ints}, \texttt{prod}, \texttt{arrw}}\rb. An  expression \T\ of the type above represents the following disjunctive normal form
\[
\bigvee_{k\in\texttt{\scriptsize Kinds}}(\T.k\wedge\Any_{k})
\]
Different fields for different kinds of atom yield a more compact representation of the types. But the real gain of such an organization is that, since
atoms of different kinds do not mix, then all set-theoretic
operations can be implemented component-wise.
\iflmcs%
 In other terms, \S$\wedge$\T, \S$\vee$\T,
and $\S\setminus\T$ can be respectively implemented as:\begin{alltt}
\{ tags = \S.tags\ensuremath{\wedge}\T.tags;   \{ tags = \S.tags\ensuremath{\vee}\T.tags;   \{ tags = \S.tags\ensuremath{\setminus}\T.tags;
  ints = \S.ints\ensuremath{\wedge}\T.ints;     ints = \S.ints\ensuremath{\vee}\T.ints;     ints = \S.ints\ensuremath{\setminus}\T.ints;
  prod = \S.prod\ensuremath{\wedge}\T.prod;     prod = \S.prod\ensuremath{\vee}\T.prod;     prod = \S.prod\ensuremath{\setminus}\T.prod;
  arrw = \S.arrw\ensuremath{\wedge}\T.arrw      arrw = \S.arrw\ensuremath{\vee}\T.arrw      arrw = \S.arrw\ensuremath{\setminus}\T.arrw
\}                         \,\}                         \,\}
\end{alltt}
\else
In other terms, \S$\wedge$\T\
and $\S\setminus\T$ can be respectively implemented as:
\begin{alltt}
\{ tags = \S.tags\ensuremath{\wedge}\T.tags;     \{ tags = \S.tags\ensuremath{\setminus}\T.tags;
  ints = \S.ints\ensuremath{\wedge}\T.ints;       ints = \S.ints\ensuremath{\setminus}\T.ints;
  prod = \S.prod\ensuremath{\wedge}\T.prod;       prod = \S.prod\ensuremath{\setminus}\T.prod;
  arrw = \S.arrw\ensuremath{\wedge}\T.arrw        arrw = \S.arrw\ensuremath{\setminus}\T.arrw
\}                           \,\}
\end{alltt}
and similarly for the union $\S\vee\T$.
\fi
Therefore, not only we have smaller
data structures, but also the operations are ``partitioned'' on these smaller data
structures, and thus executed much more efficiently.

To conclude the presentation of the implementation of types I still
have to show how to represent the disjunctive normal forms contained
in each field and how to implement recursive types.

For the
fields \texttt{prod} and \texttt{arrw}, corresponding to product types
and arrow types the use of BDDs with lazy unions to represent the
unions in~\eqref{dnfprod} and~\eqref{dnfarrow} is the obvious
choice. For base types, \texttt{tags} and \texttt{ints}, we can use instead
a specific representation. In particular, it is not difficult
to prove that any union of the form~\eqref{dnftag} can be equivalently
expressed either as a union of pairwise distinct atoms
$(a_1\vee\ldots\vee a_n)$ or its negation $\neg(a_1\vee\ldots\vee
a_n)$ (I leave this proof as an exercise for the reader [EX8]). The
same representation can be used also for disjunctive normal forms of
intervals, which can thus be expressed as a union of disjoint
intervals or its negation (if the intervals are maximal---i.e., adjacent intervals are merged---then this representation is unique). In
conclusion, a disjunctive normal form of base types can be expressed as
a set of atoms and a positive/negative flag indicating whether this
set denotes the union of the atoms or its complement (of course, the
implementation of the set-theoretic operations for these fields must
be specialized for this specific representation).

Finally, to represent recursive types it suffices to allow the records
representing types to be recursively defined. By construction
the recursion can occur only in the types forming an atom of a BDD in
the \texttt{prod} or \texttt{arrw} fields. This means that the
contractivity conditions of Section~\ref{typesyntax} are satisfied by
construction.

\paragraph{Emptiness (i.e., subtyping).}
I end the presentation of data structures by showing how the subtyping
algorithm described in Section~\ref{subalgo} specializes to these data
structures. As expected with these data structures the algorithm is
much simpler (the representation handles all steps of normalization) and consists just of two steps. Let \S\ and \T\ be two
types represented by a record of four fields as described above.  In order
to verify whether \S\Sub\T\ holds do:
\begin{description}
\item[Step 1] Compute $\S\minus\T$;
\item[Step 2] Check that all the fields of $\S\minus\T$ are the empty type.
\end{description}
Checking the emptiness of the base type fields \texttt{tags}
and \texttt{ints} is immediate: they must be an empty set of atoms
with a positive flag. For the fields \texttt{prod} and \texttt{arrw},
if the BDD that they contain is not $\zero$ or $\one$, then we apply the
respective decomposition rule described in \textbf{\emph{Step 4}} in
Section~\ref{subalgo}, memoize, and recurse.

As a final exercise for this part, the reader can try to define a
function \texttt{norm} that takes as argument a type produced by the
grammar given at the beginning of Section~\ref{typesyntax} and returns
the record representing its disjunctive normal form [EX9].

\subsection{Typing algorithms}

I conclude the presentation of my electrical blueprint by specifying the algorithms for typing some expressions that are commonly found in programming languages.

\subsubsection{Products}\label{se:products}

Perl 6 includes list expressions and  element selection. For the sake of simplicity I will just consider products, since lists can then be encoded as nested products.

In what follows I consider expressions of
the form \texttt{($e_1$,$e_2$)} which returns the pair formed by the
values returned by \texttt{$e_1$} and \texttt{$e_2$}, as well as the
expressions \texttt{$e$[0]} and \texttt{$e$[1]} denoting respectively
the first and second projection of $e$.

The algorithm for typing pairs is straightforward: if \texttt{$e_1$}
is of type $\T_1$ and \texttt{$e_2$} of type $\T_2$,
then \texttt{($e_1$,$e_2$)} is of type $\pair{\T_1}{\T_2}$.

The algorithm for typing projections, instead, is more complicated, since projections can be
applied to any expression of type $\Any\sprod$, that is, any
expression whose type has a normal form as in~\eqref{dnfprod}. So imagine
that we want to type the expressions \texttt{$e$[0]} or \texttt{$e$[1]}
where $e$ is of some type \T. The first thing to do is to verify that
$e$ will return a pair, that is, that $\T\Sub\Any\sprod$ holds. If it is so,
then \T\ is equivalent to the following normal form:
\begin{equation}\label{dnfprod2}
\bigvee_{i\in I}(\bigwedge_{p\in P_i}\pair{\S_p}{T_{p}}~\And\bigwedge_{n\in
N_i}\Not{\pair{\S_n}{\T_{n}}})
\end{equation}
in which case the type of the expression \texttt{$e$[0]} can be approximated by:
\begin{equation}\label{proj1}
\bigvee_{i\in I}\bigvee_{N'\subseteq N_i}(\bigwedge_{p\in P_i}\S_p~\And\bigwedge_{n\in N'}\Not{\S_n})
\end{equation}
and, likewise, the type of the expression \texttt{$e$[1]} can be approximated by:
\begin{equation}\label{proj2}
\bigvee_{i\in I}\bigvee_{N'\subseteq N_i}(\bigwedge_{p\in P_i}\T_p~\And\bigwedge_{n\in N'}\Not{\T_n})
\end{equation}
with the convention that an empty intersection of atoms denotes the top type of the corresponding kind.

Let me explain how to pass from~\eqref{dnfprod2} to~\eqref{proj1}
and~\eqref{proj2}. The idea is simple and consists to transform the
union in~\eqref{dnfprod2} into a union of products (i.e., no intersection of
products and no negated product) by using two
simple observations. First, an intersection of products is equivalent to the product
of the intersections: $\bigwedge_{p\in P}\pair{\S_p}{T_{p}}$ is equivalent to
$\pair{\bigwedge_{p\in P}\S_p}{\bigwedge_{p\in P}T_{p}}$. Second, the intersection of a product
with a negated product can be distributed inside the product:
$\pair{\S_1}{\S_2}\And\Not{\pair{\T_1}{\T_2}}$ is equivalent to
$\pair{\S_1\And\Not{\T_1}}{\S_2}\Or\pair{\S_1}{\S_2\And\Not{\T_2}}$. For multiple intersections such as
\[
\bigwedge_{p\in P}\pair{\S_p}{T_{p}}~\And\bigwedge_{n\in
N}\Not{\pair{\S_n}{\T_{n}}}
\]
the two transformations above yield the following equivalent type
\begin{equation}\label{prodsimple}
\bigvee_{N'\subseteq N}\pair{\bigwedge_{p\in P}\S_p~\And\bigwedge_{n\in N'}\Not{\S_n}}{\bigwedge_{p\in P}\T_p~\And\bigwedge_{n\in N\setminus N'}\Not{\T_n}}
\end{equation}
it is then easy from this type to deduce the types~\eqref{proj1} and~\eqref{proj2} of the projections, simply by observing that the projection of a union of products is the union of the projections of each product.

Finally, I said that the types in~\eqref{proj1} and~\eqref{proj2} are
approximations. Indeed in some cases it is possible to define more precise types for the projections (thus allowing more programs to be typed). Consider again the products forming the union
in~\eqref{prodsimple}. If for some $N'$ either $\bigwedge_{p\in
P}\S_p~\And\bigwedge_{n\in N'}\Not{\S_n}$ is empty or $\bigwedge_{p\in
P}\T_p~\And\bigwedge_{n\in N\setminus N'}\Not{\T_n}$ is empty, then so
is their product. Therefore this $N'$ can be eliminated from the union
in~\eqref{prodsimple} and, thus, also from the unions in~\eqref{proj1}
and~\eqref{proj2}. So the way to proceed to type projections is to
transform the type of $e$ into the form of~\eqref{prodsimple},
eliminate the empty products and take the union of the remaining
projects. Formally, if $\T$ is equivalent to the type in~\eqref{dnfprod2}
then the type of \texttt{$e$[0]} is
\begin{equation}
\bigvee_{i\in I}~~\bigvee_{N'\in\;\textsf{\scriptsize N}_1}(\bigwedge_{p\in P_i}\S_p~\And\bigwedge_{n\in N'}\Not{\S_n})
\end{equation}
where
\[
{\textsf{N}_1}=\lb N' ~|~   N'\subseteq N \textsf{ and }  \bigwedge_{p\in P}\T_p~\not\!\!\Sub\bigvee_{n\in N\setminus N'}\T_n\rb
\]
and, likewise the type of  \texttt{$e$[1]} is:
\begin{equation}
\bigvee_{i\in I}~~\bigvee_{N'\in\; \textsf{\scriptsize N}_2}(\bigwedge_{p\in P_i}\T_p~\And\bigwedge_{n\in N'}\Not{\T_n})
\end{equation}
where
\[
{\textsf{N}_2}=\lb N' ~|~  N'\subseteq N \textsf{ and }  \bigwedge_{p\in P}\S_p~\not\!\!\Sub\bigvee_{n\in N'}\S_n\rb
\]

\subsubsection{Subroutines}
The technique to type subroutines should be pretty clear by now. Given a
definition either of the form

\texttt{\k{sub} (T$_1$ \$x$_1$,$\ldots$, T$_n$ \$x$_n$) \{ $e$ \}}

\noindent or of the form

\texttt{\k{sub} (T$_1$ \$x$_1$,$\ldots$, T$_n$ \$x$_n$) \k{returns} \S\ \{ $e$ \}}

\noindent it has type \texttt{(T$_1$,$\ldots$, T$_n$)\To{}S} if under the hypothesis that \texttt{\$x$_1$} has type \texttt{T$_1$} and that  \texttt{\$x$_n$} has type \texttt{T$_n$} it is possible to deduce that $e$ has type \S\ (the difference between the two cases is that in the former the type \S\ is the one returned by the algorithm for $e$, while in the latter the algorithm checks whether the type returned for $e$ is a subtype of the type \S\ specified by the programmer). Likewise

\texttt{\k{sub} (T$_1$ \$x$_1$,$\ldots$,T$_n$ \$x$_n$\,\k{;;}\,T$_{n+1}$ \$x$_{n+1}$,$\ldots$,T$_{n+k}$) \{ $e$ \}}

\noindent
has type \texttt{(T$_1$,$\ldots$,
T$_n$)\To(T$_{n+1}$,$\ldots$,T$_{n+k}$)\To{}S} if under the hypotheses
that \texttt{\$x$_i$} has type \texttt{T$_i$} (for $i=1..n+k$) it is
possible to deduce that $e$ has type \S.

Finally, given a multi-subroutine composed of $n$ definitions, if the
$i$-th definition has type $\S_i\To\T_i$ and all these definitions
form a set that is both specialization sound and free of ambiguity (cf.\ Definitions~\ref{specialization} and~\ref{ambiguity}), then the
multi-subroutine has type  $\bigwedge_{i=1}^n\S_i\To\T_i$.

Notice that I defined type-checking only for subroutines whose
parameters are explicitly typed. The definition of a type system that
infers also the types of subroutine parameters (technically, this is
called a ``type reconstruction'' system) and that infers polymorphic
types as done in the languages of the ML family is possible. However,
its technical development is complex: it would need twice the
space taken by this paper and is clearly outside the scope of our
presentation. I invite the interested reader to consult the references
given in Section~\ref{theory} on the subject.

\subsubsection{Applications}

Now that we can type subroutines, it is time to type their
applications. The typing of an application is the same,
independently of whether the applied subroutine is multi or not. In both cases the
function value is typed by an intersection of arrows, formed by just
one arrow when the subroutine is not multi. Suppose we have two
expressions \texttt{e$_1$} and \texttt{e$_2$} which are well typed
respectively with type \T\ and \S. We want to check $(a)$ whether the
application \texttt{e$_1$e$_2$} is well typed and, if so, $(b)$ deduce
the best possible type for this application.

In order to verify $(a)$ the algorithm proceeds in two steps: first it
checks that  \texttt{e$_1$} is a function (i.e., it returns neither a constant nor a
pair). This is done by checking that the type \T\ of \texttt{e$_1$} is
a subtype of $\Any\sarrw$, that is $\T\Sub\Empty\To\Any$. If this
holds, then it checks that the type $\S$ of the argument is a subtype of the domain \dom\T\
of the function, that is $\S\Sub\dom\T$, where the domain is defined
as follows:
\[
\dom{\bigvee_{i\in I}(\bigwedge_{p\in P_i}\S_p\To\T_{p}~\And\bigwedge_{n\in
N_i}\Not{\S_n\To\T_{n}})}\eqdef\bigwedge_{i\in I}\bigvee_{p\in
P_i}\S_p
\]
The definition of domain is given for a normal form as
in~\eqref{dnfarrow} since this is the disjunctive normal form of any
type smaller than $\Any\sarrw$. To compute the domain only the
positive arrows are used.%
\footnote{\label{negarrows} The intuition is
that negative arrows characterize the output of functions but not their
input. For instance, a function of type
$(\Any\To\Any)\And\Not{\Int\To\Bool}$ can be applied to any argument;
the negated type $\Not{\Int\To\Bool}$ does not mean that the
function cannot be applied to an integer: it means that for some
integer arguments---but not necessarily all integer arguments---the
function will return results not in \Bool{}. This also explains why every
application of this function will by typed by \Any{} and not by, say, \Not\Bool. See Section 4.8
of~\cite{AlainThesis} for technical details.}
The domain of an intersection of arrows is
the union of their domains, since a function of that type accepts
arguments that are compatible with at least one arrow in the intersection: whence the inner union
(e.g., a function of type (\Bool\To\S)\And(\Int\To\T) can be applied
both to Boolean and to integer arguments). If a function is typed by
a union of function types, then it accepts only arguments that
are compatible with \emph{all} arrows in the union: whence the outer intersection (e.g., an
expression of type \texttt{([1..4]\To\S)\Or([2..6]\To\T)} can be
applied only to arguments in \texttt{[2..4]}: since we cannot
statically know which of the two types the function returned by the
expression will have, then we must accept only arguments that are
compatible with both arrows).

Once the algorithm has verified that \S\ and \T\ satisfy the two
conditions above, then the application is well
typed. It is then time to compute the type of this application.

Since $\T\Sub\Any\sarrw$---i.e., \T\ is a function type---, then
\begin{equation}\label{typeT}
\T=\bigvee_{i\in I}(\bigwedge_{p\in P_i}\S_p\To\T_{p}~\And\bigwedge_{n\in
N_i}\Not{\S_n\To\T_{n}})
\end{equation}
for some $I$, $P_i$ and $N_i$. Then the type of the application of a function of the type \T\  defined in~\eqref{typeT} to an argument of type \S\ is
\begin{equation}\label{typeapp}
\bigvee_{i\in I}\left(\bigvee_{Q\subsetneq P_i
\textrm{ \scriptsize s.t. }\texttt{\scriptsize S}\not\leq\bigvee_{q\in Q}\texttt{\scriptsize S}_q}\left(\bigwedge_{p\in P_i\setminus
Q} \T_p\right)\right)
\end{equation}
Let me decrypt this formula for you, but feel free to skip directly to Section~\ref{tyclasses} if you are not interested in these details. First, consider the case of~\eqref{typeT}---and, thus,~\eqref{typeapp}---where $I$ is a singleton. Since negated arrow types (i.e., those in $N_i$) do not play any role in
the typing expressed by formula~\eqref{typeapp} (see Footnote~\ref{negarrows}), then we can for simplicity consider that the type \T\ of
the function is
\[\bigwedge_{p\in P}\S_p\To\T_{p}\]
and that the type \S\ of the argument is a subtype of the domain
of \T, that is $\S\,\Sub\bigvee_{p\in P}\S_p$.

If the argument can return a value in the domain $\S_i$ (for some $i\in P$) of some arrow
(i.e., if $\S\And\S_i$ is not empty), and this actually happens, then the
result returned by the application (if any) will be in $\T_i$. If the
argument can return a value in domain of \emph{two} arrows say $\S_i$
and $\S_j$ for $i,j\in P$ (i.e., $\S\And\S_i\And\S_j$ is not empty), then
the result returned by application (if any) in that case will be a value in $\T_i\And\T_j$. Of
course we want to deduce the most precise type for the result type. So
in this case we will deduce as a type for our result $\T_i\And\T_j$
rather than just $\T_i$ or $\T_j$. Also, it may be the case that
$\T_i\And\T_j$ does not cover all possible cases for the result: not only
$\S$ may intersect intersections of the domains smaller than $\S_i\And\S_j$ (e.g., the intersection of \emph{three} arrow domains), but also this does not
cover the case in which the argument returns a result that falls outside $\S_i\And\S_j$, that is, a result in
$\S\setminus(\S_i\Or\S_j)$, whenever this set is not empty. The return
type for this case must be computed separately and then added to the
final result type.

So the algorithm proceeds as follows. First $(a)$ it computes all the
possible intersections of the types $\S_i$ for $i\in P$ and keeps
only those whose intersection with $\S$ is not empty; then $(b)$ for
every intersection kept that is also \emph{minimal} it computes the
intersection of the corresponding codomains; finally $(c)$ it takes
the union of the computed intersections of codomains.

This is exactly what the formula in~\eqref{typeapp} does. First it
considers all possible combinations of the domains $\S_p$ for $p\in
P$, that is, all non-empty subsets of $P$. Then, for each of these
subsets it keeps the largest subsets such that the intersections of
$\S$ and the corresponding domains is not empty. To do that it takes
every strict subset $Q$ of $P$ (strict, since we want $P\setminus Q$
to be non-empty) such that $S\not\leq\bigvee_{q\in Q}\S_q$. Now take
any \emph{maximal} subset $Q$ that satisfies $S\not\leq\bigvee_{q\in
Q}\S_q$. Being maximal means that if we add to $Q$ any index $p$ in
$P\setminus Q$ then by adding the corresponding $\S_p$ we cover the
whole $\S$. This means that all elements of $\S$ that are missing in
$\bigvee_{q\in Q}\S_q$---that is all elements in
$S\setminus\bigvee_{q\in Q}\S_q$---are in \emph{all} the $\S_p$ for $p\in
P\setminus Q$. Therefore the intersection $\S\wedge\bigwedge_{p\in
P\setminus Q}\S_p$ is not empty, since it contains
$S\setminus\bigvee_{q\in Q}\S_q$. The maximality of $Q$ implies the
minimality of the intersection of the domains in $P\setminus Q$. So we
take the intersection of the corresponding codomains, that is
$\bigwedge_{p\in P\setminus Q}\T_p$, and we union all these
intersections (notice that the union in~\eqref{typeapp} also adds some
smaller intersections which corresponds to the subsets $Q$ contained
in the maximal sets: this does not matter since these intersection are
already contained in the previous ones, and thus do not change
the result).

To complete the decryption of our formula~\eqref{typeapp}, it remains
the case in which $I$ is not a singleton: when we have a union of
function types, then we can apply a function of this type only to
arguments that are in the intersection of the domains of the types
that form the union and, therefore, we are sure that whatever the
actual type of the function will be the argument will be a subtype of
the domain of the actual type. However we do not know which result
type to pick among the possible ones: since it may be any of them,
then we take all of them, that is their union. This explains the outermost union in~\eqref{typeapp}. For instance, if we
have a function whose static type is a union of two arrows:
\begin{equation}
\texttt{(\,\,(Even,Int)\,\To\,0\,\,) \Or (\,\,(Int,Odd)\,\To\,1\,\,)}
\end{equation}
then it may dynamically be a function of either of the two types. Thus
the static typing must cover all possible cases: such a function must
be applied to arguments that are compatible with both arrows---i.e.,
arguments that are in the intersection of the two
domains: \texttt{(Even,Odd)}---; since the result of this application
can be in either of the result types, then the application of such a
function to an argument in the intersection of the domains will be
typed by their union, that is \texttt{0\Or 1}. 

\subsubsection{Classes}
\label{tyclasses}

I already discussed how to type-check classes in Section~\ref{covcon}:
one can use very sophisticated type systems for classes (there is a plethoric literature
on the subject) but the bare minimum is that $(i)$  methods have the type
they declare,  $(ii)$ that overriding methods are typed by a subtype of the type of the
methods they override,  and $(iii)$ that the type of read and write instance
variables must be the same in all subclasses.

In my discussion I considered just the very basic features of class-based object-oriented programming, since a more detailed treatment would have led me far away from the purpose this work. There is however a feature that I omitted in the discussion and that seems worth mentioning, namely the case of multiple inheritance. In many languages (Perl 6 included) it is possible to define a class as a subclass of several distinct classes: for instance, I could have defined the \texttt{ColPoint} class in this way:
\begin{alltt}
\k{class} Point \{
     \k{has} $.x \k{is} rw;
     \k{has} $.y \k{is} rw;\bigskip
     \k{method} equal(Point $p) \{
         ($.x==$p.x)&&($.y==$p.y)
     \}
 \};

\end{alltt}
\begin{alltt}
 \k{class} Color \{
     \k{has} Str $.c \k{is} rw;\bigskip
     \k{method} iswhite() \{
         \k{return} ($.c=="white");
     \}
 \};

\end{alltt}
\begin{alltt}
 \k{class} ColPoint \k{is} Point \k{is} Color \{\bigskip
    \k{method} move(Int $dx, Int $dy) \{
         \k{if} not(\k{self}.iswhite()) \{
            $.x += $dx;
            $.y += $dy;
         \}
         \k{return} \k{self};
     \}
 \};
\end{alltt}
A problem with multiple inheritance is how to deal with methods that
are defined in more than one superclass. For instance, which
method should be selected for \texttt{\$a.equal(\$a)} if \verb|$a| is
an instance of the class \ColPoint\ and the class \texttt{Color} had
defined a method \texttt{equal\,:\,Color\,\To\,Bool}? The interpretation
given in Section~\ref{synsugar} to classes and the definition of
ambiguity give us an immediate answer to this, since therein \texttt{equal}
would be the following multi-subroutine
\begin{alltt}
  \k{multi sub} equal(Point \s{$self} \k{;;} Point $p) \{
    (\s{$self}.x==$p.x)&&(\s{$self}.y==$p.y)  \};

  \k{multi sub} equal(Color \s{$self} \k{;;} Color $p) \{
    \s{$self}.c==$p.c \};
\end{alltt}
Adding the class \texttt{ColPoint} corresponds to adding a common subtype
to \Point\ and \texttt{Color}. As a consequence, the
type \Point\And\texttt{Color} is no longer empty (it contains all
the \texttt{ColPoint} objects) and therefore the multi-subroutine
above no longer is free from ambiguity. The solution in this case is
straightforward: to satisfy the ambiguity condition the
class \ColPoint\ must provide its own definition of the \texttt{equal}
method (in Perl 6 this is what it is done with ``roles''). This is
however problematic when multi-subroutines are not the result of
class definitions but are defined directly: defining a class by
multiple inheritance from some other classes means creating a new
subtype common to all these classes; therefore a
multiple subroutine whose parameters are typed by some of those classes and that was ambiguity-free might,
because of the new class definition, break the ambiguity condition. This means that the ambiguity condition must thus be rechecked for all
such multi subroutines, thus hindering the compositionality and
modularity of the type system.

\subsection{Records}\label{tyrecords}

Although records (also known as ``structs'' in C(++), ``structures''
in Visual Basic, objects in JavaScript, \ldots) are pervasive in
programming, I did not explain how to handle them. It is time to
remedy to this omission by this section%
\ifLONG%
\ (which is the most technical section of the presentation and can be skipped at first reading)%
\fi
.  In their simplest form records are finite
maps from ``labels'' to values which are equipped with a selection
operation that returns the value associated to a label by the
record. In Perl\,6 labels are strings and records (i.e., ``hashes'' in
Perl parlance) are defined by the syntax
\texttt{\{\,$\ell_1$ => $e_1$,\ldots,$\ell_n$ => $e_n$\,\}} which denotes
the record that associates the label $\ell_i$ to (the value returned
by) the expression $e_i$, for $i\in [1..n]$. For instance, if we
write
\begin{alltt}
   \k{my} $x = \{
        "foo" => 3,
        "bar" => "foo"
      \};
\end{alltt}
then the variable \texttt{\$x} denotes the record that associates the
label/string \texttt{foo} to the integer \texttt{3} and the
string \texttt{bar} to the string \texttt{foo}. When the same label is  multiply defined, then the rightmost association is taken.
Selection in Perl\,6
is denoted by \texttt{$e$<$\ell$>} which returns the value associated
to the label $\ell$ in the record returned by the expression $e$. In
the example above \texttt{\$x<foo>} (note the absence of the double
quotes around the label) returns the integer \texttt{3}
and \texttt{\$x<bar>} returns the string \texttt{foo}. The selection of a label that is not present in the record returns  \texttt{undef}, a special value that when used in any context causes a runtime error. Perl provides syntax to initialize undefined fields and to associate new values to defined ones. To model (and type) these operations, as well as the definition of records with multiple fields, I build records starting by single-field records that are merged by the record concatenation operator ``\texttt{+}'' that, in case of multiply defined labels, gives priority to the rightmost fields. For completeness, I will also consider field deletion, that is, I will use the record expression $e$\texttt{\char`\\}$\ell$ whose result is the record value returned by $e$ in which the field labeled by $\ell$, if any, is undefined.
The record expressions that I consider henceforth are then defined by the following grammar:
\begin{equation}\label{recexp}
e\;\quad ::=\quad \; \texttt{\{\,$\ell$\,=>\,$e$\,\}}\quad | \quad e\texttt{+}e\quad | \quad e\texttt{\char`\\}\ell\quad | \quad e\texttt{<$\ell$>}
\end{equation}
For instance, the record associated to the variable \texttt{\$x} in the previous expression is written in this syntax \texttt{\{\,foo\,=>\,3\,\}\,+\,\{\,bar\,=>\,"foo"\,\}}. However, for the sake of simplicity in examples I will still use the syntax  \texttt{\{\,foo\,=>\,3\,,\,bar\,=>\,"foo"\,\}}. 

In what follows I show how to type this simple form of
records. However, in many programming languages records have much a
richer set of operations and features. I will discuss some of them at
the end of this section.

In Perl\,6 records are implemented by the class \texttt{Hash} which
provides a very rich set of features for them. This means that Perl\,6
lacks specific types for records. So I will depart from what I did so
far in this article, and instead of borrowing the syntax of Perl types I will use my own syntax for record types. More precisely, to type
fields I will use either the syntax $\ell\col\T$ or (for optional
fields) the syntax $\ell\,\texttt{?:}\,\T$. The former means that
selecting the label $\ell$ will return a value of type $\T$, the
latter that the same operation will return either a value of type $\T$
or \texttt{undef}. This syntax will be used in the record types which
come in two flavors: closed record types whose values are records with
exactly the fields specified by the type, and open record types whose
values are records with \emph{at least} the fields specified by the
type. I use $\patrec{f_1,\ldots,f_n}$ for the former and
$\patorec{f_1,\ldots,f_n}$ for the latter (where the $f_i$'s denote
either of the two forms of field typing I described before) and use \R\
to range over either kinds of record type expressions. For instance,
the record \texttt{\{\,foo\,=>\,3\,,\,bar\,=>\,"foo"\,\}} of our 
example can be given several distinct types. Among them we
have: \texttt{\{\,foo\,:\,Int\,,\,bar\,:\,Str\,\}}, the closed record
type that includes all the records with exactly two fields, one that maps
the label \texttt{foo} to integer values and the other that maps the
label \texttt{bar} to string values; \patorec{\texttt{foo\col{}Int}}
the open record type formed by all the records that contain at least a
field \texttt{foo} containing integers; \patorec{\texttt{foo\,?:\,Int
, pol\,?:\,Bool\,}} the open record type with two optional
fields; \patrec{\texttt{foo\,:\,Int, bar\,?:\,Str, pol\,?:\,Bool\,}}
the closed record type in which one of the two fields that are present is
declared optional. Finally the record of our example has also
type \toprec{}, that is, the type that contains only and all record values (\toprec{} is the top type for record kind, that is,
$\Any_{\mathtt{recd}}$).

The goal of this section is to illustrate the algorithms that assign
types to expressions of records, as illustrated in the example
above. Deducing that \texttt{\{\,foo\,=>\,3\,,\,bar\,=>\,"foo"\,\}} 
has type \texttt{\{\,foo\,:\,Int\,,\,bar\,:\,Str\,\}} is
straightforward. Instead, to deduce that it has
type \patorec{\texttt{foo\col{}Int}} or to infer the types of the operations of record
selection and concatenation  is more difficult. The former
deduction is obtained by subsumption\footnote{~Subsumption characterizes the fact that every expression of a given type can be also given any super type of that type.}  (likewise for the deduction of \toprec{} type), the two latter deductions need the definition
of some operations on record \emph{types}. Thus I need first to explain the
semantics of record types, then their subtyping relation, and finally
some operations on them.  At the beginning of this section I
introduced records, as customary, as finite maps from an infinite set of labels $\mathcal{L}$ to values. In what follows I use
a slightly different interpretation and consider records as specific total maps
on $\mathcal{L}$, namely, maps that are constant but on a finite subset of the
domain. So in this interpretation the
record \texttt{\{\,foo\,=>\,3\,,\,bar\,=>\,"foo"\,\}} maps the 
label \texttt{foo} into the value \texttt{3}, the label \texttt{bar}
into the value \texttt{"foo"}, and for all the other labels it is the 
constant function that maps them into the special
value $\bot$ (i.e., the \texttt{undef} value of Perl).

Formally, let $Z$ denote some set, a function $r:\mathcal{L}\to Z$ is
\emph{quasi-constant} if there exists $z\in Z$ such that the set
$\lbrace\ell{\in}\mathcal{L}\mid r(\ell)\not=z\rbrace$ is finite; in this case
I denote this set by $\textsf{dom}(r)$ and the element $z$ by
$\textsf{def}(r)$ (i.e., the \textsf{def}ault value). I use $\mathcal{L}\rightarrowtriangle Z$ to denote
the set of quasi-constant functions from $\mathcal{L}$ to $Z$ and the
notation $\lbrace\ell_1=z_1, \ldots,\ell_n=z_n,\,\underline{~~}=z\rbrace$ to denote the
quasi-constant function $r:\mathcal{L}\rightarrowtriangle Z$ defined
by $r(\ell_i)=z_i$ for $i=1..n$ and $r(\ell)=z$ for
$\ell\in\mathcal{L}\setminus\lbrace\ell_1,\ldots,\ell_n\rbrace$. Although this
notation is not univocal (unless we require $z_i\not=z$ and the $\ell_i$'s to be pairwise distinct), this is largely sufficient for the
purposes of this section.

Let $\bot$ be a
distinguished constant (one not in \Any). I single out two particular sets of quasi-constant functions:
the set $\texttt{string}\rightarrowtriangle\textbf{Types}\cup\lbrace\bot\rbrace$, called the set
of \emph{quasi-constant typing functions} ranged over by $r$; and
$\texttt{string}\rightarrowtriangle\textbf{Values}\cup\lbrace\bot\rbrace$
the set of record values. The constant $\bot$ represents the value of the fields
of a record that are ``undefined''. To ease the presentation I use
the same notation both for a constant and the singleton type that
contains it: so when $\bot$ occurs in
$\texttt{string}\rightarrowtriangle\textbf{Values}\cup\lbrace\bot\rbrace$ it denotes a value,
while in $\texttt{string}\rightarrowtriangle\textbf{Types}\cup\lbrace\bot\rbrace$ it denotes
the singleton type that contains only the value $\bot$.

Given the definitions above, it is clear that the record type
expressions I defined earlier in this section are nothing but
specific notations for some quasi-constant functions in
$\texttt{string}\rightarrowtriangle\textbf{Types}\cup\lbrace\bot\rbrace$. More
precisely, the open record type
$\patorec{\ell_1\col \T_1,\ldots,\ell_n\col \T_n}$ denotes the
quasi-constant function
$\lbrace\ell_1= \T_1,\ldots,\ell_n=\T_n,\underline{~~}=\pator{\Any}{\bot}\rbrace$
while the closed record type
$\patrec{\ell_1\col \T_1,\ldots,\ell_n\col \T_n}$ denotes the
quasi-constant function
$\lbrace\ell_1= \T_1,\ldots,\ell_n=\T_n,\underline{~~}=\bot\rbrace$. Similarly,
the optional field notation $\patrec{\ldots, \ell\texttt{?:}\T,\ldots}$
denotes the record typing in which $\ell$ is mapped either to $\bot$
or to the type $\T$, that is, $\lbrace\ldots, \ell
= \pator{\T\,}{\bot},\ldots\rbrace$. In conclusion, the open and closed
record types are just syntax to denote specific quasi-constant
functions in
$\texttt{string}\rightarrowtriangle\textbf{Types}\cup\lbrace\bot\rbrace$.

\paragraph{Subtyping.}

The first thing to do is to define subtyping for record types, that
is, to extend the algorithm I described in Section~\ref{subalgo} to
handle record type expressions. The algorithm proceeds as in
Section~\ref{subalgo} with the difference that the simplification
performed in \textbf{\emph{Step 3}} may yield besides the
cases~\eqref{atomtag}--\eqref{atomfun} the following case:
\begin{equation}\label{atomrec}
 \displaystyle\bigwedge_{p\in P} \!\R_p\quad~\And~\bigwedge_{n\in N} \!\!\Not{\R_n}
\end{equation}
where, I remind, every type $\R_i$ in this formula is a record type
atom, that is, it is either of the form
$\patrec{\ell_1\col\T_1\texttt{,}\ldots\texttt{,}\ell_n\col\T_n}$ or of
the form
$\patorec{\ell_1\col\T_1\texttt{,}\ldots\texttt{,}\ell_n\col\T_n}$. Then \textbf{\emph{Step
4}} simplifies the instances of this new case by using the containment
property for quasi-constant functions (see Lemma~9.1
in~\cite{AlainThesis}) specialized for the case in~\eqref{atomrec}
(remember that the \R's in the formula above denote special cases of
quasi-constant functions.) In particular to decide the emptiness of
the type in~\eqref{atomrec} we decompose the problem into checking
that for every map $\iota: N\to L\cup\lbrace\underline{~~}\rbrace$
\begin{equation}
\hspace*{-3mm}\begin{array}{ll}
\displaystyle\left(\exists \ell \in L. (\;\bigwedge_{p\in P} \!\R_p(\ell)~~~\Sub\!\!\bigvee_{n\in N | \iota(n)=\ell}\!\!\!\!\R_n(\ell)\;)\right)\textsf{ or }\\
\displaystyle\left(\exists n_\circ\in N .(\;(\iota(n_\circ )= \underline{~~})\textsf{ and } (\bigwedge_{p\in P} \!\textsf{def}(R_p)\;\Sub\;\textsf{def}(\R_{n_\circ}))\;)\right)
\end{array}
\end{equation}
where $L=\displaystyle\bigcup_{i\in P\cup N}\textsf{dom}(\R_i)$.\\
The technique used to derive~\eqref{implem} can be adapted to this case as well (just by branching on $L\cup\lbrace\underline{~~}\rbrace$ rather than a single element).

\paragraph{Type operators.}
Let $\T$ be a type and $\R_1, \R_2$ two record type atoms. The \emph{merge} of  $\R_1$, and $\R_2$ with respect to $\T$, noted $\oplus_{\texttt{T}}$ and used infix, is the quasi-constant typing function defined as follows:
\[
(\R_1\oplus_{\texttt{T}} \R_2)(\ell) \eqdef \left\lbrace
    \begin{array}{ll}
       \R_1(\ell) &\textrm{if } \patand{\R_1(\ell)}{\T}\leq\Empty\\
       \pator{(\R_1(\ell)\minus \T)}{\R_2(\ell)} & \textrm{otherwise}
     \end{array}\right.
\]
Record types are akin to product types (a classical interpretation of records is that of products indexed over a finite domain, e.g.~\cite{BL90}). As products commute with intersections  (cf.\ the discussion at the end of Section~\ref{se:products}) so do record types and, therefore, every type containing only record values can be expressed as a union of just record type atoms. In other terms if $\T\leq\toprec{}$, then $\T$ is equivalent to a finite union of the form
$
\bigvee_{i\in I}\R_i
$ (see Lemma~11 in~\cite{BCNS13} for details).
So the definition of \emph{merge} can be easily extended to all record types as follows
\[
(\bigvee_{i\in I}\R_i)\oplus_{\texttt{T}}(\bigvee_{j\in J}\R'_j)\eqdef\bigvee_{i\in I,j\in J}(\R_i\oplus_{\texttt{T}}\R'_j)
\]
The merge operator can be used to define concatenation and field deletion on record types. Let $\T_1$, $\T_2$, and $\T$ range over record types (i.e., types smaller than $\toprec$). Define:
\begin{eqnarray}
\T_1 + \T_2 & \eqdef & \T_2 \oplus_\bot \T_1\label{concat}\\
\T\setminus \ell & \eqdef& \lbrace\ell =\bot , \underline{~~} = c_\circ\rbrace \oplus_{c_\circ} \T\label{deletion}
\end{eqnarray}
where $c_\circ$ is some constant/tag different from $\bot$ (the semantics of the operator does not depend on the choice of $c_\circ$ as long as it is different from $\bot$).

Notice in particular that the result of the concatenation of two record type atoms $\R_1+\R_2$ may result for each field $\ell$ in three different outcomes:
\begin{enumerate}
\item if $\R_2(\ell)$ does not contain $\bot$ (i.e., the field $\ell$ is surely defined), then we take the corresponding field of $\R_2$: $(\R_1+\R_2)(\ell) = \R_2(\ell)$
\item if $\R_2(\ell)$ is undefined (i.e., $\R_2(\ell)=\bot$), then we take the corresponding field of $\R_1$: $(\R_1+\R_2)(\ell) = \R_1(\ell)$
\item  if $\R_2(\ell)$ \emph{may} be undefined (i.e.,  $\R_2(\ell)=\pator{\T}{\bot}$ for some type $\T$), then we take the union of the two corresponding fields since it can results either in $\R_1(\ell)$ or  $\R_2(\ell)$ according to whether the record typed by $\R_2$ is undefined in $\ell$ or not:  $(\R_1+\R_2)(\ell) = \pator{\R_1(\ell)}{(\R_2(\ell)\setminus\bot)}$.
\end{enumerate}
This explains some weird behavior of record concatenation such as $\patrec{a\col\Int,b\col\Int}+\patrec{a\texttt{\;?:\;}\Bool}$ $= \patrec{a\col\pator{\Int}{\Bool},b\col\Int}$ since ``$a$'' \emph{ma}y be undefined in the right hand-side record while ``$b$'' \emph{is} undefined in it, and $\patrec{a\col\Int}+\patrec{\textbf{..}}= \patrec{a\col\Any,\textbf{..}}$ stating that at least the field ``$a$'' is defined (since ``$a$'' in the right hand-side record is defined as $a\mapsto\pator{\Any}\bot$, then the third case of the above outcomes applies). As an aside, notice the type $\patrec{a\texttt{\;?:\;}\Empty,\textbf{..}}$ which states that at least the field ``$a$'' is undefined; typically, this is obtained when deleting the field $a$ (e.g., if $e$ is of type $\patrec{\textbf{..}}$, then $e\minus a$ has type  $\patrec{a\texttt{\;?:\;}\Empty,\textbf{..}}$).

\paragraph{Typing.}
All it remains to do is to give the typing rules for the expressions of grammar~\eqref{recexp}. Thanks to the definitions~\eqref{concat} and~\eqref{deletion} this is really straightforward:
\begin{description}
\item[\emph{record}] if $e$ has type $\T$, then  \texttt{\{\,$\ell$\,=>\,$e$\,\}} has type $\patrec{\ell\col\T}$;
\item[\emph{concatenation}] if $e_1$ has type $\T_1$, $e_2$ has type $\T_2$, and both $\T_1$ and $\T_2$ are subtypes of $\toprec$, then   $e_1\texttt{+}e_2$ has type $\T_1+\T_2$;
\item[\emph{deletion}] if $e$ has type $\T$ and $\T$ is a subtype of $\toprec$, then  $e\minus\ell$  has type $\T\setminus\ell$;
\item[\emph{selection}] if $e$ has type $\T$ and $\T$ is a subtype of $\toprec$, then $\T$ is equivalent to a union type of the form $\bigvee_{i\in I}\R_i$ and the type of  e\texttt{<$\ell$>} is  $\bigvee_{i\in I}\R_i(\ell)$.
\end{description}

\medskip\noindent
To conclude this section on records let me hint at the richer forms of records
you may encounter in other programming languages. A first improvement
you can find in several programming languages and in Perl is that the labels of a record expression may be computed by other expressions rather than being just constants. For instance,  it is possible to define:
\begin{alltt}
   \k{my} $y = \{
        $x<bar> => 3,
      \};
\end{alltt}
which associates to \texttt{\$y} the record that maps the
string \texttt{foo} to the integer \texttt{3} (where \texttt{\$x} is
defined as at the beginning of the section).  While this extension
greatly improves the versatility of records, it also greatly
complicates their static typing since in general it is not possible to
statically determine the label that will be returned by a given
expression. It is possible to give a rough approximation by using
union types when an expression is known to produce only a
finite set of strings (for an example of this technique the reader can
refer to Section~{4.1} of~\cite{BCNS13}) but the precision of such a
technique is seldom completely satisfactory.

A further improvement is to allow  the label in a selection
expression to be computed by an expression, too. In Perl\,6 this is done by
the expression \texttt{$e_1$\{$e_2$\}} which returns the value
associated to the string returned by $e_2$ in the record returned by
$e_1$. Back in the example of the beginning of this section we
have that the four
expressions \texttt{\$x<foo>}, \texttt{\$x\{"foo"\}}, \texttt{\$x\{\$x<bar>\}}, 
and \texttt{\$x\{\$x\{"bar"\}\}} all return the integer \texttt{3} (as 
a matter of fact, in Perl\,6 \texttt{$e$<$s$>} is syntactic sugar
for \texttt{$e$\{"$s$"\}}). Such an extension makes records 
equivalent, to all intents and purposes, to arrays indexed over a
finite set of strings (rather than an initial segment of
integers). These in other languages are called associative
maps/arrays/lists. Once more the extension improves versatility while
complicating static typing and one can try to adapt the techniques
suggested for the previous extension also to this case. Finally, some
languages (e.g., MongoDB) consider the fields of a record to be
ordered and allow the fields of a record to be selected by their
order number, as if they were an array. The techniques presented in
this section can be hardly adapted to such an extension.

\ifLONG%
\subsection{Summary for the electrical blueprint}
\fi

The algorithms I presented in this part are essentially those defined
by Alain Frisch in his PhD.\ thesis~\cite{AlainThesis} and implemented
in the compiler of \cduce. These algorithms are directly derived from
the set-theoretic interpretation of types. This fact, not only allows
us to precisely define the semantics of the types and of the subtyping
relation, but also makes it possible to optimize the algorithms by
applying classic set-theoretic properties. Furthermore, by using
well-known and robust data structures to represent Boolean
functions such as the BDDs, it was possible to modularize the
implementation according to the kinds of type constructors. This makes
it possible to avoid expensive normalization phases and makes it much
easier to extend the system to include new type constructors.

Although I gave a pretty complete presentation of the algorithms and
data-structures that efficiently implement a type system based on
semantic subtyping, this is not the whole story. The actual
implementation has to use hash-tables for efficient memoization,
hash-consing for efficient manipulation of types, determine the best
policy for node sharing in BDDs, tailor representations of basic data
according to the application domain of the language (for instance in
\cduce{} strings use a representation tailored for a lazy implementation of basic operators), and so on. However, even a naive
implementation of the algorithms of this section should display pretty
decent performances.

While I described in details how to check types and subtypes, I
completely omitted any discussion about what to do when these checks
fail, that is, I did not discuss the art (or black magic) of producing
meaningful error messages. I did not discuss it since this would lead
us quite far away, but I want at least to stress that the
set-theoretic interpretation of types comes quite handy in these
situations. Type-checking fails only when a subtyping check does: the
programmer wrote an expression of some type \S\ where an expression of
type \T, not compatible with \S, was expected. In order to produce a
useful error message the system can compute the type $\S\minus\T$
and show to the programmer some default value in this type. This is an
example of value that might be produced by the expression written by
the programmer and make the program fail. Our experience
with \cduce{} shows that in many occasions producing such a sample value is well
worth any other explanation.


%% file: theory.tex
\ifLONG
A survey on the ``Types'' mailing list\footnote{Cf.\ the thread \url{https://www.seas.upenn.edu/~sweirich/types/archive/1999-2003/msg01314.html}. For the ``Types'' mailing list see \url{https://lists.seas.upenn.edu/mailman/listinfo/types-list}.} traces the idea of interpreting 
types as sets of values back to Bertrand Russell and Alfred Whitehead's
Principia Mathematica. Closer to our interests it seems that the idea
independently appeared in the late sixties early seventies and later
back again in seminal works by Roger Hindley, Per Martin-L\"of, Ed
Lowry, John Reynolds, Niklaus Wirth and probably others. More
recently, it was reused in the context of XML processing languages by
Hosoya, Pierce, and Vouillon~\cite{hosoya00regular,hosoya01patter,hosoya01thesis,xduce_toit}.
At this point of the presentation I can confess that I have been slightly cheating, since in Section~\ref{type} I presented the fact that a type is a set of values as an  unshakable truth while, in reality, such an interpretation holds, in the semantic subtyping framework, only for strict languages. As a matter of fact, the type system I presented in this work is unsound for non-strict languages, insofar as the application of a well-typed function to a diverging expression of type \Empty{} is always well typed (I leave as exercise to the reader [EX10] to show why this is unsound in a lazy language). In my defense, I would like to say that this is something  I realized just recently and that can be fixed with minimal effort, as shown in~\cite{types18b}.

\fi

The idea of using multiple definitions of methods to implement
covariant specialization of binary methods was first proposed in \myC
PhD thesis~\cite{Cas94phd,oobook} and the covariance and contravariance paper
\Ih wrote 20 years ago~\cite{Cas94}. This technique was dubbed
\emph{encapsulated multi-methods} in~\cite{BruEtAl96} and implemented
in different flavors for O$_2$~\cite{Cas95} and Java~\cite{BC97}. It
was based on the type theory
\ifANONYMOUS
Castagna, Ghelli, and Longo
\else
Giorgio Ghelli, Giuseppe Longo, and I
\fi
developed in~\cite{CGL92,CGL95} which was the first formal type
theory for multiple-dispatching: the conditions
of \emph{specialization soundness} (Definition~\ref{specialization})
and \emph{ambiguity freedom} (Definition~\ref{ambiguity}) were first
introduced there and are nowadays used by several multiple dispatching
programming languages %
\ifLONG%
such as MultiJava~\cite{multijava},
Fortress~\cite{fortress}, Cecil~\cite{cecil},
Dubious~\cite{dubious}, and Doublecpp~\cite{doublecpp}.

\fi

In this essay I revisited those ideas in the framework
of \emph{semantic subtyping}, that is, a type theory with a full set
of Boolean type connectives whose characterization is given in terms
of a semantic interpretation into sets of values. The first works
to use a semantic interpretation of types as sets of values
\ifLONG
in recent research in programming languages, are those
by Hosoya and Pierce
already cited above. Hosoya and Pierce used the semantic
interpretation essentially to characterize unions of possibly
recursive types, but were not able to account for higher order
functions.
\fi
The \emph{semantic subtyping} type system is the first and,
at this moment of writing, most complete work that accounts for a
complete set of Boolean connectives as well as for arrow types. It was
defined by
\ifANONYMOUS%
Frisch, Castagna, and Benzaken
\else
Alain Frisch, V\'eronique Benzaken, and myself
\fi
in~\cite{semsub02,FCB08}. A gentle introduction to the main concepts
of semantic subtyping can be found in the article for the joint
keynote talk \IC gave at ICALP and PPDP~\cite{CF05-ppdp} while the most
comprehensive description of the work is by far Alain Frisch's PhD
thesis~\cite{AlainThesis}%
\ifLONG
, a remarkable piece of work that I strongly recommend
if you are interested in this topic (and if you can read
French)
\fi
.

The use of semantic subtyping to revisit the covariance vs.\
contravariance problem brings two important novelties with respect to
the theory \IC used in the original co-/contra-variance
paper~\cite{Cas94}. First, the use of intersection types brings
a clear distinction between what belongs to the realm of the type theory
and what to the design of the language, specifically, the definition
of formation rules for expressions. In particular, we have seen that
types, their theory, and their subtyping relation are defined
independently of the particular language we apply them to. I tried
to clearly stress that conditions such as those of ambiguity
(Definition~\ref{ambiguity}) and specialization
(Definition~\ref{specialization}) concern the definition of the
expressions: they are given to ensure that expressions have an unambiguous semantics as well as definitions that match the programmer's intuition,
but they do not concern the theory of types. In~\cite{Cas94}, instead,
this difference was blurred, since these conditions were given for the
formation of types rather than for the formation of expressions. So,
for instance, in~\cite{Cas94} a type such as~\eqref{equaltype} was
considered ill formed while in the semantic subtyping framework this
type is legitimate (it is rather the multi-subroutine declared of having
such a type that is likely to be ill-formed.) A second, more technical
difference is that in~\cite{Cas94} it was not possible to compare an
arrow with an intersection of arrows (arrows and intersections were
considered different type constructors) and this posed problems of
redundancy (what is the difference between a function and an
overloaded function with just one arrow?) and modularity (it is not
possible to specialize the type of a function by adding new code unless it
is already
\ifLONG
overloaded; this, transposed to the context of this paper means that multi-methods can only override other multi-methods, thus hindering the solution of Section~\ref{synsugar}%
\fi).

\ifLONG
A big advantage of semantic subtyping is that it comes with robust and
optimal algorithms that work well in practice. They are implemented in
the language \cduce, whose distribution is free and
open-source~\cite{cduce} and the reader can use the interactive toplevel of that
language to play and experiment with set-theoretic types.  For
instance, to test the subtyping relation of equation~\eqref{sei} for
types \Int\ and \texttt{Char} one can use the \texttt{debug} directive as
follows (the hash symbol \texttt{\#} is the prompt of the interactive toplevel while
italics is used for the toplevel's answer):
\begin{alltt}\small
# debug subtype ((Int -> Char) & (Char -> Int))
                ((Int | Char) -> (Int | Char));;
\color{darkblue}\em[DEBUG:subtype]
Char -> Int & Int -> Char <= (Char | Int) -> (Char | Int)
: true
\end{alltt}
These algorithms, which are the ones I outlined in Section~\ref{electric}, are precisely defined
in~\cite{FCB08}. There the reader will find the formal proofs that the
decompositions used in {\emph{Step 4}} are sound and complete (see in
particular Section~6.2 of~\cite{FCB08}). The two decompositions that transform the formula~\eqref{atomprod} into~\eqref{prodsubtyping} and the formula~\eqref{atomfun} into~\eqref{arrowsubtyping} are
the generalizations to type connectives of the classic subtyping rules
for products and arrows, respectively, as they can be found in subtyping system with syntax-oriented definitions.
This point can better be grasped by considering the particular
cases of the intersections in~\eqref{atomprod}
and~\eqref{atomfun} when both $P$ and $N$ contain exactly one
type. Then checking the emptiness of~\eqref{atomprod}
and~\eqref{atomfun} corresponds to checking the following two
subtyping relations
\begin{eqnarray*}
\pair{\S_1}{\S_2}&\Sub&\!\!\pair{\T_1}{\T_2}\\
{\S_1}\To{\S_2}&\Sub&{\T_1}\To{\T_2}
\end{eqnarray*}
and I leave as exercises to the reader [EX11, EX12] to check that in these cases the two decomposition formulas~\eqref{prodsubtyping} and~\eqref{arrowsubtyping} of \emph{Step 4} become, respectively:
\[
\begin{array}{c}
(\S_1\Sub\Empty)\textsf{ or }(\S_2\Sub\Empty)\textsf{ or } (\S_1\Sub\T_1\textsf{ and }\S_2\Sub\T_2)\\[2mm]
(\T_1\Sub\Empty)\textsf{ or } (\T_1\Sub\S_1\textsf{ and }\S_2\Sub\T_2)
\end{array}
\]
These are nothing but the classic subtyping rules specialized for the
case in which types may be empty.

For defining the functions $\Phi$ and $\Phi'$ I drew my inspiration
from the algorithms defined in Chapter~7 of Alain Frisch's PhD
thesis~\cite{AlainThesis} and those used in the compiler
of \cduce.

For an alternative presentation of the material of
Section~\ref{electric}, I warmly recommend the on-line
tutorial \emph{Down and Dirty with Semantic Set-theoretic
Types}~\cite{SST-Tutorial} in which Andrew M.\ Kent reexplains
the implementation details of Section~\ref{electric} in his own style.

The theory of quasi-constant functions and its application to record
types is described in Chapter~9 of Alain Frisch's PhD
thesis~\cite{AlainThesis}, where the proof of soundness and
completeness of the decomposition rules I gave in
Section~\ref{tyrecords} can also be found.

The data structures I described are  those used in the implementation of
the language \cduce\ and described in details in Chapter~11 of Alain
Frisch's PhD thesis~\cite{AlainThesis}. The only difference is that
the structures used in the compiler of \cduce\ to implement BDDs and types, contain
some extra fields, typically for storing hashes (to perform efficient
comparison and hash-consing), and for pretty printing and error
messages. Also the structure representing types includes more fields to
account for extra kinds of atoms, namely, unicode characters, record
types, and XML types.

\fi

Semantic subtyping is a general and streamlined theory that can
be adapted to other settings.
\ifLONG%
One of the most recent results about semantic subtyping
\fi
is that it can be extended with parametric
polymorphism. It is possible to add type variables to the types I presented in
\ifLONG%
this paper (namely, those of the grammar in~\eqref{concretetypes})
\fi
and leave to the type system the task to deduce how to
instantiate them, as it is done in languages such as OCaml and
Haskell.
\ifLONG
Explaining here how to do it would have lead us too far.
\fi
The interested
reader can refer to the article \emph{Polymorphic Functions with Set-Theoretic Types}, published in two
parts~\cite{polyduce1,polyduce2}, that describes how to define
polymorphic functions, to type them and to implement them in an efficient
way.
\ifANONYMOUS
\else
The work on polymorphic functions is based on the extension of
semantic subtyping to polymorphic types, which was defined
in~\cite{CX11}. To learn how to do ML-like type reconstruction in languages with polymorphic
set-theoretic types, I warmly recommend the PhD thesis manuscript of
Tommaso Petrucciani~\cite{Pet19phd} which is written in a clear and
pedagogical style. There you will find a gentle introduction to
polymorphic semantic subtyping, the definition of constraint
generation and constraint resolution algorithms for type
reconstruction, as well as an account on the integration of gradual typing
with set-theoretic polymorphic types. For more information about the
latter and a detailed description on how to perform gradual typing in the presence of set-theoretic
types the reader can refer to~\cite{CLPS19}.
\fi


%% file: conclusion.tex
I wrote this work as a challenge: to introduce sophisticated type
theory to average functional programmers and to do it by using a
popular programming language such as Perl that was not conceived with
types in mind. The goal was to show that when a type system is well
designed, it can be explained to programmers in very simple terms,
even when its definition relies on complex theories that are
prerogative of specialists: hopefully what programmers must retain of this system should all sum up to the 6+2 rules I gave
in Sections~\ref{lesson1} and~\ref{lesson2}.
\ifLONG%
I pushed the experiment further and in
Section~\ref{electric} I tried also to explain to potential language
designers, the main implementation techniques for these types. Once
more, I aimed at demonstrating that it is not necessary to be a
researcher to be able to implement this kind of stuff: so instead of
explaining why, say, the decomposition rules in \emph{Step 4} of the
subtyping algorithm are correct, I preferred to explain how to implement
these decompositions in a very efficient way. Whether I succeeded in this challenge or
not, is not up to me to say. I just hope that by reading this paper
some eventual language designers will have learned a few basic notions
and techniques so as not to start the design of their language from
scratch.
\fi

Personally, what I learned from this work is that you should fit
programming languages to types and not the other way round, insofar as
a type theory should be developed pretty much independently of the
language (but, of course, not of the problem) it is to be applied to.  This
observation is quite arguable and runs contrary to common practice according to which type
theories are developed and fitted to overcome some problems in
particular languages (even though it is what I have been doing for the
last ten years%
\ifANONYMOUS\else
\ with the semantic subtyping approach%
\fi
). I reached such a
conclusion not because this paper adapts a type theory (semantic
subtyping) to a language for which it, or any other type theory, was
not conceived (Perl), but because it shows the limitations of my own work, that is, the type
system (but was it really a type system?)
\ifANONYMOUS
I and other colleagues contributed to develop
\else
my colleagues and I developed
\fi
twenty years
ago for the covariance vs.\ contravariance problem. In that system you
were not allowed to have a type $(\S_1\To\T_1)\And(\S_2\To\T_2)$ with
$\S_2\Sub\S_1$ and $\T_2{\not\!\!\Sub}\T_1$: such a type was
considered ``ill-formed'' and thus forbidden. With (twenty years of) hindsight that
was a mistake. What the theory of semantic subtyping shows is that a
type as the above is and must be admissible as long as it is clear
that a function with that type applied to an argument of type $\S_2$
will return results in $\T_1\And\T_2$. It now becomes a problem of
language design to devise function definitions that make it clear to
the programmer that the functions she/he wrote have this property. A
way to do that is to design the language so that whenever
$\S_1\Sub\S_2$ the type printed for and thought by the programmer for
a function of type $(\S_1\To\T_1)\And(\S_2\To\T_2)$ is instead
$(\S_1\To\T_1)\And(\S_2\To\T_1\And\T_2)$---which, by the way, is an equivalent type, in the sense that both types denote the same set of functions. This
is what the covariance condition does: when adding a new code for
$\S_2$ inputs to a function of type $\S_1\To\T_1$ with $\S_2\Sub\S_1$,
it forces the programmer to write this code so that the return type
$\T_2$ declared by the programmer for this code is a subtype of $\T_1$, so that the types $\S_2\To\T_1\And\T_2$
and $\S_2\To\T_2$ are exactly the same. Likewise, a type
$(\S_1\To\T_1)\And(\S_2\To\T_2)$ with $\S_1\And\S_2$ non-empty is a
perfectly fine type (while in the type-system of the original
covariance vs.\ contravariance paper,
\ifANONYMOUS
it was banned
\else
we banned it
\fi
because it did not
respect the ambiguity free condition).  Again, it is a language design
problem to ensure that whenever we have a function of that type,
then the code executed for each combination of the types of the arguments is not only
unambiguously defined, but also easily predictable by the
programmer.

In conclusion the, deliberately provocative, lesson of
this work is that in order to solve type-related problems, you must first
conceive the types and only after you can think of how to design a language
that best fits these types.


%% file: acks.tex
The first version of this work was made available on my web page in
2013. Since then, this work benefited from the reading and suggestions
of many people, and I regularly updated it to take this feedback into
account (which partially explains the unreasonable number of footnotes
and parenthetic sentences present in this text). In particular, I want
to thank the students of my course in advanced programming languages
at the \emph{\'Ecole Normale Sup\'erieure de Cachan} as well as the
members of the \texttt{Perl6-language} mailing list who found errors
and gave several suggestions on how to improve the presentation, among
whom Brent Laabs who suggested the encoding of
Footnote~\ref{encoding}, Yary Hluchan who suggested the code in
Footnote~\ref{interesting} and Todd Hepler. Andrew M.\ Kent found some
really subtle errors in Section~\ref{electric} and suggested the right
fixes for some of them: to fix the others I profited of the invaluable
help of Tommaso Petrucciani. Last but not least, I want to thank
Anita Badyl, Juliusz Chroboczek, Jim Newton, Kim Nguyen, Luca
Padovani, and the anonymous reviewers who provided useful and detailed
feedback.

Despite all the help listed above, the version published on LMCS on February
2020 still contained some errors. A subtle one in the definition of
difference of BDDs with lazy unions was spotted by James Clark when
implementing it for the Ballerina programming language who proposed
the right definition that is given in the corrected version of December 2021.


%% file: exercises.tex
\paragraph{[EX1] Problem:} find a function that is in~$(\S_1\Or\S_2)\To(\T_1\Or\T_2)$  and not in $(\S_1\To\T_1)\,\And\,(\S_2\To\T_2)$.

\smallskip\noindent
\emph{Solution:} this is possible only if both $\S_1\To\T_1\not\!\!\!\Sub\;\S_2\To\T_2$ and vice versa. So let us suppose that both $\S_1{\minus}\S_2$---i.e., $\S_1\And\Not{\S_2}$, that is the difference between $\S_1$ and $\S_2$---and  $\T_2{\minus}\T_1$ are not empty types. Then, any function in $(\S_1\To\T_1)\,\And\,(\S_2\To\T_2)$ that maps at least one value in $\S_1\minus\S_2$ into a value in $\T_2\minus\T_1$ is a solution of the problem.

\bigskip
\paragraph{[EX2] Problem:} prove the relations in Footnote~\ref{extrarules} and their converse

\smallskip\noindent
\emph{Solution:} Every function in $(\S\To\T)\And(\S\To\U)$ is a function that when applied to a value in $\S$ it returns a result in $\T$ \emph{and}  when applied to a value in $\S$ it returns a result in $\U$; this means that when such a function is applied to  a value in $\S$ it returns a result in $\T\And\U$ and, therefore, it is a function in $\S\To\T\And\U$.
Conversely, every function in $\S\To\T\And\U$ when applied to a value
in $\S$ it returns a result in $\T\And\U$; thus, in particular,
when it is applied to a value in $\S$ it returns a result in
$\T$ \emph{and} when applied to a value in $\S$ it returns a result in
$\U$, and therefore it is $(\S\To\T)\And(\S\To\U)$.

Every function in $(\S\To\U)\And(\T\To\U)$ is a function that when
applied to a value in $\S$ or to a value in $\T$ it returns a result
in $\U$; therefore when it is applied to a value in $\S\Or\T$ it
returns a result in $\U$, and thus it is in
$\S\Or\T\To\U$. Conversely, every function in $\S\Or\T\To\U$ when applied
to a value in $\S\Or\T$ it returns a result in $\U$, therefore, a
fortiori, when \emph{and} when applied to a value in $\S$ it returns a
result in $\U$ and so does for values in $\T$; therefore it is a
function in $(\S\To\U)\And(\T\To\U)$.

\bigskip
\paragraph{[EX3] Problem:} check that
\begin{alltt}
 \k{multi} \k{sub} sum(Int $x, Int $y) \{ $x + $y \}
 \k{multi} \k{sub} sum(Bool $x, Bool $y) \{ $x && $y \}
 \k{multi} \k{sub} sum(Bool $x, Int $y) \{ sum($x , $y>0) \}
 \k{multi} \k{sub} sum(Int $x, Bool $y) \{ sum($y , $x) \}
\end{alltt}
has type
\begin{equation}\label{quattrobis}\iflmcs\else\nonumber\fi
\begin{array}{rcl}
&&(\pair{\Int}{\Int}\To\Int)\\[-.4mm]\iflmcs\else\nonumber\fi
&\And& (\pair{\Bool}{\Bool}\To\Bool)\\[-.4mm]
&\And& (\pair{\Bool}{\Int}\To\Bool)\\[-.4mm] \iflmcs\else\nonumber\fi
&\And& (\pair{\Int}{\Bool}\To\Bool).
\end{array}
\end{equation}

\smallskip\noindent
\emph{Solution:} This is a recursive definition. So we have to prove under the hypothesis that the recursion variable \p{sum} has type~\eqref{quattrobis} that the function has each type in the intersection. This means that we have to prove that if \p{sum} is applied to a pair of integers, then it returns an integer, and that if any of the two arguments is a Boolean then it returns a Boolean. This can be easily verified by checking the selection rules for multi subroutines. The delicate cases are the last two \p{multi} definitions since they contain a recursive call. Let us try to type the last definition (the other is similar), that is to deduce that if \p{\$x} has type \Int\ and \p{\$y} has type \Bool,  then the result of \p{sum} is of type \Bool. The type returned by this definition will be the type of the application \p{sum(\$y , \$x)}. By hypothesis \p{sum} has the type in~\eqref{quattrobis} and it is here applied to a first argument \p{\$y} of type \Bool and to a second argument \p{\$x} of type \Int. The type in~\eqref{quattrobis} tells us that when a function of this type is applied to argument in \p{(\Bool,\Int)}, then the function returns a result in \Bool\ (third arrow of the intersection in~\eqref{quattrobis}), which is the expected result.

\bigskip
\paragraph{[EX4] Problem:} Prove that the type in~\eqref{quattro} and in Footnote~\ref{ex4} are equivalent.

\smallskip\noindent
\emph{Solution:} The exercise in Footnote~\ref{extrarules} states that $\S\To\T\And\U$ is equivalent to   $(\S\To\U)\And(\T\To\U)$ for all types \S, \T, and \U. So in particular,  \texttt{\Bool\Or\Int\To\Bool} is equivalent to \texttt{(\Bool\To\Bool)\ \And\ (\Int\To\Bool)} from which the result follows.

\bigskip
\paragraph{[EX5] Problem:} find a function that is in~\eqref{ty0} and not in~\eqref{ty1}.

\smallskip\noindent
\emph{Solution:} a simple example is the constant function \p{1}:
\begin{alltt}
 sub (Int $x , Int $y) \{ 1 \}
\end{alltt}

\bigskip
\paragraph{[EX6] Problem:} Give a linear function to compute subtyping of product types.

\smallskip\noindent
\emph{Solution:}
Since an intersection of products is the product of the component-wise
intersections, then without loss of generality we can consider the
case in which the left hand side intersection of~\eqref{atomprod} is formed
by a single product. Then we have the following recursive definition for the product case of the subtyping relation:
\[
\begin{array}{l}
\displaystyle\pair{\S_1}{\S_2}\;\Sub\bigvee_{\spair{T}\in N} \!\!\!\!{\pair{\T_1}{\T_2}} \qquad\iff
\qquad \S_1\Sub\Empty\;\;\textsf{or}\;\;\S_2\Sub\Empty\;\;\textsf{or}\;\;\Phi(\S_1,\S_2,N)
\end{array}
\]
where
\[
\begin{array}{rcl}
\Phi(\S_1\;,\;\S_2\;,\;\varnothing)&  = & \texttt{false}\\[2mm]
\Phi(\S_1\;,\;\S_2\;,\;N\cup\lb\pair{\T_1}{\T_2}\rb) &= &
((\S_1\Sub\T_1)\ \textsf{ or }\ \Phi(\S_1\setminus\T_1\;,\;\S_2\;,\;N))\quad\textsf{and }\\
&&((\S_2\Sub\T_2)\ \textsf{ or }\ \Phi(\S_1\;,\;\S_2\setminus\T_2\;,\;N))
\end{array}
\]
The justification of the above definition and several possible
optimizations for this algorithm can be found in Section 7.3.1
of~\cite{AlainThesis}.

\smallskip
\paragraph{[EX7] Problem:} modify the definitions of $\Phi$ and $\Phi'$ so that they do not perform any check for~\eqref{arrowsubtyping} when $P'=P$.

\smallskip\noindent
\emph{Solution:} The case $P'=P$ corresponds to a call
$\Phi(\T_1\;,\;\T_2\;,\;P^\circ\;,\;P^+\;,\;P^-)$ in which both
$P^\circ$ and $P^-$ are empty, that is $P^+=P$. In the formulation given for~\eqref{bof} this corresponds to the case in which the last parameter of $\Phi$ was never modified and, therefore, is still \Any. Concretely, this corresponds to adding to the subsequent definition a further case of termination of the form
\[
\begin{array}{l}
\Phi(\T_1\;,\;\T_2\;,\;\varnothing\;,\;\D\;,\;\Any) = \texttt{true}
\end{array}
\]
or, equivalently, to modifying the termination case as follows
\[
\begin{array}{l}
\Phi(\T_1\;,\;\T_2\;,\;\varnothing\;,\;\D\;,\;\C) = (\Any\Sub\C)\textsf{ or } (\T_1\Sub \D)\textsf{ or } (\C\Sub \T_2)
\end{array}
\]
where the clauses are evaluated from left to right.

For what concerns $\Phi'$, the case for $P'= P$ corresponds to a call in which the second parameter was never modified (i.e., it still is the $\Not{\T_2}$ of the initial call in~\eqref{implem}) and the last parameter is empty. In order to verify the first condition, we may add an extra parameter to $\Phi'$ that indicates whether the second parameter was modified or not. If this is not case and we are at the end of the recursive calls (i.e., just one element remainins to be selected), then we have to perform just one of the two recursive calls, that is:
\[
\begin{array}{l}
\Phi'(\T_1\;,\;\T_2\;,\;\lb\S_1^\circ\To\S_2^\circ\rb) =\Phi'(\T_1\;,\;\T_2\And\S_2^\circ\;,\;\varnothing)
\end{array}
\]
As we see, this optimization is mildly interesting, since it just spares the check $\Phi'(\T_1\setminus\S_1^\circ\;,\;\T_2\;,\;\varnothing)$ (in practice, it avoids performing the difference $\T_1\setminus\S_1^\circ$) which is why it is not implemented by the compiler of \cduce.

\bigskip
\paragraph{[EX8] Problem:} prove that a disjunctive normal form of tags can always be expressed by either $(t_1\vee\ldots\vee t_n)$  or $\neg(t_1\vee\ldots\vee t_n)$.

\smallskip\noindent
\emph{Solution:} First of all notice that \Any\ can be represented as
$\neg()$ the negation of an empty union of tags. Next, take any
intersection $\bigwedge_{p\in P} t_p~\And~\bigwedge_{n\in
N} \Not{t_n}$. If $P$ is not a singleton then the intersection is
empty (the intersection of two distinct tags is always empty); if
there exist $p\in P$ and $n\in N$ such that $t_p=t_n$, then the
intersection is empty; if there exist $p\in P$ and for all $n\in N$ we
have $t_p\not=t_n$, then the intersection is $t_p$ (since it is
contained in all the negations of distinct tags). Therefore we are
left with just two cases: either $(i)$ $P$ is a singleton and $N$ is
empty or $(ii)$ $P$ is empty: in all the other cases the intersection
$\bigwedge_{p\in P} t_p~\And~\bigwedge_{n\in N} \Not{t_n}$ reduces
either to \Empty\ or to \Any. In conclusion any intersection in a
disjunctive normal form of tags is equivalent either to a single
positive tag or to an intersection of negated tags.  Now consider
again the latter case, that is when the intersection is just formed by
negated tags. The tags that are contained in $\bigwedge_{n\in
N} \Not{t_n}$ are exactly all the tag values, apart from all $t_n$ for
$n\in N$. That is, the set of tags in $\bigwedge_{n\in N} \Not{t_n}$
are exactly those in the co-finite union $\Not{\bigvee_{n\in N}
t_n}$. Now take any union formed only of such types (i.e., either a
single positive tag or a co-finite union of tags). Such a union, our
disjunctive normal form, is equivalent to either a finite union of
tags (when the disjunctive normal forms unites only positive tags) or
a co-finite union of tags (when the disjunctive normal form unites at
least one co-finite union: all the positive tags in the disjunctive
normal form are removed from it and all the other co-finite unions,
and the cofinite unions are merged into a single one).

\bigskip
\paragraph{[EX9] Problem:} define \texttt{norm}

\smallskip\noindent
\emph{Solution:} I present the solution without recursive types. These can be easily included by using references, for instance.
I use the notation \texttt{\{ $r$ with $\ell$ = $v$ \}} to denote the record in which the field $\ell$ contains $v$ and the remaining fields are as in the record $r$.
\begin{alltt}

let empty = \{
  tags = \ensuremath{\zero} ;              {\color{gray}// \textrm{\it or } (\lb\ \rb,\textrm{\it positive}) }
  ints = \ensuremath{\zero} ;              {\color{gray}// \textrm{\it or } (\lb\ \rb,\textrm{\it positive}) }
  prod = \ensuremath{\zero} ;
  arrw = \ensuremath{\zero} ;
\}

let any  =  \{
  tags = \ensuremath{\one} ;              {\color{gray}// \textrm{\it or } (\lb\ \rb,\textrm{\it negative}) }
  ints = \ensuremath{\one} ;              {\color{gray}// \textrm{\it or } (\lb\ \rb,\textrm{\it negative}) }
  prod = \ensuremath{\one} ;
  arrw = \ensuremath{\one} ;
\}

let norm = function
 | Empty -> empty
 | Any   -> any
 | \ensuremath{t}     \,-> \{ empty with tags = (\lb\ensuremath{t}\rb,\textrm{\it positive}) \}
 | [\ensuremath{i}..\ensuremath{j}]\,\,-> \{ empty with ints = (\lb[\ensuremath{i}..\ensuremath{j}]\rb,\textrm{\it positive}) \}
 | (S,T) -> \{ empty with prod = (S,T)?\one:\zero:\zero \}
 | S\To{}T -> \{ empty with arrw = S\To{}T?\one:\zero:\zero \}
 | S|T   -> (norm S)\ensuremath{\vee}(norm T)
 | S&T   -> (norm S)\ensuremath{\wedge}(norm T)
 | not(T)-> any\ensuremath{\setminus}(norm T)

\end{alltt}
According to the above definition, the types that form the atoms of BDDs are not normalized. An alternative solution is to store them already normalized, that is returning \texttt{(norm S,norm T)?\one:\zero:\zero} instead of \texttt{(S,T)?\one:\zero:\zero} (and similarly for arrows).

\bigskip
\paragraph{[EX10] Problem:} Show that the type system is unsound for lazy languages since the application of a well-typed function to a diverging expression of type \Empty{} is well typed.

\smallskip\noindent
\emph{Solution:} Consider the function  \p{\textbf{sub}}\verb| doublefirst(Int $x, Int $y){ x+x }|. This function is of type $\pair\Int\Int\To\Int$. By subtyping it has also the type $\Empty\To\Int$ (contravariance on the domain). So we can apply it to any argument of type $\Empty$. Now consider the pair \p{(True,$e$)} where $e$ is any diverging function of type $\Empty$. This pair has also type \Empty, since the product with an empty set is itself an empty set. Therefore, the application \p{doublefirst(true,$e$)} is well typed with type \Int. In a strict language this does not pose any problem since the evaluation of the argument will never terminate, and so will not the application. But in a lazy language this application reduces to \p{True+True}, thus yielding a type error at run-time (or, at least, it should: in Perl 6 it returns \p{2}).

\bigskip
\paragraph{[EX11] Problem:} Prove that by applying the \emph{Step 4} of the subtyping algorithm of Section~\ref{subalgo} to $\pair{\S_1}{\S_2}\And\Not{\pair{\T_1}{\T_2}}$ we obtain $(\S_1\Sub\Empty)\textsf{ or }(\S_2\Sub\Empty)\textsf{ or } (\S_1\Sub\T_1\textsf{ and }\S_2\Sub\T_2)$.

\smallskip\noindent
\emph{Solution:} We have to check two cases, that is for $N'=\varnothing$ and $N'=N$. These yield:
\[(\S_1\Sub\Empty\textsf{ or }\S_2\Sub\T_2)\textsf{ and
}(\S_1\Sub\T_1\textsf{ or }\S_2\Sub\Empty)\] By distributing the
``and'' we obtain:
\[\begin{array}{l}
(\S_1\Sub\Empty\textsf{ and }\S_1\Sub\T_1)\textsf{ or }\\
(\S_1\Sub\Empty\textsf{ and }\S_2\Sub\Empty)\textsf{ or }\\
(\S_2\Sub\T_2\textsf{ and }\S_2\Sub\Empty)\textsf{ or }\\
(\S_2\Sub\T_2\textsf{ and }\S_1\Sub\T_1)
\end{array}\]
By observing that $\S_1\Sub\Empty$ implies $\S_1\Sub\T_1$ and that $\S_2\Sub\Empty$ implies $\S_2\Sub\T_2$ we obtain
\[\begin{array}{l}
(\S_1\Sub\Empty)\textsf{ or }\\
(\S_1\Sub\Empty\textsf{ and }\S_2\Sub\Empty)\textsf{ or }\\
(\S_2\Sub\Empty)\textsf{ or }\\
(\S_2\Sub\T_2\textsf{ and }\S_1\Sub\T_1)
\end{array}\]
and since $(\S_1\Sub\Empty\textsf{ and }\S_2\Sub\Empty)$ implies both
$\S_1\Sub\Empty$ and $\S_2\Sub\Empty$ we can remove this literal from
the clause obtaining the result.

\bigskip
\paragraph{[EX12] Problem:} Prove that by applying the \emph{Step 4} of the subtyping algorithm of Section~\ref{subalgo} to ${\S_1}\To{\S_2}\And\Not{{\T_1}\To{\T_2}}$ we obtain $(\T_1\Sub\Empty)\textsf{ or } (\T_1\Sub\S_1\textsf{ and }\S_2\Sub\T_2)$

\smallskip\noindent
\emph{Solution:} We have check that two conditions are satisfied, namely the condition on the domains and the ``or'' for the case $P'=\varnothing$. These yield:
\[(\T_1\Sub\S_1)\textsf{ and }(\T_1\Sub\Empty \textsf{ or } \S_2\Sub\T_2)\]
By distributing the ``and'' we obtain
\[(\T_1\Sub\S_1\textsf{ and }\T_1\Sub\Empty)\textsf{ or } (\T_1\Sub\S_1\textsf{ and }\S_2\Sub\T_2)\]
By observing that $\T_1\Sub\Empty$ implies $\T_1\Sub\S_1$ we obtain the result.
